%% file: massen.tex
\documentclass[aps]{revtex4}
\usepackage{epsfig, amssymb, latexsym, amsfonts, amsmath, amsthm}

\newcommand{\tw}{ {\text{tw}} }

\newcommand{\twz}{ {\text{tw2}} }

\def \d{{\mathrm{d}}}
\newcommand{\h}{\mathbf{h}}
\newcommand{\U}{\mathbf{u}}

\newcommand{\D}{\mathbf{d}}
\newcommand{\pd}{ \partial }

\newcommand {\kln}[1]{\left( #1 \right)}
\newcommand {\Kln}[1]{\bigl( #1 \bigr)}
\newcommand {\KLn}[1]{\Bigl( #1 \Bigr)}

\newcommand {\kls}[1]{\left\{ #1 \right\}}

\newcommand {\KLs}[1]{\Bigl\{ #1 \Bigr\}}

\newcommand {\klso}[1]{\left\{ #1 \right.}

\newcommand {\okls}[1]{\left. #1 \right\}}

\newcommand {\kle}[1]{\left[ #1 \right]}
\newcommand {\Kle}[1]{\bigl[ #1 \bigr]}

\newcommand {\kleo}[1]{\left[ #1 \right.}

\newcommand {\okle}[1]{\left. #1 \right]}

\newcommand {\im}{{\text{i}}}
\newcommand {\e}{{\text{e}}}

\newcommand {\ra} {\rightarrow}

\begin{document}


\title{Target mass corrections for virtual Compton scattering at twist-2 and \\
generalized, non-forward Wandzura-Wilczek and Callan-Gross relations\\
\phantom{relation}}

\author{ Bodo Geyer }
\affiliation{ Institute of Theoretical Physics,
Leipzig University, Augustusplatz~10, D-04109~Leipzig, Germany}

\author{ Dieter Robaschik }
\affiliation{BTU Cottbus, Fakult\"at 1, Postfach 101344,
D-03013~Cottbus, Germany }

\author{ J\"org Eilers }
\affiliation{ Center for Theoretical Studies, Leipzig
University, Augustusplatz~10, D-04109~Leipzig, Germany}

\date{\today}

\begin{abstract}
\vspace*{.8cm}

\noindent Abstract: The off-cone Compton operator of twist-2 is
Fourier transformed using a general procedure which is applicable,
in principle, to any QCD tensor operator of definite (geometric)
twist. That method allows, after taking the non-forward matrix
elements, to separate quite effectively their imaginary part and
to reveal some hidden structure in terms of appropriately defined
variables, including generalized Nachtmann variables. In this way,
without using the equations of motion, generalizations of the
Wandzura-Wilzcek relation and of the mass-corrected Callan-Gross
relation to the non-forward scattering, having the same shape as
in the forward case, are obtained. In addition, new relations for
those structure functions which vanish in the forward case are
derived. All the structure functions are expressed in terms of
iterated generalized parton distributions of $n$-th order. In
addition, we showed that the absorptive part of twist-2 virtual
Compton amplitude is determined by the non-forward extensions of
$g_1, W_1$ and $W_2$ only.

\vspace*{0.5cm}


\end{abstract}

\maketitle

    \input{M_Intro3} %
    \input{M_Four3}  %
    \input{M_Tasym3}

\input{M_Tsym3}  %
    \input{M_Conclusions3}
    \input{M_Append3}   %

\input{M_Bib3}

\end{document}

%% file: M_Intro3.tex
 \section{Introduction}
\renewcommand{\theequation}{\thesection.\arabic{equation}}
\setcounter{equation}{0}

 The Compton amplitude for the scattering of a virtual photon off a hadron,
$ \gamma^*_1 + P_1 \rightarrow \gamma_2^{'*} + P_2, $ provides one
of the basic tools to understand the short--distance behavior of
the nucleon and to test Quantum Chromodynamics (QCD) at large
space--like virtualities. In that kinematic regime, usually
denoted as generalized Bjorken region, the Compton amplitude is
dominated by the singularities on the light--cone and the Compton
amplitude factorizes into universal, non-perturbative distribution
amplitudes (DA) -- commonly denoted also as generalized parton
distributions (GPD) -- and perturbative hard scattering amplitudes
of the participating partons. The DA's parametrize the matrix
elements of appropriate bilocal quark-antiquark, gluon-gluon as
well as non-local higher order light-ray operators. For deep
inelastic lepton-hadron scattering (DIS) the parton distributions,
modulo kinematical factors, are given as forward matrix elements
of non-local light-ray operators occuring in the light-cone
expansion of the time-ordered product of appropriate hadronic
currents, and for (deeply) virtual Compton scattering (DVCS) and
meson production the generalized distribution amplitudes,
respectively, are given by corresponding non-forward matrix
elements. Thereby, various processes are governed by one and the
same (set of) light-ray operators.

There exist various different approaches to attack the problem of
determining the virtual Compton amplitude in leading as well as
beyond leading order. Usually, the light--cone expansion, either
the local one \cite{BPF} or the non-local one
\cite{AS,BR80,K83,BB89}, is applied
\cite{LeiVCS,XJ,RAD,BGR99,BR00,BEGR02}; for a review see,
e.g.,~\cite{MRGHD,VGG99,BKM,D03}. Here, we prefer the approach
using the non-local light-cone expansion. That approach also has
been used for the treatment of diffractive processes \cite{BR01}.
Differently, also the method of light-cone quantization
\cite{KS70,BL89,LCQ} has been used,~e.g.,~in DIS \cite{BL}, DVCS
\cite{BDH} and large-angle Compton scattering \cite{DFJK01}.

Besides the leading contributions, being determined in the early
days of QCD \cite{AP}, there occur corrections from operators of
higher (geometric) twist and/or containing more and more primary
fields, from the radiative corrections of these operators as well
as taking into consideration transverse (parton) momenta
\cite{MUL}. The main concern of considering these sub-leading
contributions is to determine more accurately the various
phenomenological (multi-variable) distribution amplitudes and
their $Q^2-$evolution in a well-defined quantum field theoretical
frame, to study their possible interrelations and to determine the
influence of target mass and, eventually, quark mass corrections.

Concerning higher twist, there are to be distinguished not only
different approaches,~e.g.,~local and non-local ones, but also
different definitions. The notion of (geometric) twist $\tau = $
(scale) dimension $-$ (Lorentz) spin has been introduced first by
Gross and Treiman for the local light-cone operators \cite{GT71};
accordingly it has a group theoretical meaning being related to
the (collinear) conformal group \cite{OHR}. From this one has to
distinguish the notion of dynamical twist $t$, being related to
the introduction of `good' and `bad' components of quark spinors
in the infinite momentum frame \cite{KS70}, and counting, in fact,
the inverse powers $(M/Q)^{t-2}$ appearing in the phenomenological
distribution functions or amplitudes \cite{JJ91}. Whereas
dynamical twist can be related to matrix elements only geometric
twist is directly related to the operators and has an invariant
meaning.

The decomposition of {\it non-local} light-ray operators (finite
on-cone decomposition) with respect to that geometric twist, as
far as they are relevant for hadronic processes, have been given
in Refs.~\cite{GLR99,GL00} (for earlier partial results, see,
\cite{BB89}); the extension to non-local operators off the
light-cone (infinite off-cone decomposition) has been considered
recently in Refs.~\cite{GLR01,LAZAR} and quite generally in
Ref.~\cite{E04}. Loosely speaking, non-local operators of definite
geometric twist are determined by towers of local operators
spanning irreducible tensor representations of the Lorentz group
which are characterized by well-defined symmetry type (under index
permutations of the traceless tensors). The various symmetry types
are determined by corresponding Young patterns having, in the case
of the Lorentz group, at most three lines, $({\underline m})=
(m_1, m_2, m_3)$. For operators which locally are characterized by
totally symmetric traceless tensors, there exists a well-defined
group theoretical framework \cite{BT77} which works on the
light-cone as well as, by unique harmonic extension, also off the
light-cone. The off-cone operators of definite twist are given for
the local as well as the non-local case. Thereby, local operators
of given $n$ are represented by (a finite series of) Gegenbauer
polynomials $C^\nu_n(z),\, \nu \geq 1$, whereas the non-local
operators, being obtained by re-summation with respect to $n$, are
represented by (a related series of) Bessel functions
$J_{\nu-\frac{1}{2}}(z)$.

In the case of DIS, i.e., forward matrix elements on the
light-cone, it has been shown that there exist one-to-one
relations between the quark distributions functions of
well-defined geometric twist and those of dynamical twist
\cite{GL01}; analogous relations also hold for the meson
distribution functions \cite{LB}. From them, without using the
equations of motion, the well-known Wandzura-Wilczek (WW)
relations \cite{WW77} between the distribution functions $g_2$ and
$g_1$ were derived together with additional ones for the
distribution functions $g_3, f_4, h_L$ and $h_3$; analogous
relations occur also in the case of $\rho-$meson wave functions.
Concerning the case of non-forward matrix elements a first
derivation of a generalized WW- and CG-relation was given in
Ref.~\cite{BR00}. Later on, the discussion of generalized
WW-relations led to different points of view concerning the use of
equations of motion and current conservation \cite{dWW,RW}. In
this paper we are able to show how the generalizations of the
WW-relation as well as of the Callan-Gross(CG) relation
\cite{CG69} occur in case of non-forward matrix elements of the
Compton operator. And this comes out without the the use of the
equations of motion and fully taking into account the target mass
contributions.

The influence of target masses in case of forward scattering has
been studied for unpolarized DIS \cite{N73, GP76} as well as
polarized DIS \cite{BE76,W77,MU80,KU95}. A short review of these
attempts was given in Ref.~\cite{GRW79} containing also the not so
well-known related exact result about the representation of a
causal, Lorentz invariant structure function $W(x,p)$ as globally
convergent series of Gegenbauer polynomials \cite{RTW78} using the
Jost-Lehmann-Dyson representation \cite{DJL}.

Using the local operator product expansion Georgi and Politzer
\cite{GP76} showed precocious $\xi-$scaling in deep inelastic
electron-proton scattering, where $\xi = {2x}/[{1 + \sqrt{1 + 4
x^2 M^2/Q^2}}]$ is the well-known Nachtmann variable with $x\equiv
x_{\rm Bj}$ being the Bjorken variable and $M$ the target (proton)
mass; for the structure functions $W_1(Q^2,x)$ and $\nu
W_2(Q^2,x)/M$ they obtained the following connection with the
quark distribution function $F(\xi)$:
\begin{align}
\label{WW0} 2x W_1 &=  \frac{x^2}{(1 + 4 x^2 M^2/Q^2)^{1/2}}F(\xi)
        +\frac{2x^3 M^2/Q^2}{1 + 4 x^2 M^2/Q^2}\!\int_\xi^1\! d\xi' F(\xi')
        +\frac{4x^4 M^4/Q^4}{(1 + 4 x^2 M^2/Q^2)^{3/2}}\!\int_\xi^1\! d\xi'
         \!\int_{\xi'}^1\! d\xi''F(\xi'')\,,
\nonumber\\
\frac{\nu}{M} W_2 &=  \frac{x^2}{(1 + 4 x^2 M^2/Q^2)^{1/2}}F(\xi)
        +\frac{6x^3 M^2/Q^2}{1 + 4 x^2 M^2/Q^2}\!\int_\xi^1\! d\xi' F(\xi')
        +\frac{12x^4 M^4/Q^4}{(1 + 4 x^2 M^2/Q^2)^{3/2}}\!\int_\xi^1\! d\xi'
         \!\int_{\xi'}^1\! d\xi''F(\xi'')\,;
\end{align}
and similarly also for $W_1(Q^2,x)$ in neutrino scattering. That
method has been extended later on for studying the target mass
contributions in polarized DIS due to different twist
\cite{PR98,BT99}. In the latter work, analogous relations have
been found for the structure functions $G_1(Q^2,x), \ldots,
G_5(Q^2,x)$, especially there holds the following:
\begin{align}
\label{CG0} G_1 &=  \frac{x/\xi}{(1 + 4 x^2 M^2/Q^2)^{3/2}}
 \left[{\hat F}(\xi)
 +\frac{x(x+\xi) M^2/Q^2}{(1+4 x^2 M^2/Q^2)^{1/2}}
 \!\int_\xi^1\!d\xi'{\hat F}(\xi')
 - \frac{x\xi M^2}{2Q^2}
 \frac{2 - M^2/Q^2}{1 + 4 x^2 M^2/Q^2}\!\int_\xi^1\! d\xi'
         \!\int_{\xi'}^1\! d\xi''{\hat F}(\xi'')\right],
\nonumber\\
G_2 &=  \frac{- x/\xi}{(1 + 4 x^2 M^2/Q^2)^{3/2}} \left[{\hat
F}(\xi)
 -\frac{1 - x\xi M^2/Q^2}{(1 + 4 x^2 M^2/Q^2)^{1/2}}
 \!\int_\xi^1\! d\xi' {\hat F}(\xi')
 - \frac{3}{2}
 \frac{x\xi M^2/Q^2}{1 + 4 x^2 M^2/Q^2}\!\int_\xi^1\! d\xi'
         \!\int_{\xi'}^1\!d\xi''{\hat F}(\xi'')\right],
\end{align}
leading to the well-known WW-relations and extending the
CG-relation. However, that method is tailored to the forward case
and cannot be extended to the non-forward one.

Quite differently, the group theoretical method for the
determination of target mass corrections using harmonic scalar
operators of definite spin and the corresponding matrix elements
in terms of Gegenbauer polynomials has been introduced for the
first by Nachtmann~\cite{N73} and continued by Baluni and Eichten
\cite{BE76}. Remarkably, this method can be generalized for the
consideration of target mass resp. power corrections to virtual
Compton scattering in non-forward case \cite{BM01,GLR01}. Thereby,
in order to get information about the target mass contributions
one is forced to consider the (geometric) twist decomposition
off-cone, taking into account all the trace terms leading to power
suppressed expressions depending on $M^2/Q^2$.

Concerning the twist-2 part Belitsky and M{\"u}ller \cite{BM01}
were able to completely sum up the mass corrections to a closed
form in terms of (di)logarithms depending on unique variable
${\cal M}^2$ approaching $4 x^2 M^2/Q^2$ in the forward case. On
the other hand, in the previous work \cite{GLR01} it has been
shown how the well-known twist decomposition of non-local scalar
off-cone operators according to the method of harmonic extension
can
 be used for the non-local vector and skew-symmetric tensor quark-antiquark
 operators, leading to power corrections of the related off-forward
double distributions as well as the vector meson distribution
amplitudes in $x-$space. But, the application to virtual Compton
scattering which requires to carry-out the Fourier transformation
of the operator matrix elements times the coefficient functions
(in Born approximation), due to its complexity, remained open.

In a previous consideration a straightforward procedure of
performing the Fourier transformation has been presented for the
off-cone matrix elements of scalar non-local operators
\cite{EG02}. Here, we extend it to the case of (axial) vector
operators appearing as part of the Compton operator. Despite it is
already known for a long time that current conservation in
non-forward case can be arrived in a sufficient approximate way
only when non-leading corrections, specially contributions of
appropriate twist-3 operators are included \cite{BR00,RW,BM01}, we
nevertheless think that it is important to consider the full
content of the twist-2 contributions separately as a first step.
Therefore, like in Ref.~\cite{BM01}, we consider the contributions
of the twist-2 operators only.

The straightforward procedure of Ref.~\cite{BM01} used a
representation of the matrix elements of the non-local twist-2
operator by an infinite sum of matrix elements of {\em local}
twist-2 operators, transformed each term separately and finally
re-summed these expressions.
Here, we proceed in a similar manner. However, the Fourier
transform of the Compton operator is performed first and
afterwards the matrix elements are taken. In the case of scalar
non-local operators of definite twist the general procedure for
determining their Fourier transform makes use of the relation
between Bessel functions and Gegenbauer polynomials \cite{EG02}.
Now, that procedure which, in principle, applies to arbitrary
tensor operators of definite twist, is applied to the case of
vector operators (Appendix A). Since the existence and convergence
of the corresponding re-summations is by no means obvious we
checked our calculation in case of the trace part of the Compton
operator by Fourier transforming its closed non-local expression
directly (Appendix B).

In fact, we re-obtained the result of Ref.~\cite{BM01} in a
straightforward manner, also closing some gaps in their
representation. But, more interesting than this is the fact, that
our procedure leads to a much deeper insight into the structure of
the virtual Compton amplitude which, especially, allows for a
straightforward generalization of the Wandzura-Wilczek relation
and of the Callan-Gross relation to the non-forward case. More
precisely, introducing appropriate variables $\xi_\pm,\zeta$, we
are able derive non-forward generalizations of $W_1, W_2$ and
$G_1, G_2$ in terms of these variables which obey relations of the
same functional dependence as their counterparts in forward case
above! Therefore, they constitute non-forward generalizations of
the WW- and CG-relations. In addition, new generalized
distribution amplitudes and corresponding relations occur which
disappear in the forward case. Here, we should emphasize again
that these results are obtained without using any dynamical input
but taking into account only the correct twist-2 representation of
the Compton operator off the light-cone.

 The paper is organized as follows. In Sec.~2 we
remember some basic formulae concerning Born term contributions
and definitions of operators. In Sec.~3 we Fourier transform the
Compton operator using the full off-cone content of the twist-2
light-cone operator. The calculation is outlined for the trace
part and the final results are presented for the symmetric and
antisymmetric part of the Compton operator separately. The
corresponding calculations are shifted into Appendices C and D. In
Secs.~4 to 6 we investigate the matrix elements for the trace
part, the antisymmetric and the symmetric part of the Compton
operator. Thereby, we introduce a general kinematic decomposition
of the matrix elements. For each kinematic structure appears one
basic generalized parton distribution function. We insert these
findings into the partial results of Sec.~3. Thereby the remaining
auxiliary parameter integrations can be managed in different ways.
We perform it in such a way that we obtain a representation of the
Compton amplitude in terms of iterated generalized distribution
functions which relay on the basic generalized parton distribution
functions. Such representations are already known for the case of
forward Compton scattering by \cite{GP76}. Moreover, these
representations allow in a very simple manner the extraction of
the absorptive part. At this stage it is possible to compare our
results with the results obtained for the forward case in
Refs.~\cite{GP76} and \cite{BT99}. Finally we look for
generalizations of the Wandzura-Wilczek and the Callan-Gross
relations to non-forward scattering together with additional
relations concerning those structure functions which vanish in the
case of forward scattering.


\section{Leading contribution of the virtual Compton Amplitude}
\renewcommand{\theequation}{\thesection.\arabic{equation}}
\setcounter{equation}{0}

The virtual Compton amplitude is given by
\begin{eqnarray}
\label{CA_nonf}
 T_{\mu\nu}(P_i,q; S_i)
 = \im \int \d^4 \! x \; \e^{\im qx}
 \langle P_2,S_2|RT [J_\mu(x/2)J_\nu(-x/2)\,\mathcal{S}]|P_1,S_1\rangle \; ,
\end{eqnarray}
where $P_1 (P_2)$ and $S_1 (S_2)$ are the momenta and spins of the
incoming (outgoing) hadrons, $q=(q_2+q_1)/2$,  $p_+ =P_1 + P_2$
and $p_- = P_2 - P_1 = q_1 - q_2\, $ denotes the momentum
transfer; $R$ denotes the renormalization procedure and $\cal S$
is the (renormalized) $S-$matrix.
Here, we consider the Compton amplitude in the generalized Bjorken
region,
\begin{eqnarray}
\nu = q p_+ \longrightarrow \infty,
 \qquad
 Q^2 = - q^2 \longrightarrow \infty,
\end{eqnarray}
with the variables
\begin{eqnarray}
 x = \frac{Q^2}{qp_+}
 \quad {\rm and} \quad
 \eta = \frac{qp_-}{qp_+}
\end{eqnarray}
keeping fixed.

Obviously, the Compton amplitude appears as Fourier transform of
the non-forward matrix elements of the Compton operator,
\begin{eqnarray}
\label{int_input}
 \widehat T_{\mu\nu}(x)
 &\equiv&
 R\,T \kle{J_\mu\kln{\frac{x}{2}}\,J_\nu\kln{-\frac{x}{2}}{\cal S}} \,.
\end{eqnarray}
In the Bjorken region, the physical processes are determined by
the singularities of the Compton operator on the light-cone and,
therefore, the operator product expansion can be applied. A
general study of it, using the non-local light-cone expansion
\cite{AS,MRGHD}, has been given, e.g., in
Refs.~\cite{BGR99,GLR01}, cf.~also~Refs.~\cite{XJ,RAD}. As is
well-known, in leading order the Compton operator simply reads
\begin{eqnarray}
\label{CA_nonf1}
 \widehat T_{\mu\nu}(x)
 &\approx&
 \frac{1}{2\,\pi^2\kln{ x^2 - \im \epsilon}^2}\,
 \KLn{{S_{\mu\nu|}}^{\alpha\beta}\,
 x_\alpha\, {\bf O}_\beta\kln{\kappa x, -\kappa x}
 -\im\,{\epsilon_{\mu\nu}}^{\alpha\beta}
 x_\alpha\, {\bf O}_{5\,\beta}\kln{\kappa x, -\kappa x}} \,,
\end{eqnarray}
where the tensor $S_{\mu\nu|\alpha\beta} \equiv
S_{\mu\alpha\nu\beta} = g_{\mu\alpha} g_{\nu\beta} + g_{\mu\beta}
g_{\nu\alpha} -
 g_{\mu\nu} g_{\alpha\beta} $ is symmetric in $\mu\nu$ and $\alpha\beta$
and $\kappa = 1/2$. This expression corresponds to the handbag
diagram; contributions from the quark masses and higher order
terms, e.g., four-quark and quark-gluon operators, are neglected.
In order to omit the subtleties of operator mixing, the
non-singlet case is taken and the flavor structure has been
suppressed.

The hermitean, chiral-even vector and axial vector operators,
 ${\cal O}_{\alpha}(\kappa_1 x,\kappa_2 x)$ and
 ${\cal O}_{5\,\alpha}(\kappa_1 x,\kappa_2 x)$,
respectively, at the first, are renormalized on the light-cone,
${\tilde x}^2 = 0$, \cite{AS,BR80,K83} and then, in order to allow for an appropriate
target mass expansion, extended off-cone by replacing
$\tilde x \rightarrow x$. They are given by
\begin{eqnarray}
\label{str_oprab}
 {\bf O}_{\alpha}(\kappa_1 x,\kappa_2 x)
 &=&
 \im \big(O_{\alpha}(\kappa_1 x,\kappa_2 x)
 - O_{\alpha}(\kappa_2 x,\kappa_1 x)\big),
 \\
\label{str_oprbb}
 {\bf O}_{5\,\alpha}(\kappa_1 x,\kappa_2 x)
 &=&
 O_{5\,\alpha}(\kappa_1 x,\kappa_2 x)
 +
 O_{5\,\alpha}(\kappa_2 x,\kappa_1 x),
 \end{eqnarray}
 with
\begin{eqnarray}
 O_{\alpha}(\kappa_1 x,\kappa_2 x)
 &:\,=&
 R \, T \Kle{:\bar\psi\kln{\kappa_1 \tilde x} \gamma_\alpha\,
 U\!\kln{\kappa_1 \tilde x,\kappa_2 \tilde x}\,
 \psi\kln{\kappa_2 \tilde x}\!:{\cal S}}\Big|_{\tilde x \rightarrow x},
 \\
 O_{5\,\alpha}(\kappa_1 x,\kappa_2 x)
 &:\,=&
 R \, T \Kle{:\bar\psi\kln{\kappa_1 \tilde x}  \gamma_5 \gamma_\alpha\,
 U\!\kln{\kappa_1 \tilde x,\kappa_2 \tilde x}\,
 \psi\kln{\kappa_2 \tilde x}\!:{\cal S}}\Big|_{\tilde x \rightarrow x}.
 \end{eqnarray}
The path-ordered phase factors, which are given by
\begin{eqnarray}
U\kln{\kappa_1 x,\kappa_2 x}
 := {\cal P} \exp\left\{- \im g
 \int_{\kappa_2}^{\kappa_1} \d\tau\, x^\mu A_\mu(\tau x)\right\}\,,
 \nonumber
\end{eqnarray}
in the following will be omitted, i.e., the Schwinger-Fock gauge,
$x^\mu A_\mu(x)=0$, will be assumed. The Compton operators
${\bf O}_{(5)\beta}\!\kln{\kappa x,-\kappa x}$
are introduced such that their matrix elements lead to real
(analytic) distribution amplitudes. Occasionally, they
will be considered in their unsymmetrized form
 $O_{(5)\beta}\!\kln{\kappa x,-\kappa x}$; their symmetric and
 antisymmetric form is re-obtained by observing their behavior
under the exchange $\kappa \ra - \kappa$.

The bi-local off-cone vector operators, when decomposed w.r.t.
their geometric twist, contain contributions from all the twists
$\tau = 2, 3, \ldots \infty$. Let us emphasize that for the
procedure of twist decomposition \cite{GLR99,GL00,LAZAR,E04} only
the tensorial structure of the operators is essential and not
their behavior under renormalization. The vector operators of
definite geometric twist already have been determined in
$x$--space \cite{EG03}. Now, their Fourier transform has to be
determined. However, in order to present the straightforward
procedure of determining the Fourier transform of such QCD
operators, which has been introduced for the first time in the
case of scalar operators \cite{EG02}, we investigate the
contribution of the twist-2 parts only. In fact, this also is the
most interesting contribution.

%% file: M_Four3.tex
\section{Fourier transform of the twist-2 vector operator}
\renewcommand{\theequation}{\thesection.\arabic{equation}}
\setcounter{equation}{0}

In this Section, according to Eqs.~(\ref{CA_nonf}),
(\ref{int_input}) and (\ref{CA_nonf1}), we perform a Fourier
transformation of the non-local vector and axial vector operator
of leading twist-2, $O^{\rm tw 2}_{\beta} \kln{ \kappa x , -\kappa
x }$ and $O^{\rm tw 2}_{5\beta} \kln{ \kappa x , -\kappa x }$,
multiplied with the leading part ${x_\alpha}/\left[2\pi^2(x^2 -\im
\epsilon)^{2}\right]$ of the propagator. This will be done by the
same general procedure as has been introduced for scalar operators
in Ref.~\cite{EG02}.
 It essentially uses the relation between the Fourier transform of
 $(ux)^n/\left[2\pi^2(x^2 -\im \epsilon)^{2}\right]$
and Gegenbauer polynomials in the variable $z=(uq)/\sqrt{u^2 q^2}$,
$u$ being any vector, together with the fact that (local) tensor
operators of definite (geometric) twist $\tau$ are given in terms
of  harmonic tensor polynomials. Let us emphasize that this
method, in principle, applies to tensor {\em operators} of any
twist. Matrix elements, forward or non-forward ones, are to be
taken after Fourier transformation according to the general
description introduced in Ref.~\cite{GLR01}.

Below, we first briefly review the general procedure and apply it
to the trace part of the Compton operator, already adopted to the
case under consideration, and, afterwards, we present the results
for the QCD vector operators (\ref{str_oprab}) and
(\ref{str_oprbb}); the explicit computations together with the
necessary prerequisites are postponed to Appendix A. After taking
non-forward matrix elements and performing the remaining
intermediate integrations we are able to confirm the results of
Ref.~\cite{BM01} in a straightforward, less artistic manner. In
addition, we have proven that this result which is obtained via
re-summation of the local expressions may be obtained also
directly by the Fourier transformation of matrix elements of the
corresponding {\it non-local} operators; a sample calculation for the
trace of the Compton amplitude is given in Appendix B.

\subsection{The twist-2 vector operators and general definitions}

In order to determine power corrections, e.g., the target mass
corrections, to hard, light-cone dominated hadronic processes one
has to consider the twist decomposition of the corresponding QCD
operators off-cone \cite{GLR01} which, for non-local operators, is
infinite. For scalar operators this decomposition is well-known
\cite{BB89} (cf., also \cite{GLR99,GLR01}), in the case of vector
operators it has been given recently \cite{EG03} and in the
general case of tensor operators it is under consideration
\cite{E04}. However, when restricting onto the light-cone, $x
\rightarrow \tilde x, {\tilde x}^2 =0$, the twist decomposition is
finite for all the non-local QCD-operators (see, e.g.,
Ref.~\cite{GLR99}). Especially, for scalar operators only the
leading twist occurs, i.e., $\tilde x^\mu O_{(5)\mu
}\kln{\kappa\tilde x, -\kappa\tilde x} = \tilde x^\mu O^{\rm tw
2}_{(5)\mu} \kln{\kappa\tilde x,-\kappa\tilde x}$, while higher
twists occur for vector and tensor operators \cite{GL00}.

Here, we consider only the leading non-local twist-2 vector
operators off-cone, which are given as follows (cf., \cite{GLR01},
Eq.~(4.25)):
\begin{eqnarray}
\label{NLO}
 O^{\rm tw 2}_{(5)\alpha} \kln{\kappa x,-\kappa x}
 &=&
 \pd_\alpha \int_0^1 dt
 \int \D^4 u\, O_{(5)\mu}(u) \left\{x^\mu (2 + \kappa\pd_\kappa)
  - \hbox{\large$\frac{1}{2}$}\im\kappa t\, u^\mu x^2\right\}
 (3 + \kappa\pd_\kappa) \mathcal{H}_2 (u, \kappa t x)
\end{eqnarray}
with
\begin{eqnarray}
\mathcal{H}_\nu (u, x) = \sqrt{\pi} \,
\Big(\sqrt{(ux)^2 - u^2 x^2}\Big)^{1/2-\nu}
J_{\nu-1/2}\Big(\hbox{$\frac{1}{2}$}
\sqrt{(ux)^2-u^2 x^2}\Big)
e^{\im (xu)/2 }\,,
\end{eqnarray}
whose local components are
\begin{eqnarray}
 O_{(5)\alpha\ n}^{\tw 2}\kln{x}
&=&
 \frac{1}{n+1}\,\partial^x_\alpha\left[ H_{n+1} \kln{x^2, \Box_x}
 \int \D^4u\, O_{(5)\mu n}\kln{u}x^\mu \,(xu)^n\right]\,,
\end{eqnarray}
with the operator
\begin{equation}
\label{def_Hn}
 H_n\kln{x^2, \Box_x} = \sum_{k=0}^{\kle{ \frac{n}{2} }}
 \frac{ \kln{-1}^k \, \kln{n-k}! }{ 4^k \, k! \, n! } \kln{x^2}^k \Box_x^k
\end{equation}
 projecting any (scalar) homogeneous polynomial of degree $n$ onto
 traceless scalar harmonics of degree $n$.
In these relations a (formal) Fourier transformation of the
non-local operators followed by an expansion into local operators
has been used which are defined as follows:
\begin{eqnarray}
\label{O^Gint} O_{(5)\mu}(\kappa x, -\kappa x) &=& \int \D^4 u
\,O_{(5)\mu}(u)\, \e^{i\kappa xu} =\sum_{n=0}^\infty \frac{(\im
\kappa)^n}{n!}\,O_{(5)\mu\, n}(x),
\\
\label{O^Gloc} O_{(5)\mu\,n}(x)&=&
 \int \D^4 u \,O_{(5)\mu}(u)\, (xu)^n
  = (-\im)^n \frac{\pd^n}{\pd\kappa^n}
  O_{(5)\mu}(\kappa x, -\kappa x)\big|_{\kappa=0}.
\end{eqnarray}
Here and in the following, for simplicity of notation, we
understand the factor $1/(2\pi)^4$ to be included into the measure
of the Fourier transformation, $\D^4 u \equiv \d^4 u / (2\pi)^4 $.

Now, because of the symmetry of $(ux)^n$, leading to the
equality,
\begin{eqnarray}
\partial^x_\alpha \, H_{n+1} \kln{x^2, \Box_x} \; x_\mu (xu)^n
=
\partial^u_\mu \, H_{n+1} \kln{u^2, \Box_u} \; u_\alpha (xu)^n\,,
\label{EQU}
\end{eqnarray}
according to Eqs. (\ref{CA_nonf}) and (\ref{CA_nonf1}) we have to
determine the Fourier transforms of
${x_\alpha}(xu)^n/\left[2\pi^2(x^2 - \im \epsilon)^{2}\right]$
only.

As shown in Appendix A, it holds
\begin{eqnarray}
 \int \frac{\d^4\!x}{\pi^2} \; \e^{\im q x} \; \frac{\kln{ux}^n}{\kln{x^2-\im\epsilon}^2}
&=&
 -  \, \im^{n+1} \; n! \;\; \h_n^{0}\kln{u,q},
\label{f1}\\
 \int \frac{\d^4\!x}{\pi^2} \; \e^{\im q x} \; \frac{x_\alpha  \kln{ux}^n}{\kln{x^2-\im\epsilon}^2}
&=&
 \, \im^{n} \; n! \; \kln{q_\alpha \; \h_n^{1}\kln{u,q} - u_\alpha \; \h_{n-1}^{1}\kln{u,q} },
 \label{f2}
\end{eqnarray}
where the following set of functions has been introduced:
\begin{alignat}{2}
\label{h0}
 \h_n^\nu\kln{u,q}
&:=
 \frac{2^{n+\nu} \, \Gamma\kln{\nu}}{\kln{q^2 + \im\epsilon}^{n+\nu}} \;
 \kln{\hbox{$\frac{1}{2}$}\sqrt{q^2 u^2}}^n \;
 C_n^{\nu}\bigg(\frac{uq}{\sqrt{q^2 u^2}}\bigg)
 \qquad
& \text{for}& \qquad n \geq 0, \quad 
\kln{\nu,n} \, \neq \kln{0,0}\,,
\\
 \h^0_0\kln{q}
&:=
 - \ln\kln{q^2/q^2_0}
 \phantom{\frac{2^{n+\nu} \, \Gamma\kln{\nu}}{\kln{q^2 + \im\epsilon}^{n+\nu}}}
 &
 \text{for}&
 \qquad \kln{\nu,n} = \kln{0,0}\,,
\label{h00}\\
\h_n^\nu\kln{u,q} &:=
 0 &
 \text{for}& \qquad  n < 0\,;
 \label{hneg}
\end{alignat}
here, $q_0$ is some (physically irrelevant) reference momentum and
$C^\nu_n\kln{z}$ are the Gegenbauer polynomials (cf.,
Ref.~\cite{PBM}, Appendix II.11) which may be introduced according
to (cf.,~Ref.~\cite{PBM}, Eq.~5.13.1.1)
\begin{eqnarray}
\label{GBdef}
 \sum_{n=0}^\infty t^n C_n^\nu(z) = (1-2zt + t^2)^{-\nu}
  \qquad\qquad {\rm for} \qquad |t| < 1.
\end{eqnarray}

These functions obey the following relations (see also Eqs.~(\ref{h1})
-- (\ref{u7})):
\begin{eqnarray}
 H_n\kln{u^2,\Box_u} \, \h^0_n\kln{u,q}
&=&
 \frac{q^2}{2 \, n} \, \h^1_n\kln{u,q} \qquad\qquad\qquad {\rm for}\qquad
 n\geq 1\,,
\label{h3}
\\
 H_n\kln{u^2,\Box_u} \, \h_n^1\kln{u,q}
&=&
 \h_n^1\kln{u,q}\,,
\label{h4}
\\
 \sum_{n=0}^\infty \kln{\pm\kappa}^n \, \h_n^\nu\kln{u,q}
&=&
 2^\nu \, \Gamma\kln{\nu}
 \left[\kln{q \mp \kappa \, u}^2  + \im \epsilon\right]^{-\nu}\,.
 \label{h7}
 \end{eqnarray}
The last relation is an immediate consequence of (\ref{GBdef})
thereby having in mind that $|q^2|$ is large, $Q^2 \ra \infty$.

Let us also note another property of the projection operator
$H_{n}\kln{u^2,\Box_u}$ which makes it useful to introduce the operators
$ \U^n_\alpha(u, \partial^u)$ (already being used by Baluni and Eichten \cite{BE76}), namely
\begin{eqnarray}
 H_{n}\kln{u^2,\Box_u} \, u_{\alpha}
&=&
 \U^{n}_{\alpha}\,
 H_{n-1}\kln{u^2,\Box_u}\,,
\label{u3}\\
\text{with}\qquad \U^n_\alpha(u, \partial^u) &\equiv&
 u_\alpha - \frac{u^2}{2 \, n} \, \partial_\alpha^u\,.
\label{u1}
\end{eqnarray}

\subsection{Fourier transform of the trace part}

Now, let us demonstrate how the procedure works in case of the trace
part of the Compton operator \cite{EG02}. First, introducing the local
operators and using relation (\ref{EQU}) one obtains
\begin{eqnarray}
\hat T^{\twz}_{\rm trace}(q)
&=& -\,2\im
 \int \frac{\d^4 \! x}{2\pi^2} \; \e^{\im qx} \frac{1}{\kln{x^2 - \im\epsilon}^2} \;
 x^\alpha \,\left(O^\twz_\alpha\kln{\kappa x, - \kappa x} -
 {O}^\twz_\alpha\kln{-\kappa x,  \kappa x}\right)
\nonumber\\
&=& -\,2\im \sum_{n=0}^\infty \frac{\kln{\im\kappa}^n
(1-(-1)^n)}{\kln{n+1}!} \int \D^4 \! u \;
 O_\alpha\!\kln{u}\; \partial^\alpha_u \;
H_{n+1}\kln{u^2,\Box_u} \int \frac{\d^4 \! x}{2\pi^2} \;
\e^{\im qx} \frac{\kln{ux}^{n+1}}{\kln{x^2 - \im\epsilon}^2}\,,
\nonumber
\end{eqnarray}
then, relations (\ref{f1}) and (\ref{h3}) together with
$ \partial_\alpha^u \h_n^\nu =  q_\alpha \h_{n-1}^{\nu+1}
- u_\alpha \h_{n-2}^{\nu+1}$, cf.,~Eq.~(\ref{h1}), are used to get
\begin{eqnarray}
\hat T^{\twz}_{\rm trace}(q)
&=&
 -\, \im \,\sum_{n=0}^\infty \left(\kln{-\kappa}^n-\kappa^n\right)
 \int \D^4 u \;
O_\alpha\!\kln{u} \; \partial^\alpha_u \; H_{n+1}\kln{u^2,\Box_u}
\; \mathbf{h}_{n+1}^0\kln{u,q}
\nonumber\\
&=&
 -\, \im\, \frac{q^2}{2} \sum_{n=0}^\infty
 \frac{\left(\kln{-\kappa}^n-\kappa^n\right)}{n+1}
 \int \D^4 u \;O_\alpha\!\kln{u} \;
 \left(q^\alpha \mathbf{h}_{n}^2\kln{u,q} -u^\alpha
 \mathbf{h}_{n-1}^2\kln{u,q}\right)\,;
\nonumber
\end{eqnarray}
finally, using the representation $1/(n+1) = \int_0^1 d\tau \tau^n$
together with Eq.~(\ref{h7}), we  get
\begin{eqnarray}
\label{FTrace} \hat T^{\twz}_{\rm trace}(q)
 &=& - 2 \int_0^1 \d\tau
 \; \int \D^4 u  \; \im
 \left(O_\alpha\!\kln{u} - O_\alpha\!\kln{-u}\right) \; \frac{q^2\kln{q^\alpha +
 \kappa\tau u^\alpha}}{\big[\kln{q+\kappa\tau u}^2 +
 \im\epsilon\big]^2}\;.
\end{eqnarray}
Performing the $\tau$--integration one obtains
\begin{eqnarray}
\hat T^{\twz}_{\rm trace}(q)
&=&
 - \frac{q^2}{2} \int \frac{\D^4 u}{\kappa^4}  \;
 \im\left(O_\mu\!\kln{\frac{u}{\kappa}} - O_\mu\!\kln{-\frac{u}{\kappa}}\right)
 \frac{u^\mu q^\nu - u^\nu q^\mu}{(uq)^2 -u^2q^2}
\nonumber\\
&& \times
 \left\{
 \frac{q_\nu+u_\nu}{(q+u)^2 +\im\epsilon}
 - \frac{q_\nu}{q^2+\im\epsilon}
 + u_\nu \int_0^1 \d\tau \; \frac{1}{(q+\tau u)^2+\im\epsilon}
 \right\}\,,
\label{FTrace1}
\end{eqnarray}
where, under the assumption $(uq)^2 - u^2 q^2 \geq 0$, we get
\begin{eqnarray}
\label{I_00}
\hspace{-.8cm}
\int_0^1 \d\tau \; \frac{1}{(q+\tau u)^2+\im\epsilon}
&=&
 \frac{1}{2\sqrt{(qu)^2 -u^2q^2}}
 \ln \Bigg|\frac{q^2+uq + \sqrt{(qu)^2 -u^2q^2}}{q^2+uq - \sqrt{(qu)^2 -u^2q^2}}
 \Bigg| \quad
 \stackrel{(uq)^2= u^2q^2}{\longrightarrow} ~
 \frac{1}{q^2 + uq}\;.
\end{eqnarray}

Here, some remarks are in order.

After taking matrix elements according to the description of
Ref.~\cite{GLR01}, which is sketched in the next Section,
Eqs.~(\ref{gda0}) -- (\ref{abk2}), the expression (\ref{FTrace1})
not only allows for a simpler extraction of the imaginary part of
the Compton amplitude, it also reveals some structure of the
Compton amplitude which can be taken over, after appropriately
introducing generalized distribution amplitudes, to relations
between them. This will be shown in detail in the main part of the
paper, Sections IV - VI. Of course, such relations must be hidden
also in the expression (\ref{FTrace3}) below but we don't see how
to extract it in a simple way.

Furthermore, reminding the definition of the non-local twist-2
vector operators (\ref{NLO}) and a corresponding expression  for
the scalar ones (see, Ref.~\cite{GLR01}, Eq.~(2.30)), it is of
interest to perform the Fourier transformation of these non-local
operators also directly. This, however, is much more difficult
than by the procedure which has been used above where the Fourier
transformation has been performed before summing up the local
expressions. In order to be convinced that the procedure of
Fourier transformation and summation could be exchanged we checked
this explicitly for the trace part, see, Appendix B.

\subsection{The complete Compton operator in momentum space}

The same procedure as for the trace part has been applied to the complete Compton operator, Eqs.~(\ref{CA_nonf}) and (\ref{CA_nonf1}); the explicit computations are postponed to Appendix A. Let us present the result for
the antisymmetric and the symmetric part separately.

For the antisymmetric part we obtained
\begin{eqnarray}
 \hat T^{\twz}_{[\mu\nu]}(q)
 &=&
 \epsilon_{\mu\nu}^{~~\alpha\beta}
 \int \frac{\d^4 \! x}{2\pi^2} \;
  \frac{\e^{\im qx}\, x_{\alpha} }{\kln{x^2 - \im\epsilon}^2} \;
 \Big(O_{5\beta}^{\twz} \kln{\kappa x, - \kappa x}
 +
 O_{5\beta}^{\twz} \kln{- \kappa x, \kappa x}\Big)
\nonumber\\
&=&
 {q^2} \int_0^1 \d\tau_1 \int _0^1 \d\tau_2 \int \D^4 u
 \;\Big( O_{5\rho} \kln{{u}} + O_{5\rho} \kln{{-u}}\Big) \,
 \epsilon_{\mu\nu}^{~~\alpha\beta}\,
 \partial^\rho_u \kln{ \frac{q_{\kleo{\alpha}}u_{\okle{\beta}}}
 {[(q + \kappa\tau_1\tau_2u)^2 + i\epsilon]^2 }}\,.
\label{OAS1}
\end{eqnarray}
The $\tau$--integrations may be performed with the result
\begin{eqnarray}
\hat T^{\twz}_{[\mu\nu]}(q)
 &=&
 \epsilon_{\mu\nu}^{~~\alpha\beta}
 \int \D^4 u
\; \left( O_{5\rho} \kln{{u}} + O_{5\rho} \kln{{-u}}\right) \,
\partial^\rho_u \KLn{ q_{\kleo{\alpha}}u_{\okle{\beta}} \; {\cal G}_1 /q^2}
 \label{OAS2}
\end{eqnarray}
 with
\begin{equation*}
{\cal G}_1 =  \frac{1}{4\kappa X \kln{1-M^2}\sqrt{1-M^2}} \kln{ \kln{1-M^2} \,
L_- + \sqrt{1-M^2} \, L_+ + M^2 \, L }
\end{equation*}
and
\begin{eqnarray*}
L_\pm &=& \ln\kln{ 1+\kappa X \kln{ 1+\sqrt{1-M^2} } } \pm \ln\kln{
1+\kappa X \kln{ 1-\sqrt{1-M^2} } }
\\
L &=& \text{dilog} \kln{ 1+\kappa X \kln{ 1+\sqrt{1-M^2} } } -
\text{dilog} \kln{ 1+\kappa X \kln{ 1-\sqrt{1-M^2} } }
\end{eqnarray*}
where the following abbreviations are used:
\begin{eqnarray}
 M^2 := \frac{u^2 \, q^2}{\kln{uq}^2}
 \qquad \text{and} \qquad
 X := \frac{\kln{uq}}{q^2}\,.
 \label{BM}
\end{eqnarray}

For the symmetric part we obtained
\begin{eqnarray}
 \hat T^{\twz}_{\{\mu\nu\}}(q)
 &=&
  {S_{\mu\nu|}}^{\alpha\beta} \int \frac{\d^4 \! x}{2\pi^2} \;
  \frac{\e^{\im qx}\,x_{\alpha}}{\kln{x^2 - \im\epsilon}^2} \,
 \im\,\KLn{ O_{\beta}^\twz\kln{\kappa x, -\kappa x} -
       O_{\beta}^\twz\kln{-\kappa x, \kappa x} }
 \nonumber\\
 &=&
  2 \, \int_0^1 \frac{\d\tau}{\tau} \; \kln{1-\tau
 + \tau \ln\tau} \int \frac{\D^4 u}{\kappa^4\tau^4} \;
 \im\,\left(O_\rho \kln{\frac{u}{\kappa\tau}} -
       O_\rho \kln{\frac{-u}{\kappa\tau}}\right)
 \, \partial^\rho_u
\nonumber\\
&&
 \qquad \times  \Bigg[\,
 \frac{2 }{[(q + u)^2 + i\epsilon]^3}
 \Big\{\kln{ q^2 \, g_{\mu\nu} - q_\mu q_\nu }  \,
  \left[(u q)^2- u^2q^2\right]
 + \,  \kln{ u_\mu {q^2}- {q_\mu \kln{uq}} }
       \kln{ u_\nu {q^2}- {q_\nu \kln{uq}} }\Big\}\;
\nonumber\\
&&
 \qquad\quad +\,
 \frac{u^2}{[(q + u)^2 + i\epsilon]^2}\kln{ q^2 \, g_{\mu\nu} - q_\mu q_\nu }
 \Bigg].
 \label{OS1}
 \end{eqnarray}
 Concerning the $\tau$ integration we remark that, implicitly, it
 as a triple-integration (see, relation (\ref{tau4})).
Again, performing the $\tau$--integration one obtains
\begin{eqnarray}
\hat T^{\twz}_{\{\mu\nu\}}(q)
  = - \frac{\im}{q^2} \, \int \D^4 u \;
  \left( O_\rho\kln{ u } - O_\rho\kln{ -u }\right)\, \partial^\rho_u  \,
 \KLs{ \kln{ q^2 \, g_{\mu\nu} - q_\mu q_\nu } \;  {\cal F}_1
 + \, \Kln{ u_\mu - X \, q_\mu } \Kln{ u_\nu - X \, q_\nu } \; {\cal F}_2 }
 \label{OS2}
\end{eqnarray}
with
\begin{eqnarray*}
{\cal F}_1 &=& \frac{1}{4\kappa\kln{1-M^2}\sqrt{1-M^2} } \kln{ \kln{ 1 - 2 M^2
- \kappa X M^4} \, L_- + \sqrt{1-M^2} \, L_+ + M^2 \, L }
\\
{\cal F}_2 &=& \frac{1}{4\kappa\,X^2\kln{1-M^2}^2\sqrt{1-M^2} } \kln{ \kln{ 1
 4 M^2 - 3\,\kappa X M^4} \, L_- + \sqrt{1-M^2} \kln{1+2\,M^2} \, L_+ +
3 \, M^2 \, L }
\end{eqnarray*}

Taking the trace of that result one gets 
\begin{eqnarray}
\label{FTrace3} \hat T^{\twz}_{\rm trace}(q) &=&
\frac{\im}{\kappa} \int{\D^4 u}  \;
 \left(O_\mu\!\kln{u} - O_\mu\!\kln{-u}\right)
 \partial^\mu_u \left(\frac{1}{\sqrt{1-M^2}}\, L_- + \sqrt{1-M^2}\, L_+ \right)\,,
\end{eqnarray}
where still a further derivation of the functions $(1-M)^{\pm 1/2}L_\pm(u,q)$
has to be performed. Contrary to this, the corresponding expression
following from (\ref{FTrace1}) and (\ref{I_00}),
\begin{eqnarray}
\hat T^{\twz}_{\rm trace}(q)
&=&
 -\,\frac{\im}{2 q^2} \int \frac{\D^4 u}{\kappa^4}  \;
 \left(O^\mu\!\kln{\frac{u}{\kappa}}
  - O^\mu\!\kln{-\frac{u}{\kappa}}\right)
 \frac{1}{X^2 (1-M^2)}
\nonumber\\
 && \times
 \left\{ \frac{u_\mu \left(q^2 + (uq)\right)
 - q_\mu \left((uq) + u^2\right)}{(q+u)^2 +\im\epsilon} -
 \frac{u_\mu q^2 - q_\mu uq}{q^2+\im\epsilon}
 + \frac{u_\mu (uq)- q_\mu u^2}{2 (uq)\sqrt{1-M^2}}\, L_-
 \right\}\,,
\label{FTrace2}
\end{eqnarray}
is more explicit.

Furthermore, after taking matrix elements, choosing $\kappa\ra
-\kappa= -1/2$, $X = - 1/\Xi $ and $M^2 = - {\cal  M}^2$,
observing $\mathrm{dilog}(z) = \mathrm{Li}_2 (1-z)$ as well as
changing $L_\pm \ra \mp L_\pm$ and $L \ra - L$ one reproduces the
result of Ref.~\cite{BM01}. Unfortunately, in that form,
Eqs.~(\ref{OAS2}) and (\ref{OS2}), not only the derivations w.r.t.
$u$ are finally to be performed but, more unpleasant, the result
is not well suited for extracting the imaginary part of the
Compton amplitude. Therefore, in our further considerations the
expressions (\ref{OAS1}) and (\ref{OS1}) will be used.

%% file: M_Tasym3.tex
\section{Non-forward Compton amplitude: Trace part (nonlocal)}
\renewcommand{\theequation}{\thesection.\arabic{equation}}
\setcounter{equation}{0}

Now, let us begin to study the Fourier transform of the
non-forward Compton amplitude using an appropriate definition of
the generalized distribution amplitudes and, thereafter,
introducing variables which are adjusted to the Bjorken region.
The trace part of the Compton amplitude, Eq.~(\ref{FTrace}), is
the simplest one. Therefore, let us start with it thereby already
introducing the general procedure, described in \cite{GLR01}, of
parametrizing the non-forward matrix element and of taking their
Fourier transform. Since we like to investigate especially their
imaginary part it is advantageous to start with the expression
(\ref{FTrace}). In fact, the $\tau$--integration must not be
performed completely since part of it can be put into the
definition of the generalized distribution amplitudes (see,
below).

Let us take the matrix element of (\ref{FTrace}) together with the
definition of the generalized parton distribution amplitude
according to Ref.~\cite{GLR01}, Eqs.~(2.3) -- (2.11), as follows
\begin{eqnarray}
\langle P_2,S_2 |\hat T^{\twz}_{\rm trace}(q)|P_1,S_1 \rangle
 &=&
 -\,2\im\int \frac{\d^4 \! x}{2\pi^2} \;  \frac{\e^{\im qx}}{\kln{x^2 - \im\epsilon}^2} \;
 x^\alpha \,\langle P_2,S_2 |\Big(
 O^\twz_\alpha\kln{\kappa x, - \kappa x} -
 O^\twz_\alpha\kln{-\kappa x,  \kappa x}
 \Big)|P_1,S_1 \rangle
\nonumber\\
 &=&
 -\,2\im \int_0^1 \d\tau \; \int \d^4 u  \;
 \langle P_2,S_2 |\Big(
 O_\rho\!\kln{u} - O_\rho\!\kln{-u}
 \Big)|P_1,S_1 \rangle \;
 \frac{q^2\kln{q^\rho + \kappa\tau u^\rho}}
 {\big[\kln{q+\kappa\tau u}^2 + \im\epsilon\big]^2}
 \nonumber\\
 &=&
 -\,2\int_0^1 d\tau
 \int D {\mathbb Z} \; %
 \Phi_a ({\mathbb Z}, \mu^2)\,
  {\cal K}_\rho^a({\mathbb P},{\mathbb S})\,
 \frac{q^2 (q^\rho +  \tau \Pi^\rho )}
 {\big[\left( q + \tau \Pi \right)^2  + \im \epsilon\big]^2}\,. %
 \label{T_as0}
\end{eqnarray}
Here, the matrix elements of $O^\twz_\alpha\!\kln{u}$ are
obtained, as has been demonstrated quite generally in
Ref.~\cite{GLR01}, by the correspondence, holding under the
$u-$integral,
\begin{eqnarray}
  \langle P_2,S_2 |\,\im \big({O}_\alpha(u)
  - {O}_\alpha(-u)\big)|P_1,S_1 \rangle
  =
  {\cal  K}_\alpha^a({\mathbb P},{\mathbb S})
  \int D {\mathbb Z}\, \delta^4(u - {\mathbb{PZ}})\,
  \Phi_a ({\mathbb Z}, P_i P_j; \mu^2)\,,
  \label{gda0}
 \end{eqnarray}
 with the kinematic form factors ${\cal  K}_\alpha^a({\mathbb P},{\mathbb S})=
 \{\bar u( P_2,S_2 )  \gamma_\alpha  u(P_1,S_1 ),\,
 \bar u( P_2,S_2 )  \sigma_{\alpha\beta} p_-^\beta  u(P_1,S_1 ),\, \ldots \}$ being
 the Dirac and Pauli form factor etc., and the generalized distribution amplitudes
  $\Phi_a({\mathbb Z},P_i P_j; \mu^2)$ depending on two variables $z_1, z_2$,
  sometimes also called double distributions (Their dependence on the momenta will
  be omitted in the following).
In addition, we introduced the following abbreviations (remind
$\kappa = {1}/{2}$):
\begin{eqnarray}
&& \Pi_\mu = \kappa {\mathbb{P_\mu Z}}, 
\qquad
{\mathbb{PZ}} = P_1 z_1 + P_2 z_2 = p_+ z_+ + p_- z_-,
\label{abk1}\\%
&& D{\mathbb Z} = dz_1 dz_2
\theta(1-z_1)\theta(z_1+1)\theta(1-z_2)\theta(z_2+1) %
= D(-{\mathbb Z}) \,,
\label{abk2}\\ %
&& \Phi_a ({\mathbb Z}, \mu^2) = f^{\twz}_a({\mathbb Z}, \mu^2) -
f^{\twz}_a(-{\mathbb Z}, \mu^2) = - \Phi_a (- {\mathbb Z},
\mu^2)\,.
 \label{abk3}
\end{eqnarray}

 Now, in order to reduce the power of the denominator in the $\tau$ integral
 (\ref{T_as0}) by one unit we perform a partial integration.
 Using the abbreviation $R(\tau)= (q + \tau \Pi)^2$ we get
\begin{eqnarray}
    \int_0^1 d\tau
    \frac{1}{\big[R(\tau)  + \im \epsilon\big]^2}
    &=&
    \frac{-1}{2[(q\Pi)^2 - q^2\Pi^2]}
    \left\{
    \frac{q\Pi + \Pi^2}{R(1)+\im \epsilon} - \frac{q\Pi}{R(0)+\im \epsilon}
    + \Pi^2 \int_0^1 d\tau
    \frac{1}{ R(\tau)  + \im \epsilon}
    \right\}\,,
    \label{I(0,2)}
    \\
    \int_0^1 d\tau \frac{\tau}{[R(\tau)+ \im\epsilon]^2} &=&
    \frac{1}{2[(q\Pi)^2 - q^2\Pi^2]}
    \left\{
    \frac{q^2 + q\Pi}{R(1) + \im \epsilon} - \frac{q^2 }{R(0) + \im \epsilon}
    + q\Pi \int_0^1 d\tau \frac{1}{R(\tau)+ \im\epsilon}
    \right\}\,.
    \label{I(1,2)}
\end{eqnarray}
where, again, $(q\Pi)^2 - q^2 \Pi^2 \geq 0$ is assumed to hold because of the physical kinematics.
Then, in complete analogy to Eq.~(\ref{FTrace1}), we get
\begin{eqnarray}
\langle P_2,S_2 |\hat T^{\twz}_{\rm trace}(q)|P_1,S_1 \rangle
 &=&
 \int D {\mathbb Z} \; %
 \Phi^a ({\mathbb Z}, \mu^2)\,
  {\cal K}^\mu_a({\mathbb P},{\mathbb S})\,
  \frac{-\,q^2}{[(q\Pi)^2 - q^2\Pi^2]}
  \Bigg\{
   \frac{\Pi_\mu(q^2 + q\Pi) - q_\mu(q\Pi + \Pi^2)}{R(1) + \im \epsilon}
    \nonumber\\
  && \qquad\qquad\qquad
  - \,\frac{\Pi_\mu\, q^2  - q_\mu (q\Pi)}{R(0) + \im \epsilon}
 + \left(\Pi_\mu(q\Pi) - q_\mu\,\Pi^2\right)
   \int_0^1 d\tau
    \frac{1}{R(\tau) + \im \epsilon}\Bigg\}\,.
\end{eqnarray}

Introducing generalized distribution amplitudes $\Phi_a^{(n)}$ of order $n$
as follows,
\begin{eqnarray}
 \Phi_a^{(0)}({\mathbb Z}) \equiv \Phi_a({\mathbb Z})
 \quad {\rm and} \quad
 \Phi_a^{(n)}({\mathbb Z}) \equiv
 \int_0^1 \frac{d\tau_1}{\tau_1^2}
 \cdots\int_0^1 \frac{d\tau_n}{\tau_n^2}\,
 \Phi_a\!\left(\frac{\mathbb Z}{\tau_1\ldots\tau_n}\right)
 \quad {\rm for} \quad n\geq 1,
 \label{GDA}
\end{eqnarray}
after scaling the ${\mathbb Z}$--variables by $\tau$, we finally
obtain
\begin{eqnarray}
\langle P_2,S_2 |\hat T^{\twz}_{\rm trace}(q)|P_1,S_1 \rangle
 &=&
 {\cal K}^\mu_a({\mathbb P},{\mathbb S})\,\int D {\mathbb Z} \;
 \frac{-\,q^2}{[(q\Pi)^2 - q^2\Pi^2]}\Bigg\{
 \Phi_a^{(0)}({\mathbb Z}, \mu^2)\,
 \Bigg[
 \frac{\Pi_\mu(q^2 + q\Pi) - q_\mu(q\Pi + \Pi^2)}{R(1) + \im \epsilon}
 \nonumber\\
 && \qquad\qquad\qquad
  - \, \frac{\Pi_\mu\, q^2  - q_\mu (q\Pi)}{R(0) + \im \epsilon}\Bigg]
  + \Phi_a^{(1)}({\mathbb Z}, \mu^2)\,
 \frac{\Pi_\mu(q\Pi) - q_\mu\,\Pi^2}{R(1) + \im \epsilon}\Bigg\}\,.
 \label{Ftrace0}
\end{eqnarray}

Now, let us consider the imaginary part of this expression. Due to
the overall factor $q^2$ in Eq.~(\ref{Ftrace0}) it results from
the $1/R(1)$--terms only. For that reason let us rewrite
\begin{eqnarray}
 R(\tau)
 \equiv
 (q + \tau\Pi)^2
 =
 \Pi^2(\tau -\tilde\xi_+)(\tau -\tilde\xi_-)
\end{eqnarray}
with
\begin{eqnarray}
\label{xi}
    \tilde\xi_\pm
    = \frac{-\, q\Pi \pm \sqrt{(q\Pi)^2 - q^2\Pi^2}}{\Pi^2}
    = \frac{-\, q^2}{q\Pi \pm \sqrt{(q\Pi)^2 - q^2\Pi^2}}
\end{eqnarray}
to get
\begin{eqnarray}
 \label{IM}
 \frac{1}{R(\tau)+ \im\epsilon}=
 \frac{1}{2\sqrt{(q\Pi)^2 - q^2\Pi^2}}
 \left(
 \frac{1}{\tau - \tilde\xi_+ + \im \epsilon} -
 \frac{1}{\tau - \tilde\xi_- - \im\epsilon}
 \right).
\end{eqnarray}

Now, we take the imaginary part
\begin{eqnarray}
 \label{imR}
\im\, {\mathrm{Im}} \frac{1}{R(1)+ \im\epsilon}=
 -\frac{\im \pi}{2}\frac{1}{\sqrt{(q\Pi)^2 - q^2\Pi^2}}
 \left[
 \delta(1 - \tilde\xi_+ ) + \delta(1 - \tilde\xi_- )
 \right]\,.
\end{eqnarray}
Obviously, the following equalities
\begin{eqnarray}
     \left(\Pi^2 + q\Pi \right) \delta(1 - \tilde\xi_\pm)
    = \pm\,\left[\sqrt{(q\Pi)^2 - q^2\Pi^2}\right] \delta(1 - \tilde\xi_\pm)
    = -\left( q^2 + q\Pi\right) \delta(1 - \tilde\xi_\pm)
    \label{+}\,
    \end{eqnarray}
can be used to simplify some of the resulting expressions below.

 In addition, let us introduce, instead of $z_+$ and $z_-$,
 new variables, $t$ and $\zeta$, being more adequate for the
 generalized Bjorken region, namely (the arrow shows their value
 in the limiting case of forward scattering, i.e., for $P_1=P_2=P$)
\begin{eqnarray}
    x = \frac{-q^2}{qp_+} \Longrightarrow x_{\rm Bj}, \qquad
    \eta = \frac{qp_-}{qp_+} \Longrightarrow 0,\qquad
    -\rho^2 = \frac{p_-^2}{p_+^2} = \frac{M^2 - P_1P_2}{M^2 +P_1P_2}
    \Longrightarrow 0, \qquad
    p_+^2 = 2(M^2 +P_1P_2) \Longrightarrow 4 M^2,
    \nonumber
\end{eqnarray}
together with
\begin{eqnarray}
\label{newvar}
    t = z_+ + \eta z_- \Longrightarrow z, \qquad
    \zeta = z_- / t \Longrightarrow 0\,,
\end{eqnarray}
which means that $z_+$ and $z_-$ is scaled according to
\begin{eqnarray}
z_+ = t ( 1 - \eta \zeta), \qquad z_- = t \zeta\,.
\end{eqnarray}
This leads to
\begin{eqnarray}
    \Pi^\mu
    &=&\kappa t \left(p^\mu_+ (1 - \eta\zeta) + p^\mu_- \zeta\right)
    \equiv \kappa t \,{\cal P}^\mu(\eta,\zeta)\,,
    \nonumber\\
    q\Pi &=& \kappa t\, q{\cal P}\equiv  \kappa t\, qp_+
    \Longrightarrow 2\kappa\,z\,qP,
    \nonumber\\
    \Pi^2
    &=&(\kappa t)^2 p_+^2
    \left[1-2\eta\zeta+(\eta^2-\rho^2)\zeta^2\right]
    \equiv (\kappa t)^2 {\cal P}^2
    \Longrightarrow 4\kappa^2\,z^2\,M^2
\end{eqnarray}
and
\begin{eqnarray}
 \tilde\xi_\pm
 =
  \frac{x}{\kappa t} \frac{1}{1\pm \sqrt{1 + x^2 \mathcal{P}^2/Q^2}}
 \equiv
  \frac{\xi_\pm}{t}
 \Longrightarrow
  \frac{x_{\rm Bj}}{z}\frac{2}{1\pm \sqrt{1+4x_{\rm Bj}^2M^2/Q^2}}\,.
\end{eqnarray}
Obviously, $\tilde\xi_\pm$ are appropriate generalizations of the
Nachtmann variable to the non-forward case and $t$ measures the
collinear momentum, i.e. the portion of momentum in direction
$p_+$.

The variable ${\tilde t} = z_+ + z_-\,{\tilde x}p_- / {\tilde
x}p_+$ or, rather, ${ t} = z_+ + z_-\,{q}p_- / {q}p_+$ has been
originally introduced for a partial diagonalization of the
renormalization group equation for the generalized parton
distributions $f(z_+,z_-) = {\hat f}(t,z_-;\eta)$ in
Refs.~\cite{LeiVCS}, cf. also \cite{MRGHD}. As a second step, the
reparametrization $z_- = t \zeta$ leads to ${\hat{\hat
f}}(t,\zeta;\eta)$ where the external parameter $\eta$ and the
internal variable $\zeta$ parametrize the deviation from the
forward direction.

With these definitions the measure of the $\mathbb Z$--integration  gets
($0\leq |\eta| \leq 1$, see, \cite{MRGHD,BGR99})
\begin{eqnarray}
 D{\mathbb Z} &=& 2 dz_+ dz_-
 \theta(1-z_++z_-)\theta(1+z_+-z_-)
 \theta(1-z_+-z_-)\theta(1+z_++z_-)
 \nonumber\\
 &=& 2tdtd\zeta
 \theta\big(1-t+(1+\eta)t\zeta \big)\theta\big(1+t-(1+\eta)t\zeta\big)
 \theta\big(1-t-(1-\eta)t\zeta \big)\theta\big(1+t+(1-\eta)t\zeta\big)\,.
 \end{eqnarray}

 With these new variables for the imaginary part of the trace of the Compton amplitude we get:
\begin{eqnarray}
 {\mathrm{Im}}\, T_{\rm trace}\kln{q}
 &=&
 {\mathrm{Im}}\, T^+_{\rm trace}\kln{q}
 +
 {\mathrm{Im}}\, T^-_{\rm trace}\kln{q}\,,
 \nonumber\\
 {\mathrm{Im}}\, T^\pm_{\rm trace}\kln{q}
    &=&
    \frac{\pi}{2}\int D{\mathbb Z} \; %
    \frac{- q^2}{[(q\Pi)^2 - q^2 \Pi^2]^{3/2}}\,
          \delta\big(1 - \tilde\xi_\pm\big)
    \times
\nonumber\\
&& \qquad \left\{ (q{\cal K}^a) \left[(q\Pi + \Pi^2)\,
\Phi^{(0)}_a(\mathbb{Z})
 + \Pi^2\, \Phi^{(1)}_a(\mathbb{Z})\right]
 -
(\Pi{\cal K}^a)\left[(q^2 + q\Pi)\, \Phi^{(0)}_a(\mathbb{Z})
 + (q\Pi) \, \Phi^{(1)}_a(\mathbb{Z})
\right]
\right\}
\nonumber\\
&=& \pi \int d\zeta \int dt \;
  \delta\big(t - \xi_\pm(\zeta)\big) %
  \frac{4\,x}{1+x^2{\cal P}^2 / Q^2}\,
\bigg\{\pm\, \frac{q{\cal K}^a}{q{\cal P}} \left(1 +
\frac{t\,x}{2}\frac{{\cal P}^2 }{Q^2}\right) \phi^{(0)}_a(t,\zeta)
\nonumber\\
&&
\qquad\qquad\quad
 +
 \frac{x}{2}\frac{{\cal P}^2 }{Q^2}
 \left(\frac{q{\cal K}^a}{q{\cal P}}
 -
 \frac{{\cal P}{\cal K}^a}{{\cal P}^2}
 \right)
\left[\mp\,t\, \phi^{(0)}_a(t,\zeta)
 + \frac{1}{\sqrt{1+x^2{\cal P}^2/Q^2}} \, \phi^{(1)}_a(t,\zeta)
\right] \bigg\},
 \end{eqnarray}
 where (\ref{+}) has been used and
 $\Phi^{(0)}_a(\mathbb{Z}),\;\Phi^{(1)}_a(\mathbb{Z})$ has been
 replaced by
\begin{eqnarray}
\label{Phi_tr}
 \Phi^{(0)}_a(\mathbb{Z})
 \equiv \phi_a^{(0)}(t,\zeta),
 \qquad
 t\,\Phi^{(1)}_a(\mathbb{Z})
 = t \int_0^1 \frac{\d\tau}{\tau^2}\,
   \Phi_a\left(\frac{t}{\tau},\zeta \right)
 = \int_t^1 dy \,\Phi_a\left(y,\zeta\right)
 \equiv \phi_a^{(1)}(t,\zeta).
\end{eqnarray}

Now, taking into account the following relations
\begin{eqnarray}
   1 + \hbox{\large $\frac{1}{2}$}x\xi_\pm\mathcal{P}^2/Q^2
    &=&
    \pm \sqrt{1+x^2\mathcal{P}^2/Q^2}
    =
    -\left(1- 2x/\xi_\pm \right)\,,
    \label{rel1}\\
    x\frac{\partial}{\partial x}\left(
    \frac{x}{[1+x^2\mathcal{P}^2/Q^2]^{1/2}}\right)
    &=&
    \frac{x}{[1+x^2\mathcal{P}^2/Q^2]^{3/2}}\,,
    \label{rel2}\\
    x\frac{\partial}{\partial x} \xi_\pm
    &=&
    \pm  \frac{\xi_\pm}{[1+x^2\mathcal{P}^2/Q^2]^{1/2}}\,,
    \label{rel3}\\
    x\frac{\partial}{\partial x}\phi_a^{(1)}(\xi_\pm,\zeta)
    &=&
    \mp  \frac{\xi_\pm}{[1+x^2\mathcal{P}^2/Q^2]^{1/2}}
    \phi_a^{(0)}(\xi_\pm,\zeta)\,,
    \label{rel4}
\end{eqnarray}
and introducing the following functions,
\begin{align}
    & \mathcal{V}^{\pm}_{a 0}(x,\eta; \zeta)
   \equiv
    \frac{x\,\phi_a^{(0)}(\xi_\pm,\zeta)}{\sqrt{1+x^2\mathcal{P}^2/Q^2}}\,,
\label{H0}
    \\
   & \mathcal{V}^{\pm}_{a\,1}(x,\eta;\zeta)
   \equiv
    x\frac{\partial}{\partial x}\left(
    \frac{x\,\phi_a^{(1)}(\xi_\pm,\zeta)}{\sqrt{1+x^2\mathcal{P}^2/Q^2}}
    \right)
    =
    \frac{x}{1+x^2\mathcal{P}^2/Q^2}
    \left[
    \mp\,\xi_\pm\, \phi_a^{(0)}(\xi_\pm,\zeta)
    +
    \frac{1}{\sqrt{1+x^2\mathcal{P}^2/Q^2}}
    \phi_a^{(1)}(\xi_\pm,\zeta)
    \right],
\label{H1}
\end{align}
one finally finds for the imaginary part of the trace of the {\em
non-forward} Compton amplitude
\begin{eqnarray}
\label{FTraceX}
 {\mathrm{Im}}\, T_{\rm trace}\kln{q}
 &=&
 4\pi \int d\zeta \; %
 \left\{
 \frac{q{\cal K}^a}{q{\cal P}}\, \mathcal{V}_{a\,0}(x,\eta; \zeta)
 +
  \left(
 \frac{q{\cal K}^a}{q{\cal P}}
 -  \frac{{\cal P}{\cal K}^a}{{\cal P}^2}
 \right)\,
 \frac{x}{2}\,\frac{{\cal P}^2 }{Q^2}
 \mathcal{V}_{a1}(x,\eta; \zeta)
 \right\}
\end{eqnarray}
with $\mathcal{V}_{a n}(x,\eta; \zeta)= \mathcal{V}^{+}_{a n}
(x,\eta;\zeta) + \mathcal{V}^{-}_{a n}(x,\eta;\zeta),\, n = 0,1$.

In the case of {\em forward scattering}, $\eta=0$, one obtains
simply (in terms of $x = x_{\rm Bj}$)
\begin {eqnarray}
{\mathrm{Im}}\,^f T_{ \rm trace}\kln{q}
 &=&
 2\pi  \; %
 \frac{ x}{\sqrt{1+4x^2\,M^2/Q^2}}\,
 f(x)\,,
 \\
f(x)&=&
 2 \int d\zeta\, \phi_{\rm D}^{(0)}(x;\zeta) \,,
\end{eqnarray}
because only the Dirac form factor survives.

\section{Non-forward Compton amplitude: Antisymmetric part (nonlocal)}
\renewcommand{\theequation}{\thesection.\arabic{equation}}
\setcounter{equation}{0}

Now, after having introduced the procedure of revealing the structure of
non-forward amplitudes, let us study the Fourier transform of the antisymmetric part of the leading twist-2 non-forward Compton amplitude:
\begin{eqnarray}
T^\twz_{[\mu\nu]}\kln{q}
&=&
\int \frac{\text{d}^4 \!x}{2\pi^2} \;%
\frac{\e^{\im qx}} { {\kln{x^2 - \im\epsilon}^2} }\;%
\epsilon_{\mu\nu}^{\phantom{\mu\nu}\alpha\beta} x_{\alpha}\,%
\langle P_2,S_2 | %
\Big( O^{\rm tw 2}_{5\beta} \Kln{\kappa x, -\kappa x}
+ O^{\rm tw 2}_{5\beta} \Kln{- \kappa x, \kappa x} %
\Big) |P_1,S_1 \rangle
\nonumber \\ %
&=& %
\int_0^1 d\tau_1 \int_0^1 d\tau_2 \int {d^4 u}%
\langle P_2,S_2 | \Big( %
  O_{5\rho} \Kln{ {u}}+ O_{5\rho} \Kln{- {u}}%
\Big) |P_1,S_1 \rangle \,%
\partial_u^\rho %
\frac{
\epsilon_{\mu\nu}^{\phantom{\mu\nu}\alpha\beta}\,q^2\,q_{\alpha}\,u_\beta}
{\big[(q + \kappa \tau_1 \tau_2 u)^2 + \im \epsilon\big]^2} %
\nonumber \\
&=& %
\int_0^1 d\tau_1 \int_0^1 d\tau_2 %
\int D {\mathbb Z} \; %
\Phi_{5a} ({\mathbb Z}, \mu^2)\; %
\epsilon_{\mu\nu}^{\phantom{\mu\nu}\alpha\beta} q_{\alpha}\, q^2 \,%
{\cal K}^a_\rho({\mathbb P},{\mathbb S})\;\partial_{\Pi}^\rho %
  \frac{\Pi_\beta }{\big[
  (q + \tau_1 \tau_2 {\Pi})^2 + \im \epsilon \big]^2} %
\label{Tas1}
\end{eqnarray}
where, in the second line, we used the expression (\ref{OAS1})
and, in the last line, we observed the correspondence
\begin{eqnarray}
&&  \langle P_2,S_2 |
 \left(O_{5\rho} \Kln{ {u}}+ O_{5\rho}\Kln{-{u}}\right)|P_1,S_1 \rangle
  =
  {\cal  K}_\alpha^a({\mathbb P},{\mathbb S})
  \int D {\mathbb Z}\, \delta^4(u - {\mathbb{PZ}})\,
  \Phi_{5a} ({\mathbb Z}, \mu^2)\,,
  \nonumber\\
&&  \Phi_{5a} ({\mathbb Z}, \mu^2) = f^{\twz}_{5a}({\mathbb Z},
\mu^2) +
 f^{\twz}_{5a}( - {\mathbb Z}, \mu^2) = \Phi_{5a} (- {\mathbb Z}, \mu^2).
 \nonumber
 \end{eqnarray}
together with the abbreviations (\ref{abk1}) and (\ref{abk2}).

In the same manner as in the case of the trace part we have to perform
appropriate partial integrations in order to obtain an expression which
finally depends on $1/[R(1)+ \im \epsilon]$ only. Since the explicit
computation, despite being straightforward, is somewhat cumbersome we
postpone it to Appendix C. The result reads as follows:
\begin{eqnarray}
\hspace{-.8cm}
 T^\twz_{[\mu\nu]}\kln{q}
    \!&=&\!\!
    \int\! D \mathbb{Z}\;
    \epsilon_{\mu\nu}^{\phantom{\mu\nu}\alpha\beta}
    \frac{-\, q^2\,q_\alpha}{2\,[(q\Pi)^2 - q^2 \Pi^2]}\,
    \Big\{
    {{\cal K}^a_\beta}\, 
    {F^{(5)}_{a\,1}}({\mathbb Z})
    + \Pi_\beta (q{\cal K}^a) \, 
    {F^{(5)}_{a\,2}}({\mathbb Z})
    + \Pi_\beta(\Pi{\cal K}^a)\,
    {F^{(5)}_{a\,3}}({\mathbb Z})
    \Big\}
    \frac{1}{R(1)+ \im \epsilon}\,,
    \label{Tas500}
\end{eqnarray}
with
\begin{eqnarray}
F^{(5)}_{a\,1}({\mathbb Z})
    &=&
    (q\Pi+\Pi^2)\,\Phi_{5a}^{(1)}(\mathbb{Z})
    +
    \Pi^2\,\Phi_{5a}^{(2)}(\mathbb{Z})\,,
\\
F^{(5)}_{a\,2}({\mathbb Z})
    &=&
    \Phi_{5a}^{(0)}(\mathbb{Z})
    -
    \left(3\frac{ q\Pi (q\Pi + \Pi^2)}{[(q\Pi)^2 - q^2 \Pi^2]}-2\right)
    \Phi_{5a}^{(1)}(\mathbb{Z})
   - 3
    \frac{(q\Pi)\,\Pi^2 }{[(q\Pi)^2 - q^2
    \Pi^2]}\Phi_{5a}^{(2)}(\mathbb{Z})\,,
\\
F^{(5)}_{a\,3}({\mathbb Z})
    &=&
    \Phi_{5a}^{(0)}(\mathbb{Z})
    + 3 \frac{q\Pi (q^2 + q\Pi)}{[(q\Pi)^2 - q^2 \Pi^2]}
    \Phi_{5a}^{(1)}(\mathbb{Z})
    +
    \left(\!
    3 \frac{ (q\Pi)^2}{[(q\Pi)^2 - q^2 \Pi^2]}-1\!\right)
    \Phi_{5a}^{(2)}(\mathbb{Z})\,,
\label{Tas50}
\end{eqnarray}
where the generalized distribution amplitudes $\Phi_{5a}^{(n)}(\mathbb{Z})$
are defined in the same manner as $\Phi_{a}^{(n)}(\mathbb{Z})$ in
Eqs.~(\ref{GDA}).

Now, let us consider the imaginary part of that expression. Again,
only the terms with $1/[R(1)+ \im \epsilon]$ contribute, leading to the
following expression for the imaginary part of (the antisymmetric part of)
the Compton amplitude:
\begin{eqnarray}
\hspace{-.8cm}
 {\mathrm{Im}}\, T^\twz_{[\mu\nu]}\kln{q}
    &=&
    \frac{-\pi}{4}\int d\zeta \int dt\,t \;
    \epsilon_{\mu\nu}^{\phantom{\mu\nu}\alpha\beta}\,
    \frac{- q^2}{[(q\Pi)^2 - q^2 \Pi^2]^{3/2}}\,
    \left[
      \delta\big(1 - \xi_+(\zeta)/t\big)
    + \delta\big(1 - \xi_-(\zeta)/t\big)
    \right]
    \times
    \nonumber\\
 \hspace{-.8cm}
    &&
    \Bigg\{
    {q_{\alpha}\,{\cal K}^a_\beta} 
    \left[
    (q\Pi+\Pi^2)\,\Phi_{5a}^{(1)}(t,\zeta)
         + \Pi^2\,\Phi_{5a}^{(2)}(t,\zeta)
    \right]
 \nonumber\\
 \hspace{-.8cm}
    &&
    + {q_{\alpha}\,\Pi_\beta} {(q{\cal K}^a)}  
    \left[
    \Phi_{5a}^{(0)}(t,\zeta)
    -
    \left(3\frac{ q\Pi (q\Pi + \Pi^2)}{[(q\Pi)^2 - q^2 \Pi^2]}-2\right)
    \Phi_{5a}^{(1)}(t,\zeta)
   - 3
    \frac{(q\Pi)\,\Pi^2 }{[(q\Pi)^2 - q^2 \Pi^2]}\Phi_{5a}^{(2)}(t,\zeta)
    \right]
 \nonumber\\
 \hspace{-.8cm}
    &&
    + {q_{\alpha}\,\Pi_\beta}{(\Pi{\cal K}^a)}
    \left[
    \Phi_{5a}^{(0)}(t,\zeta)
    + 3 \frac{q\Pi (q^2 + q\Pi)}{[(q\Pi)^2 - q^2 \Pi^2]}
    \Phi_{5a}^{(1)}(t,\zeta)
    +
    \left(\!
    3 \frac{ (q\Pi)^2}{[(q\Pi)^2 - q^2 \Pi^2]}-1\!\right)
    \Phi_{5a}^{(2)}(t,\zeta)\,\right]\!\!
    \Bigg\}.
    \label{Tas5}
\end{eqnarray}

Using the $\delta$-functions the $t$-integration can be carried
out. Let us consider a generic expression of the corresponding
integrand:
\begin{eqnarray}
    I^{(i)}
    &=& \int D{\mathbb Z}\, \Phi_{5a}^{(i)}({\mathbb Z})\,
    G^{(i)}({\mathbb Z},{\mathbb P})\,\delta(1 - \tilde\xi_\pm)
    =\int_{-\infty}^{\infty} d\zeta
    \int_{-1}^1 dt\, t\, \Phi_{5a}^{(i)}(t,\zeta)\,
    G^{(i)}(t,\zeta)\,\delta( 1 - \xi_\pm/t)
    \nonumber\\
    &=& \int d\zeta\, \xi_\pm(\zeta)\, G^{(i)}(\xi_\pm,\zeta)\,
    \widehat\Phi_{5a}^{(i)}(\xi_\pm,\zeta)\,,
    \label{NVG}
\end{eqnarray}
where $G^{(i)}({\mathbb Z},{\mathbb P})$ is some function of the
indicated arguments and, in the case $\xi > 0$, we introduced new
distribution amplitudes (for $\xi < 0$ the integral has to be
adapted correspondingly),
\begin{eqnarray}
    \widehat\Phi^{(0)}(\xi,\zeta) &\equiv&  \xi \Phi(\xi,\zeta) 
    \qquad {\rm and} \qquad 
    \widehat\Phi^{(i)}(\xi,\zeta)
    = \int^1_\xi \frac{dy_1}{y_1}\widehat\Phi^{(i-1)}(y_1,\zeta)\,.
\label{NVF}
\end{eqnarray}
The proof of relation (\ref{NVG}) for the general case is as
follows:
\begin{eqnarray}
I^{(n)}&=& \int d\zeta \int dt\, t\,G(t,\zeta)\,
    \Phi^{(n)}(t,\zeta)\,
    \delta( 1 - \xi/t)
    \nonumber\\
    &=& \int d\zeta \int dt\,t\, G(t,\zeta)
    \int^1_0\frac{d\tau_1}{\tau_1^2} \ldots
    \int^1_0\frac{d\tau_{n-1}}{\tau_{n-1}^2}
    \int^1_0\frac{d\tau_{n}}{\tau_{n}^2}\,
    \Phi\left(\frac{t}{\tau_1\ldots\tau_{n-1}\tau_n},\zeta\right)
    \delta( 1 - \xi/t)
    \nonumber\\
    &=& \int d\zeta \int dt\,t \,G(t,\zeta)
    \int^1_0\frac{d\tau_1}{\tau_1^2}\ldots
    \int^1_0\frac{d\tau_{n-1}}{\tau_{n-1}^2}
    \int^1_{t/(\tau_1\ldots\tau_{n-1})}\frac{dy_n}{y_n}\,
    \frac{\tau_1\ldots\tau_{n-1} y_n}{t}\,
    \Phi\left(y_n,\zeta\right)\,
    \delta( 1 - \xi/t)
    \nonumber\\
     &=& \int d\zeta \int dt\,t\,G(t,\zeta)
    \int^1_t\frac{dy_1}{y_1}\ldots
    \int^1_{y_{n-2}}\frac{dy_{n-1}}{y_{n-1}}
    \int^1_{y_{n-1}}\frac{dy_{n}}{y_{n}}\,
    y_n\,\Phi\left(y_n,\zeta\right)\,
    \delta( t - \xi)
    \nonumber\\
    &=& \int d\zeta\,  \xi\, G(\xi,\zeta)
    \int^1_{\xi}\frac{dy_1}{y_1} \ldots
    \int^1_{y_{n-1}}\frac{dy_n}{y_n}\, \widehat\Phi^{(0)}\left(y_n,\zeta\right)\,,
\end{eqnarray}
where a successive change of variables
$y_m={t}/({\tau_1\ldots\tau_{m-1}\tau_m}),\; m = n, \ldots, 1$,
has been made and the support restriction of
$\Phi\left(t,\zeta\right)$ to $-1 \leq t \leq 1$ has been
observed.

Now, taking relation (\ref{NVG}) into account we obtain for the
two different contributions related to $\tilde\xi_\pm = t\xi_\pm$
the following expression:
\begin{eqnarray}
\hspace{-1cm}
&& {\mathrm{Im}}\, T^\twz_{[\mu\nu]\pm}\kln{q}
    =
    \frac{\pi}{4\kappa}\int d\zeta  \,
    \epsilon_{\mu\nu}^{\phantom{\mu\nu}\alpha\beta}\,
    \left(-\frac{2x}{\xi_\pm}\,\frac{1}{[1+x^2\mathcal{P}^2/Q^2]^{3/2}}\right)
    \times
\nonumber\\
\hspace{-1cm}
    && \qquad
    \Bigg\{
    \frac{q_{\alpha}\,{\cal K}^a_\beta}{q{\cal P}} 
    \left[\left(1 + \frac{1}{2}x\xi_\pm\mathcal{P}^2/Q^2\right)
    \widehat\Phi_{5a}^{(1)}(\xi_\pm,\zeta)\,
    + \frac{1}{2}x\xi_\pm\mathcal{P}^2/Q^2\,
    \widehat\Phi_{5a}^{(2)}(\xi_\pm,\zeta)\,
    \right]
\nonumber\\
\hspace{-1cm}
    && \qquad
    + \frac{q_{\alpha}\,\mathcal{P}_\beta}{q{\cal P}}
    \frac{(q{\cal K}^a)}{q{\cal P}}  
    \left[
    \widehat\Phi_{5a}^{(0)}(\xi_\pm,\zeta)
    \mp
    \frac{1 - x\xi_\pm\,\mathcal{P}^2/Q^2}{[1+x^2\mathcal{P}^2/Q^2]^{1/2}}
    \widehat\Phi_{5a}^{(1)}(\xi_\pm,\zeta)
    -
    \frac{3}{2}\frac{x\xi_\pm \,\mathcal{P}^2/Q^2 }{[1+x^2\mathcal{P}^2/Q^2]}
    \widehat\Phi_{5a}^{(2)}(\xi_\pm,\zeta)
    \right]
\nonumber\\
\hspace{-1cm}
    && \qquad
    + \frac{x\xi_\pm}{2}\frac{\mathcal{P}^2}{Q^2}
    \frac{q_{\alpha}\,\mathcal{P}_\beta}{q{\cal P}}
    \frac{(\mathcal{P}{\cal K}^a)}{\mathcal{P}^2}
    \left[
    \widehat\Phi_{5a}^{(0)}(\xi_\pm,\zeta)
    \mp
    \frac{3}{[1+x^2\mathcal{P}^2/Q^2]^{1/2}}
    \widehat\Phi_{5a}^{(1)}(\xi_\pm,\zeta)
    +
    \frac{2- x^2 \mathcal{P}^2/Q^2}{[1+x^2\mathcal{P}^2/Q^2]}
    \widehat\Phi_{5a}^{(2)}(\xi_\pm,\zeta)
    \right]
    \Bigg\},
    \label{Tas6}
\end{eqnarray}
where the equalities (\ref{rel1}), corresponding to Eqs.~(\ref{+})
after $t$-integration, are used. Already here the similarity to the
structure of the antisymmetric part of the Compton amplitude in the
forward case, Eqs. (\ref{CG0}), may be observed: Ignoring the variable
$\zeta$ as well as reading $\xi_\pm$ as $\xi$, the first and the second
term in the curly bracket has the structure of $G_1(\xi)+G_2(\xi)$ and $G_2(\xi)$, respectively, whereas the last one is new but would vanish
in the forward case. Indeed, taking into account relations
(\ref{rel3}) and (\ref{rel4}) together with
\begin{eqnarray}
        x\frac{\partial}{\partial x}\left(
    \frac{x}{\xi_\pm}\frac{1}{[1+x^2\mathcal{P}^2/Q^2]^{1/2}}\right)
    &=&
    \frac{x}{\xi_\pm}
    \frac{1\mp [1+ x^2\mathcal{P}^2/Q^2]^{1/2}}{[1+x^2\mathcal{P}^2/Q^2]^{3/2}}
    =
    \mp
    \frac{1}{2}\frac{x^2\mathcal{P}^2/Q^2 }{[1+x^2\mathcal{P}^2/Q^2]^{3/2}}\,,
    \nonumber
\end{eqnarray}
one finds the following equalities:
\begin{align}
   & g^\pm_{a1}(x,\eta;\zeta)
   \equiv
    x\frac{\partial}{\partial x}\left[
    x\frac{\partial}{\partial x}\left(
    \frac{x}{\xi_\pm}
    \frac{\widehat\Phi_{5a}^{(2)}(\xi_\pm,\zeta)}
    {[1+x^2\mathcal{P}^2/Q^2]^{1/2}}
    \right)\right],
    \nonumber\\
    & \qquad\qquad
    =
    \frac{x}{\xi_\pm}\frac{1}{[1+x^2\mathcal{P}^2/Q^2]^{3/2}}\times
    \nonumber\\
    & \qquad\qquad
    \left[
    \widehat\Phi_{5a}^{(0)}(\xi_\pm,\zeta)
    +
    \frac{x(\xi_\pm \pm x)\,\mathcal{P}^2/Q^2}{[1+x^2\mathcal{P}^2/Q^2]^{1/2}}
    \widehat\Phi_{5a}^{(1)}(\xi_\pm,\zeta)
    -
    \frac{1}{2}x\xi_\pm \mathcal{P}^2/Q^2
    \frac{(2-x^2\mathcal{P}^2/Q^2)}{[1+x^2\mathcal{P}^2/Q^2]}
    \widehat\Phi_{5a}^{(2)}(\xi_\pm,\zeta)
    \right],
    \label{G1}\\
\intertext{}
    &  g^\pm_{a2}(x,\eta;\zeta) \equiv - x\frac{\partial^2}{\partial x^2}x\left(
    \frac{x}{\xi_\pm}
    \frac{\widehat\Phi_{5a}^{(2)}(\xi_\pm,\zeta)}
    {[1+x^2\mathcal{P}^2/Q^2]^{1/2}}
    \right)
    \nonumber\\
    & \qquad\qquad
    =
    - \frac{x}{\xi_\pm}\frac{1}{[1+x^2\mathcal{P}^2/Q^2]^{3/2}}\times
    \nonumber\\
    & \qquad\qquad
    \left[
    \widehat\Phi_{5a}^{(0)}(\xi_\pm,\zeta)
    \mp
    \frac{1 - x \xi_\pm \mathcal{P}^2/Q^2}{[1+x^2\mathcal{P}^2/Q^2]^{1/2}}
    \widehat\Phi_{5a}^{(1)}(\pm\xi_\pm,\zeta)
    -
    \frac{3}{2}
    \frac{x \xi_\pm \mathcal{P}^2/Q^2}{[1+x^2\mathcal{P}^2/Q^2]}
    \widehat\Phi_{5a}^{(2)}(\xi_\pm,\zeta)
    \right],
    \label{G2}\\
    & g^\pm_{a1}(x,\eta;\zeta) + g^\pm_{a2}(x,\eta;\zeta) \equiv
    - x\frac{\partial}{\partial x}\left(
    \frac{x}{\xi_\pm}
    \frac{\widehat\Phi_{5a}^{(2)}(\xi_\pm,\zeta)}
    {[1+x^2\mathcal{P}^2/Q^2]^{1/2}}
    \right)
    \nonumber\\
    & \qquad\qquad
    = \frac{x}{\xi_\pm}\frac{1}{[1+x^2\mathcal{P}^2/Q^2]^{3/2}}\times
    \left[\left(1 + \frac{1}{2}x\xi_\pm\mathcal{P}^2/Q^2\right)
    \widehat\Phi_{5a}^{(1)}(\xi_\pm,\zeta)\,
    + \frac{1}{2}x\xi_\pm\mathcal{P}^2/Q^2\,
    \widehat\Phi_{5a}^{(2)}(\xi_\pm,\zeta)\,
    \right],
    \label{G12}\\
    & g^\pm_{a0}(x,\eta;\zeta) = x^2 \frac{\partial^2}{\partial x^2}
    \left( x^2
    \frac{\widehat\Phi_{5a}^{(2)}(\xi_\pm,\zeta)}
    {[1+x^2\mathcal{P}^2/Q^2]^{1/2}}
    \right)
    \nonumber\\
    & \qquad\qquad
    = \frac{x^2}{[1+x^2\mathcal{P}^2/Q^2]^{3/2}}\times
    \left[
    \widehat\Phi_{5a}^{(0)}(\xi_\pm,\zeta)
    \mp
    \frac{3}{[1+x^2\mathcal{P}^2/Q^2]^{1/2}}
    \widehat\Phi_{5a}^{(1)}(\xi_\pm,\zeta)
    +
    \frac{2- x^2 \mathcal{P}^2/Q^2}{[1+x^2\mathcal{P}^2/Q^2]}
    \widehat\Phi_{5a}^{(2)}(\xi_\pm,\zeta)
    \right],
    \label{G0}
\end{align}
where repeated use has been made of relations (\ref{rel1}). With
these definitions we finally get
\begin{eqnarray}
 {\mathrm{Im}}\, T^\twz_{[\mu\nu]}\kln{q}
 &=&
 {\mathrm{Im}}\, T^\twz_{[\mu\nu]+}\kln{q}
 +
 {\mathrm{Im}}\, T^\twz_{[\mu\nu]-}\kln{q}
 \\
 {\mathrm{Im}}\, T^\twz_{[\mu\nu]\pm}\kln{q}
    &=&
    -\frac{\pi}{2\kappa}\int d\zeta  \,
    \epsilon_{\mu\nu}^{\phantom{\mu\nu}\alpha\beta}\,
    \Bigg\{
    \frac{q_{\alpha}\,{\cal K}^a_\beta}{q{\cal P}} 
    \left[g^\pm_{a1}(x,\eta;\zeta) + g^\pm_{a2}(x,\eta;\zeta)\right]
    \nonumber\\
    && \qquad\qquad\qquad\qquad
    - \frac{q_{\alpha}\,\mathcal{P}_\beta}{q{\cal P}}
    \frac{(q{\cal K}^a)}{q{\cal P}} g^\pm_2(x,\eta;\zeta)
    + \frac{1}{2}\frac{\mathcal{P}^2}{Q^2}
    \frac{q_{\alpha}\,\mathcal{P}_\beta}{q{\cal P}}
    \frac{(\mathcal{P}{\cal K}^a)}{\mathcal{P}^2}
    g^\pm_{a0}(x,\eta;\zeta) 
    \Bigg\}\,,
    \label{Tas7}
\end{eqnarray}
where ${\cal P}= {\cal P}(\eta,\zeta)$ depends also on $\zeta$!

Obviously, the contributions to the various kinematical
expressions are related to only two independent quantities.
Namely, introducing $\mathcal{F}_a(x,\eta;\zeta)=
\mathcal{F}^+_a(x,\eta;\zeta) + \mathcal{F}^-_a(x,\eta;\zeta)$ and
$\mathcal{F}^0_a(x,\eta;\zeta)= \xi_+\mathcal{F}^+_a(x,\eta;\zeta)
+ \xi_-\mathcal{F}^-_a(x,\eta;\zeta)$ with
\begin{eqnarray}
\mathcal{F}^\pm_a(x,\eta;\zeta)
    = \left(
    \frac{x}{\xi_\pm}
    \frac{\widehat\Phi_{5a}^{(2)}(\xi_\pm,\zeta)}
    {[1+x^2\mathcal{P}^2/Q^2]^{1/2}}
    \right)
\end{eqnarray}
we may rewrite the relations (\ref{G1}) --  (\ref{G0}) as follows
\begin{eqnarray}
    g_{a1}(x,\eta;\zeta) &=&
    x\frac{\partial}{\partial x}
    x\frac{\partial}{\partial x}
    \mathcal{F}_a(x,\eta;\zeta)\,,
    \nonumber\\
    g_{a2}(x,\eta;\zeta) &=&
    - x\frac{\partial}{\partial x}
    (x\frac{\partial}{\partial x} + 1)
    \mathcal{F}_a(x,\eta;\zeta)\,,
    \nonumber\\
    g_{a 0}(x,\eta;\zeta) &=& 
    x\frac{\partial}{\partial x}
    (x\frac{\partial}{\partial x} - 1)
    x\,\mathcal{F}^0_a(x,\eta;\zeta)\,.
\label{G_rel2}
\end{eqnarray}
The first two of these relations are identical in their functional
dependence with the corresponding equations in the forward case as
given in Ref.~\cite{BT99}. Therefore, they will lead to the same
general consequences:
Observing
\begin{eqnarray}
    x\frac{\partial}{\partial x}\mathcal{F}_a(x,\eta;\zeta)
    &=&
    - \int_x^1 \frac{dy}{y}g_{a1}(y,\eta;\zeta)\,,
    \nonumber\\
    \mathcal{F}_a(x,\eta;\zeta)
    &=&
    - \int_x^1 \frac{dy}{y} \int_y^1 \frac{dy'}{y'}g_{a1}(y',\eta;\zeta)\,,
    \nonumber
\end{eqnarray}
we find that, surprisingly, the well known Wandzura-Wilczek
relation from the case of forward scattering \cite{WW77} gets
generalized also to the non-forward case, namely, the (twist-2
part of) $g_{a2}(x,\eta;\zeta)$ fulfills the same relation as does
the (twist-2 part of) $g_{a2}(x)$,
\begin{eqnarray}
    g^{(2)}_{a2}(x,\eta;\zeta)
    &=& -\, g_{a1}(x,\eta;\zeta) + \int_x^1 \frac{dy}{y}g_{a1}(y,\eta;\zeta)\,.
    \label{g22}
\end{eqnarray}
However, more surprising is the fact that exactly the same structure
occurs independent of the kinematic form factors, i.e., for the Dirac
and the Pauli form factor as well any other independent form factor.
This means
that we are confronted with a very general structure of the theory
which seems to be independent of taking matrix elements. Instead,
it might be conjectured that this structure is a property of the
(leading part of the) Compton operator (when Fourier transformed
with the leading part of the propagator).

In addition, another relation holds for the (twist-2 part of) the
new structure function $g_{a0}(x,\eta;\zeta)$ which occurs only in
the case of non-forward scattering and is suppressed by a factor
${\mathcal{P}^2}/{Q^2}$:
\begin{align}
    g_{a0}^{(2)\pm}(x,\eta;\zeta)
    &=
    (x\xi_\pm )\,g^\pm_{a1}(x,\eta;\zeta)
    - \frac{2x^2+ x\xi_\pm}{[1+x^2\mathcal{P}^2/Q^2]^{1/2}}
    \int_x^1\! \frac{dy}{y}g^\pm_{a1}(y,\eta;\zeta)
    -\frac{2x^2}{[1+x^2\mathcal{P}^2/Q^2]^{3/2}}
    \int_x^1\! \frac{dy}{y} \!\int_y^1\! \frac{dy'}{y'}g^\pm_{a1}(y'\!,\eta;\zeta)\,.
    \label{g02}
\end{align}

Let us add some remarks.

First we mention that, in principle, both roots, $\xi_+$ and
$\xi_-$, will contribute. However, because of the kinematics for
the generalized Bjorken region only the first one is of
phenomenological relevance. Under these circumstances {\it only
one independent} function ${\cal F}^+_a$ survives, and
$g_{a0}^{(2)+}(x,\eta;\zeta)$ is completely determined by
$g_{a1}^{+}(x,\eta;\zeta)$.

Second, we underline once more that these relations result without
using the equations of motion or any other dynamical input as had been
done by other authors \cite{dWW}. They are derived by using only the
correct non-local twist-2 vector operators off-cone. In this respect
their derivation is similar to the derivation of the WW-relation for
the quark distribution functions in Ref.~\cite{GL01}.

Third, we emphasize that the derivation of these relations essentially
depends on the use of appropriate coordinates $(t,\zeta)$ with $t$ being
well-adopted to the forward case and afterwards being replaced by the
generalizations $\xi_\pm$ of the Nachtmann variable. Also some special
relations had to be used which are fulfilled only when the imaginary part
of the Compton amplitude is considered.

Finally, we point to the fact that both relations hold {\em
before} the integration over $\zeta$ is performed. This means that
the above mentioned general structure will be covered by the
$\zeta-$dependence of the distribution amplitudes. In the case of
forward scattering that integration disappears and the well-known
result is obtained for the functions $g_{ai}(x)$ being independent
of $\eta$ and $\zeta$(see, e.g., \cite{BT99}).

This finishes the consideration of the imaginary part of the non-forward
Compton amplitude.

%% file: M_Tsym3.tex
\section{Non-forward Compton amplitude: Symmetric Part (nonlocal)}
\renewcommand{\theequation}{\thesection.\arabic{equation}}
\setcounter{equation}{0}
\label{sec-1}

Now, we consider the Fourier transformation of the symmetric part of the
leading twist-2 non-forward Compton amplitude being obtained by taking the matrix elements of the operator expression (\ref{OS1}):
\begin{align}
T^\twz_{\{\mu\nu\}}\kln{q}
 &=
 \im \int \frac{\text{d}^4 \!x}{2\pi^2} \;%
 \frac{\e^{\im qx}} { {\kln{x^2 - \im\epsilon}^2} }\;%
 S^{\mu\nu|\alpha\beta} x_{\alpha}%
 \langle P_2,S_2 | %
 \Big( O^{\rm tw 2}_{\beta} \Kln{\kappa x, -\kappa x}
 - O^{\rm tw 2}_{\beta} \Kln{- \kappa x, \kappa x} %
 \Big) |P_1,S_1 \rangle
\nonumber \\ %
&=
 2\im \, \int_0^1 \frac{\d\tau}{\tau} \; \kln{1-\tau
 + \tau \ln\tau} \int \frac{\d^4 \! u}{\kappa^4\tau^4} \;
 \langle P_2,S_2 |\Big(O^\rho \Big(\frac{u}{\kappa\tau}\Big) -
       O^\rho \Big(\frac{-u}{\kappa\tau}\Big)\Big)|P_1,S_1 \rangle
 \, \partial_\rho^u \Bigg[\,\frac{2}{[(q + u)^2 + i\epsilon]^3}
\nonumber\\
& \quad \times
 \Big\{\kln{ q^2 \, g_{\mu\nu} - q_\mu q_\nu }  \,
 \left[(u q)^2- u^2q^2 + \hbox{$\frac{1}{2}$} u^2 (q + u)^2 \right]
 + \,  \kln{ u_\mu {q^2}- {q_\mu \kln{uq}} }
       \kln{ u_\nu {q^2}- {q_\nu \kln{uq}} }\Big\}\;
 \Bigg]
 \nonumber\\
&= %
2\int_0^1 \frac{d\tau_1}{\tau_1} \int_0^1 d\tau_2 \int_0^1 d\tau_3
\int D {\mathbb Z} \; f^{\twz}_a({\mathbb Z}, \mu^2) \;
 {\cal K}^a_\rho({\mathbb P},{\mathbb S})\; \left[q^2\right]^3
 \frac{1}{\kappa\tau_1 \tau_2 \tau_3}\partial_{\mathbb{PZ}}^\rho
\nonumber \\
& \hspace{3cm}
\times
\left\{
  \frac{2 A_{\mu\nu}^{\mathrm T}(q,\kappa\tau_1\tau_2 \tau_3\,\mathbb{PZ})}
  {\big[(q + \kappa\tau_1\tau_2\tau_3{\mathbb{PZ}})^2 + \im\epsilon \big]^3}
  + \frac{B_{\mu\nu}^{\mathrm T}(q,\kappa\tau_1\tau_2\tau_3\,\mathbb{PZ})}
  {\big[(q + \kappa\tau_1\tau_2\tau_3{\mathbb{PZ}})^2 + \im\epsilon \big]^2}
  \right.
\nonumber\\
& \hspace{3cm}\quad
  + \left.
  \frac{2 A_{\mu\nu}^{\mathrm T}(q,\kappa\tau_1\tau_2 \tau_3\,\mathbb{PZ})}
  {\big[(q - \kappa \tau_1 \tau_2 \tau_3{\mathbb{PZ}})^2 + \im\epsilon\big]^3}
  + \frac{ B_{\mu\nu}^{\mathrm T}(q,\kappa\tau_1\tau_2\tau_3\,\mathbb{PZ})}
  {\big[(q - \kappa \tau_1 \tau_2 \tau_3{\mathbb{PZ}})^2 + \im\epsilon\big]^2}
\right\}
\nonumber \\
&=
2\int_{0}^1 d\tau \int_0^1 d\sigma \int_0^1 d\rho
\int D {\mathbb Z} \; \Phi_a({\mathbb Z}, \mu^2) \;
 {\cal K}^a_\rho({\mathbb P},{\mathbb S})\; \left[q^2\right]^3
 \partial_{\widetilde\Pi}^\rho
 \left\{
  \frac{2 A_{\mu\nu}^{\mathrm T}(q,\widetilde\Pi)}{\big[
  \tilde R(\tau) + \im \epsilon \big]^3}
  + \frac{B_{\mu\nu}^{\mathrm T}(q,\widetilde\Pi)}{\big[
  \tilde R(\tau) + \im \epsilon \big]^2}
\right\} ,
\label{Ts1}
\end{align}

Again, we used the correspondence (\ref{gda0}),
 the kinematic form factors ${\cal K}_\alpha^a({\mathbb P},{\mathbb S})
 = \{\bar u( P_2,S_2 ) \gamma_\alpha  u(P_1,S_1 ),$
   $\bar u( P_2,S_2 ) \sigma_{\alpha\beta} p^\beta_- u(P_1,S_1 ),\cdots\}$
and the generalized distribution amplitudes $\Phi_a({\mathbb Z}, \mu^2)$
introduced already for the trace part; in addition, we introduced the
following new abbreviations:
\begin{eqnarray}
&&A_{\mu\nu}^{\mathrm T}(q,\Pi)
  = g_{\mu\nu}^{\mathrm T}\Big(\frac{q\Pi}{q^2}\Big)^2
  \Big(1 - \frac{q^2 \Pi^2}{(q\Pi)^2}\Big)
  +
  \frac{1}{q^2}\Pi_\mu^{\mathrm T} \Pi_\nu^{\mathrm T} \,,
\qquad
 B_{\mu\nu}^{\mathrm T}(q,\Pi)
  = g_{\mu\nu}^{\mathrm T}\frac{\Pi^2}{[q^2]^2}\,,
 \label{abkS1}
\\
&& \widetilde\Pi_\mu = |\kappa|\,\sigma\rho\,{\mathbb{P_\mu Z}} = \sigma\rho\Pi,
\qquad
g_{\mu\nu}^{\mathrm T} := g_{\mu\nu} - \frac{q_\mu q_\nu}{q^2},
\qquad
\Pi_\mu^{\mathrm T} = g_{\mu\nu}^{\mathrm T} \Pi^\nu
 = \Pi_\mu - q_\mu \frac{(q\Pi)}{q^2}
\label{abkS2}
\end{eqnarray}
together with $\tilde R(\tau) = (q + \tau \widetilde\Pi)^2$.

Also in that case we reduce the unpleasant denominators $1/(\tilde
R(\tau) + \im \epsilon)^3$ and $1/(\tilde R(\tau) + \im
\epsilon)^2$ by partial integration to $1/(\tilde R(\tau) + \im
\epsilon)$. But here the computation is much more complicated than
in the antisymmetric case, first, because a third power appears,
and, second, because also the additional integrations over
$\sigma$ and $\rho$ have to be taken into account. The explicit
calculation will be sketched in the Appendix D. Since, finally, we
are interested solely in the imaginary part for which, due to the
appearance of $(q^2)^3$, the terms proportional to $1/( R(0) + \im
\epsilon)$ are irrelevant, we determined only those terms which
are proportional to $1/(R(1) + \im \epsilon)$ (where, again
$R(1)= (q + \Pi)^2$, as in the antisymmetric case).

The result of the calculation is as follows:
\begin{eqnarray}
T^\twz_{\{\mu\nu\}}\kln{q}
 &=&
 \frac{q^2}{2} \int D {\mathbb Z}
 \Bigg\{
 \frac{q{\cal K}_a}{q\Pi} \Big[
 g_{\mu\nu}^{\mathrm T} F^a_1({\mathbb Z})
 +
 \frac{q^2 \Pi_\mu^{\mathrm T} \Pi_\nu^{\mathrm T}}{(q\Pi)^2-q^2\Pi^2}
 F^a_2({\mathbb Z}) \Big]
 \nonumber \\
&& \qquad\qquad
 +
 \Big( \frac{q{\cal K}_a}{q\Pi} - \frac{\Pi{\cal K}_a}{\Pi^2} \Big)
 \Big[
 g_{\mu\nu}^{\mathrm T} F^a_3({\mathbb Z})
 +
 \frac{q^2 \Pi_\mu^{\mathrm T} \Pi_\nu^{\mathrm T}}{(q\Pi)^2-q^2\Pi^2}
 F^a_4({\mathbb Z}) \Big]
 \nonumber \\
&& \qquad\qquad
 +
  \Big({\cal K}_{a\mu}^{\mathrm T} \Pi_\nu^{\mathrm T}
 + \Pi_\mu^{\mathrm T} {\cal K}_{a\nu}^{\mathrm T}
 - 2 \frac{q{\cal K}_a}{q\Pi} \Pi_\mu^{\mathrm T} \Pi_\nu^{\mathrm T} \Big)
 \frac{q^2}{(q\Pi)^2-q^2\Pi^2}
 F_5^a({\mathbb Z})
\Bigg\} \frac{1}{R(1)+ \im\epsilon}
\label{Ts_nonf}
\end{eqnarray}
 and
\begin{eqnarray}
\label{TR}
T^\twz_{\rm trace}\kln{q}
 &=&
 \frac{q^2}{2} \int D {\mathbb Z}
 \Bigg\{
 \frac{q{\cal K}_a}{q\Pi}\Big(3 F_1^a({\mathbb Z})  - F_2^a({\mathbb Z}) \Big)
 \nonumber\\
 &&\qquad\qquad +
 \Big( \frac{q{\cal K}_a}{q\Pi} - \frac{\Pi{\cal K}_a}{\Pi^2} \Big)
 \Big(3 F_3^a({\mathbb Z})  - F_4^a({\mathbb Z})
 - 2 \frac{q^2\Pi^2}{(q\Pi)^2-q^2\Pi^2} F_5^a({\mathbb Z}) \Big)
 \Bigg\} \frac{1}{R(1)+ \im\epsilon}\,,
\end{eqnarray}
with the functions $ F^a({\mathbb Z},q,\Pi) $ being given as follows:
\begin{align}
\label{F1}
F^a_1({\mathbb Z})
  & =
  \Phi_a({\mathbb Z})
  + \frac{\Pi^2 (q\Pi+ \Pi^2)}{(q\Pi)^2-q^2\Pi^2}
  \int_0^1\frac{d\tau_2}{\tau^2_2}
  \Phi_a\Big(\frac{{\mathbb Z}}{\tau_2}\Big)
  + \frac{[\Pi^2]^2}{(q\Pi)^2-q^2\Pi^2}
  \int_0^1\frac{d\tau_2}{\tau^2_2}
  \int_0^1\frac{d\tau_1}{\tau^3_1}
  \Phi_a\Big(\frac{{\mathbb Z}}{\tau_1 \tau_2}\Big),
\\
\label{F2}
F^a_2({\mathbb Z})
  &=
  \Phi_a({\mathbb Z})
  + \frac{3\Pi^2 (q\Pi+ \Pi^2)}{(q\Pi)^2-q^2\Pi^2}
  \int_0^1\frac{d\tau_2}{\tau^2_2}
  \Phi_a\Big(\frac{{\mathbb Z}}{\tau_2}\Big)
  + \frac{3[\Pi^2]^2}{(q\Pi)^2-q^2\Pi^2}
  \int_0^1\frac{d\tau_2}{\tau^2_2}
  \int_0^1\frac{d\tau_1}{\tau^3_1}
  \Phi_a\Big(\frac{{\mathbb Z}}{\tau_1 \tau_2}\Big)
  \nonumber\\
  &= 3 F^a_1({\mathbb Z}) - 2 \Phi_a({\mathbb Z})\phantom{\bigg|} ,
\\
\label{F3}
F^a_3({\mathbb Z})
  &=
 \frac{\Pi^2(q\Pi+ q^2)}{(q\Pi)^2-q^2\Pi^2}
  \Phi_a({\mathbb Z})
  +
  \frac{\Pi^2(q\Pi- \Pi^2)}{(q\Pi)^2-q^2\Pi^2}
  \int_0^1\frac{d\tau}{\tau^2_3}
  \Phi_a\Big(\frac{{\mathbb Z}}{\tau_3}\Big)
  \nonumber \\
  & \quad
  -
  \frac{[\Pi^2]^2[4(q\Pi)^2 - q^2 \Pi^2] +
      (q\Pi)\Pi^2[2(q\Pi)^2 + q^2 \Pi^2]}{[(q\Pi)^2-q^2\Pi^2]^2}
  \int_0^1\frac{d\tau_3}{\tau^2_3}
  \int_0^1\frac{d\tau_2}{\tau^2_2}
  \Phi_a\Big(\frac{{\mathbb Z}}{\tau_2 \tau_3}\Big)
  \nonumber \\
  & \quad
  -
  \frac{[\Pi^2]^2[4(q\Pi)^2 - q^2 \Pi^2]}{[(q\Pi)^2-q^2\Pi^2]^2}
  \int_0^1\frac{d\tau_3}{\tau^2_3}
  \int_0^1\frac{d\tau_2}{\tau^2_2}
  \int_0^1\frac{d\tau_1}{\tau^3_1}
  \Phi_a\Big(\frac{{\mathbb Z}}{\tau_1 \tau_2 \tau_3}\Big)
  \nonumber\\
  &=
  -\int_0^1\frac{d\tau}{\tau^2}
  \bigg[F_1^a\Big(\frac{{\mathbb Z}}{\tau}\Big)
  + \frac{(q\Pi)^2}{(q\Pi)^2-q^2\Pi^2}
    F_2^a\Big(\frac{{\mathbb Z}}{\tau}\Big)\bigg]
  - \frac{\Pi^2(q+\Pi)^2}{(q\Pi)^2-q^2\Pi^2}
  \int_0^1\frac{d\tau}{\tau^2}\Phi_a\Big(\frac{{\mathbb Z}}{\tau}\Big)
  \nonumber \\
& \quad
  + \frac{2(q\Pi)}{(q\Pi)^2-q^2\Pi^2}
  \int_0^1\frac{d\tau}{\tau^2}
  \bigg(
  (q\Pi+ \Pi^2)\, \Phi_a\Big(\frac{{\mathbb Z}}{\tau}\Big)
  +\Pi^2 \int_0^1\frac{d\tau_1}{\tau^2_1}
  \Phi_a\Big(\frac{{\mathbb Z}}{\tau\tau_1}\Big)
  \bigg)
  \nonumber \\
&  \quad
  +\frac{\Pi^2}{(q\Pi)^2-q^2\Pi^2}
  \bigg(
  (q\Pi+ q^2)\, \Phi_a({\mathbb Z})
  +  q\Pi \int_0^1\frac{d\tau}{\tau^2}
  \Phi_a \Big(\frac{{\mathbb Z}}{\tau}\Big)
  \bigg),
\\
\intertext{}
\label{F4}
F^a_4({\mathbb Z})
  &=
  \frac{\Pi^2(q\Pi+ q^2)}{(q\Pi)^2-q^2\Pi^2}
  \Phi_a({\mathbb Z})
  +
  \frac{\Pi^2[q\Pi- 3\Pi^2 - 2q^2]}{(q\Pi)^2-q^2\Pi^2}
  \int_0^1\frac{d\tau_3}{\tau^2_3}
  \Phi_a\Big(\frac{{\mathbb Z}}{\tau_3}\Big)
 \nonumber \\
  & \quad
  - 3
  \frac{[\Pi^2]^2[4(q\Pi)^2 + q^2 \Pi^2] +
      (q\Pi)\Pi^2[2(q\Pi)^2 + 3 q^2 \Pi^2]}{[(q\Pi)^2-q^2\Pi^2]^2}
  \int_0^1\frac{d\tau_3}{\tau^2_3}
  \int_0^1\frac{d\tau_2}{\tau^2_2}
  \Phi_a\Big(\frac{{\mathbb Z}}{\tau_2 \tau_3}\Big)
  \nonumber \\
  & \quad
  - 3
  \frac{[\Pi^2]^2[4(q\Pi)^2 + q^2 \Pi^2]}{[(q\Pi)^2-q^2\Pi^2]^2}
  \int_0^1\frac{d\tau_3}{\tau^2_3}
  \int_0^1\frac{d\tau_2}{\tau^2_2}
  \int_0^1\frac{d\tau_1}{\tau^3_1}
  \Phi_a\Big(\frac{{\mathbb Z}}{\tau_1 \tau_2 \tau_3}\Big)
  \nonumber\\
  &=
  3 F^a_3({\mathbb Z})
  - 2  \frac{q^2\Pi^2 }{(q\Pi)^2-q^2\Pi^2}\, F^a_5({\mathbb Z})
  \nonumber \\
  & \quad
  - 2  \frac{\Pi^2}{(q\Pi)^2-q^2\Pi^2}
  \bigg(
  (q\Pi+ q^2)\, \Phi_a({\mathbb Z})
  +  q\Pi   \int_0^1\frac{d\tau}{\tau^2}
  \Phi_a \Big(\frac{{\mathbb Z}}{\tau}\Big)
  \bigg) ,
\\
F^a_5({\mathbb Z})
  &=
  \int_0^1\frac{d\tau_3}{\tau^2_3}
  \Phi_a\Big(\frac{{\mathbb Z}}{\tau_3}\Big)
  +
  3  \frac{\Pi^2[q\Pi+ \Pi^2]}{(q\Pi)^2-q^2\Pi^2}
  \int_0^1\frac{d\tau_3}{\tau^2_3}
  \int_0^1\frac{d\tau_2}{\tau^2_2}
  \Phi_a\Big(\frac{{\mathbb Z}}{\tau_2 \tau_3}\Big)
  \nonumber \\
  & \quad
  +
  3  \frac{[\Pi^2]^2}{(q\Pi)^2-q^2\Pi^2}
  \int_0^1\frac{d\tau_3}{\tau^2_3}
  \int_0^1\frac{d\tau_2}{\tau^2_2}
  \int_0^1\frac{d\tau_1}{\tau^3_1}
  \Phi_a\Big(\frac{{\mathbb Z}}{\tau_1 \tau_2 \tau_3}\Big),
\nonumber \\
  &=
  \int_0^1\frac{d\tau}{\tau^2}
  F_2^a\Big(\frac{{\mathbb Z}}{\tau}\Big)\,.
  \label{F5}
\end{align}

Using these expressions for the trace part (\ref{TR}) we get,
modulo $1/R(0)$-terms, the former result (\ref{FTraceX}),
\begin{eqnarray}
T^\twz_{\rm trace}\kln{q}
 &=&
 \int D {\mathbb Z}
 \frac{q^2}{(q\Pi)^2-q^2\Pi^2} \,\Bigg\{
 (q{\cal K}^a)
 \Big[
 (q\Pi + \Pi^2)\Phi_a({\mathbb Z})
 +
 \Pi^2 \int_0^1\frac{d\tau}{\tau^2}
 \Phi_a\Big(\frac{{\mathbb Z}}{\tau}\Big) \Big]
 \nonumber\\
 &&\qquad\qquad\qquad\qquad~
 - (\Pi{\cal K}^a)
 \Big[
 (q^2 + q\Pi)
 \Phi_a({\mathbb Z})
 +
 q\Pi \int_0^1\frac{d\tau}{\tau^2}
 \Phi_a\Big(\frac{{\mathbb Z}}{\tau}\Big) \Big]
 \Bigg\}
 \frac{1}{R(1)+ \im\epsilon}\,.
 \label{Ttr_nonf}
 \end{eqnarray}

Furthermore, we observe that the functions $F^a_i, i = 3,4,5\,,$
are expressed by the functions $F_1$ and $F_2$ as well as those
combinations of (integrals over) $\Phi_a({\mathbb Z}/\tau)$
which already appear in the trace part [and for $F_3$ a term which
vanishes when the imaginary part is taken]. Therefore, because of the
similarity in the structure of $F_1$ and $F_2$ with the structure
of $W_1$ and $W_2$ in Eqs.~(\ref{WW0}) it will be advantageous first to
introduce the new coordinates $(t,\zeta)$ and, thereby, to restrict to the
forward case (where $F^a_i, i = 3,4,5\,,$ do not contribute).

\subsection{Restriction to the forward case}

Therefore, let us consider the case of forward scattering,
i.e., $P_1 = P_2 = P = \frac{1}{2}p_+,\, p_- \equiv 0,\, S_1 = S_2
= S$. Then we get ($z_+ \equiv z$ and $z_- $ being integrated out)
\begin{eqnarray}
{\cal K}^a_\mu \equiv {\cal K}^D_\mu = 2 P_\mu,
\qquad
\Pi_\mu =  P_\mu z,
\end{eqnarray}
and the expression (\ref{Ts_nonf}) reduces as follows:
\begin{eqnarray}
^f T^{\twz}_{\{\mu\nu\}}\kln{q}
 &=&
 q^2 \int \frac{dz}{z}
 \Big\{
 g_{\mu\nu}^{\mathrm T} F_{1f}(z)
 +
 \frac{q^2 P_\mu^{\mathrm T} P_\nu^{\mathrm T}}{(qP)^2-q^2P^2} F_{2f}(z)
 \Big\}
\label{Ts_f}
\end{eqnarray}
with
\begin{eqnarray}
\label{F1f}
F_{1f} (z)
  &\! =\! &
  f(z)
  + \frac{P^2 (z qP + z^2 P^2)}{(qP)^2-q^2P^2}
  \int_0^1\frac{d\tau_2}{\tau_2}
  f\Big(\frac{z}{\tau_2}\Big)
  + \frac{ z^2 [P^2]^2}{(qP)^2-q^2P^2}
  \int_0^1\frac{d\tau_2}{\tau_2}
  \int_0^1\frac{d\tau_1}{\tau^2_1}
  f\Big(\frac{z}{\tau_1 \tau_2}\Big),
\\
\label{F2f}
F_{2f}(z)
  &\!=\!&
  f(z)
  + \frac{3P^2 (z qP+ z^2 P^2)}{(qP)^2-q^2P^2}
  \int_0^1\frac{d\tau_2}{\tau_2}
  f\Big(\frac{z}{\tau_2}\Big)
  + \frac{3 z^2[P^2]^2}{(qP)^2-q^2P^2}
  \int_0^1\frac{d\tau_2}{\tau_2}
  \int_0^1\frac{d\tau_1}{\tau^2_1}
  f\Big(\frac{z}{\tau_1 \tau_2}\Big),
\end{eqnarray}
with $f\equiv \Phi_D$. Of course, in the forward case only
a single variable $z$ occurs and, therefore, the powers of
$\tau_i$ in the various integrals are reduced by one.

Now, we introduce generalized distribution amplitudes of order $n$ adjusted
to the forward case:
\begin{eqnarray}
 f^{(0)}(z) &\equiv& \frac{1}{z} f(z),
 \nonumber\\
 f^{(1)}(z) &\equiv& \int_z^1 dy\, \frac{f(y)}{y}
                = \int_0^1\frac{d\tau_1}{\tau_1}\,f\Big(\frac{z}{\tau_1}\Big),
 \nonumber\\
 f^{(2)}(z) &\equiv& \int_z^1 dy_2 \int_{y_2}^1\, dy_1\,\frac{f(y_1)}{y_1}
   = z  \int_0^1\frac{d\tau_1}{\tau^2_1}
   \int_0^1\frac{d\tau_2}{\tau_2}\,f\Big(\frac{z}{\tau_1 \tau_2}\Big),
  \nonumber\\
 f^{(n)}(z) &\equiv& \int_z^1 dy\, f^{(n-1)}(y),
 \label{GDWf}
\end{eqnarray}
where the support restriction $f(z) = 0 $ for $z>1$ has been used.
Then, observing the overall factor of $1/z$, the integrand of
Eq.~(\ref{Ts_f}) can be written in terms of these generalized
distribution amplitudes $ f^{(n)}(z)$.

Taking the imaginary part of the amplitude (\ref{Ts_f}), observing
\begin{eqnarray}
\mathrm{Im} \frac{1}{R(1)+\im \epsilon} = \frac{-
\pi}{2z}\frac{1}{\sqrt{(qP)^2-q^2 P^2}} \left(\delta(1 -
\xi_+/z)+\delta(1 - \xi_-/z)\right)
\end{eqnarray}
with the well-known Nachtmann variable(s)
\begin{eqnarray}
\xi_\pm = \frac{2x}{1\pm\sqrt{1 + 4x^2 P^2/Q^2}}
\quad {\rm with} \quad
x \equiv x_{\mathrm Bj} = Q^2/(2 qP),
\end{eqnarray}
and using the equalities (\ref{+}) and the distributions (\ref{GDWf}), we get
for the two different contributions,
\begin{align}
\hspace{-.25cm}
 {\rm Im}\,^f T^\twz_{\{\mu\nu\}\pm}\kln{q}
 &=
 \frac{-\pi}{2} \int {dz} \;  \delta(z - \xi_\pm)
 \nonumber\\
 \hspace{-.25cm}
 &\;\times\;
 \bigg[
 \frac{q^2 g_{\mu\nu}^{\mathrm T} }{[(qP)^2 - q^2 P^2]^{1/2}}
 \Big\{
      {f^{(0)}(z)}
  \pm \frac{P^2}{[(qP)^2 - q^2 P^2]^{1/2}} f^{(1)}(z)
  + \frac{(P^2)^2}{[(qP)^2-q^2P^2]} f^{(2)}(z)
 \Big\}
 \nonumber\\
\hspace{-.25cm}
 &\;+ \;
 \frac{[q^2]^2\,P_\mu^{\mathrm T} P_\nu^{\mathrm T}}{[(qP)^2 - q^2 P^2]^{3/2}}
 \Big\{
  {f^{(0)}(z)}
  \pm 3 \frac{P^2}{[(qP)^2 - q^2 P^2]^{1/2}} f^{(1)}(z)
  + 3 \frac{(P^2)^2}{[(qP)^2-q^2P^2]} f^{(2)}(z)
 \Big\}
 \bigg]\phantom{\Bigg|}
 \nonumber\\
 \hspace{-.25cm}
 &=
  \pi\bigg[
 \frac{x g_{\mu\nu}^{\mathrm T}}{[1+4x^2 M^2/Q^2]^{1/2}}
 \Big\{
    f^{(0)}(\xi_\pm)
  \pm \frac{2x {M^2}/{Q^2}}{[1+4x^2 M^2/Q^2]^{1/2}}f^{(1)}(\xi_\pm)
  + \frac{4x^2 ({M^2}/{Q^2})^2}{[1+4x^2 M^2/Q^2]}f^{(2)}(\xi_\pm)
 \Big\}
 \nonumber\\
 \hspace{-.25cm}
 &\; -
 \frac{4 x^3 P_\mu^{\mathrm T} P_\nu^{\mathrm T}/Q^2}{[1+4x^2 M^2/Q^2]^{3/2}}
 \Big\{
  f^{(0)}(\xi_\pm)
  \pm \frac{ 6 x {M^2}/{Q^2}}{[1+4x^2 M^2/Q^2]^{1/2}}f^{(1)}(\xi_\pm)
  + \frac{12 x^2 ({M^2}/{Q^2})^2}{[1+4x^2 M^2/Q^2]}f^{(2)}(\xi_\pm)
 \Big\}\bigg]
 \label{Ts_fim}
\end{align}
and
\begin{eqnarray}
\frac{1}{\pi}{\rm Im}\,^f T^\twz_{{\rm trace}\,\pm}\kln{q}
 &=&
 \frac{ 2  x }{[1+4x^2 M^2/Q^2]^{1/2}}\,f^{(0)}(\xi_\pm)\,.
\end{eqnarray}

Taking into account the definition of the structure functions
$W_1$ and $W_2$ according to
\begin{eqnarray}
\frac{1}{\pi}{\rm Im}\,^f T^\twz_{\{\mu\nu\}}\kln{q}
 &=&
 - g_{\mu\nu}^{\mathrm T}\, W_1 (x, Q^2)
 + \frac{P_\mu^{\mathrm T} P_\nu^{\mathrm T}}{M^2} \,W_2(x, Q^2)
\end{eqnarray}
we observe that both contributions independently coincide with the
result of Georgi and Politzer \cite{GP76} leading to the well known
consequences ($G(\xi) \equiv f^{(2)}(\xi), \xi\equiv\xi_\pm $):
\begin{eqnarray}
2xW_L &=& - \frac{4 x^2 M^2}{Q^2}\,x\frac{\pd}{\pd x}
\bigg(\frac{x}{\xi}\frac{G(\xi)}{\sqrt{1+4x^2 M^2/Q^2}}\bigg),
\label{for1}\\
\frac{\nu W_2}{M} \equiv \frac{W_2\,Q^2}{2x\,M^2} &=& x^2
\frac{\pd^2}{\pd x^2}
\bigg(\frac{x^2}{\xi^2}\frac{G(\xi)}{\sqrt{1+4x^2 M^2/Q^2}}\bigg),
\label{for2}\\
2xW_1 &=& (1+4x^2 M^2/Q^2) \frac{\nu W_2}{M} - 2xW_L
\nonumber\\
&=& \frac{\nu W_2}{M} + \frac{4 x^2 M^2}{Q^2}
\bigg[\bigg((x\partial)\frac{x}{\xi} + ((x\partial)
-1)(x\partial)\frac{x^2}{\xi^2} \bigg) \frac{G(\xi)}{\sqrt{1+4x^2
M^2/Q^2}} \bigg]\,. \label{for3}
\end{eqnarray}
For the kinematic situation considered in Ref.~\cite{GP76} the terms
related to $\xi_-$ do not contribute.

\subsection{The symmetric part of the Compton amplitude in non-forward case}

Now, let us consider the imaginary part of the symmetric Compton amplitude
in the nonforward case already introducing the variables $(t,\zeta)$ and
using $\Pi_\mu = \kappa t {\cal P}_\mu$ :
\begin{eqnarray}
{\rm Im}\,T^\twz_{\{\mu\nu\}}\kln{q}
 &=&
 \frac{- \pi}{2 \kappa^2} \int d\zeta \int \frac{dt}{t}\,
 \frac{q^2}{[(q{\cal P})^2 - q^2 {\cal P}^2]^{1/2}}
 \Big[\delta(1-\xi_+(\zeta)/t) + \delta(1-\xi_-(\zeta)/t)\Big]\times
 \nonumber \\
&& \qquad\qquad
 \Bigg\{
 \frac{q{\cal K}_a}{q{\cal P}}\; \Big[
 g_{\mu\nu}^{\mathrm T} F^a_1(t,\zeta)
 +
 \frac{q^2 {\cal P}_\mu^{\mathrm T} {\cal P}_\nu^{\mathrm T}}
      {(q{\cal P})^2-q^2{\cal P}^2}
 F^a_2(t,\zeta) \Big]
 \nonumber \\
&& \qquad\qquad
 +
 \Big( \frac{q{\cal K}_a}{q{\cal P}}
     - \frac{{\cal P}{\cal K}_a}{{\cal P}^2} \Big)
 \Big[
 g_{\mu\nu}^{\mathrm T} F^a_3(t,\zeta)
 +
 \frac{q^2 {\cal P}_\mu^{\mathrm T} {\cal P}_\nu^{\mathrm T}}
      {(q{\cal P})^2-q^2{\cal P}^2}
 F^a_4(t,\zeta) \Big]
 \nonumber \\
&& \qquad\qquad
 +
 \Big({\cal K}_{a\mu}^{\mathrm T} {\cal P}_\nu^{\mathrm T}
 + {\cal P}_\mu^{\mathrm T} {\cal K}_{a\nu}^{\mathrm T}
 - 2 \frac{q{\cal K}_a}{q{\cal P}}
          {\cal P}_\mu^{\mathrm T} {\cal P}_\nu^{\mathrm T} \Big)
 \frac{q^2}{(q{\cal P})^2-q^2{\cal P}^2}
 F^a_5(t,\zeta)
\Bigg\}
\label{Ts_nonfim}
\end{eqnarray}
with the functions $F^a_i({\mathbb Z})\equiv
F^a_i(t,\zeta;q,\kappa t{\cal P}),\,i=1,\ldots,5,$ being obtained
from Eqs.~(\ref{F1}) -- (\ref{F5}) thereby making use of equalities (\ref{+}).

Let us first consider $F^a_1$ and $F^a_2$ which survive in the forward
case:
\begin{eqnarray}
\label{F1x}
F^a_1(t,\zeta)
  & \equiv &
  \Bigg[\Phi_a(t,\zeta)
  \pm \frac{\kappa t{\cal P}^2}{[(q{\cal P})^2-q^2{\cal P}^2]^{1/2}}
  \int_0^1\frac{d\tau}{\tau^2}
  \Phi_a\Big(\frac{t}{\tau}, \zeta \Big)
  + \frac{\kappa^2 t^2 [{\cal P}^2]^2}{(q{\cal P})^2-q^2{\cal P}^2}
  \int_0^1\frac{d\tau}{\tau^2}
  \int_0^1\frac{d\tau_1}{\tau^3_1}
  \Phi_a\Big(\frac{t}{\tau_1 \tau}, \zeta \Big)\Bigg]\,,
\nonumber\\
\label{F2x}
F^a_2(t,\zeta)
  & \equiv &
  \Bigg[\Phi_a(t,\zeta)
  \pm \frac{3\kappa t{\cal P}^2}{[(q{\cal P})^2-q^2{\cal P}^2]^{1/2}}
  \int_0^1\frac{d\tau}{\tau^2}
  \Phi_a\Big(\frac{t}{\tau},\zeta\Big)
  + \frac{3\kappa^2 t^2 [{\cal P}^2]^2}{(q{\cal P})^2-q^2{\cal P}^2}
  \int_0^1\frac{d\tau}{\tau^2}
  \int_0^1\frac{d\tau_1}{\tau^3_1}
  \Phi_a\Big(\frac{t}{\tau_1 \tau},\zeta\Big)\Bigg] \,,
\nonumber
\end{eqnarray}
where the sign $\pm$ is related to the sign of $\xi_\pm$.

Now, let us use the distribution amplitudes (\ref{Phi_tr})
as being introduced for the trace part and extend them to
arbitrary order $n$:
\begin{eqnarray}
\phi_a^{(0)}(t,\zeta) &\equiv& \Phi_a (t,\zeta),
\qquad
\phi_a^{(n)}(t,\zeta) \equiv
    \int_t^1 dy\,\phi_a^{(n-1)}(y,\zeta)
\quad {\rm for} \quad n \geq 1 \,,
 \nonumber
\end{eqnarray}
where the same symbol $\Phi_a$ has been used also after changing
variables $z_+,z_-$ into $t,\zeta$; this leads to
\begin{eqnarray}
\phi_a^{(1)}(t,\zeta) =
     t \int_0^1 \frac{d\tau}{\tau^2}\Phi_a\Big(\frac{t}{\tau},\zeta\Big),
    \qquad 
\phi_a^{(2)}(t,\zeta) =
    t^2 \int_0^1 \frac{d\tau_1}{\tau_1^3}\int_0^1 \frac{d\tau}{\tau^2}
    \Phi_a \Big(\frac{t}{\tau_1\tau},\zeta\Big).
    \nonumber
\end{eqnarray}
Therefore, the contribution of $F^a_1$ and $F^a_2$ to the symmetric
non-forward Compton amplitude is given by
\begin{align}
 {\rm Im}\,^0 T^\twz_{\{\mu\nu\}\,\pm}\kln{q}
 =&\;
 \frac{- \pi}{2\kappa^2} \int d\zeta \int dt
  \frac{q^2}{[(q{\cal P})^2-q^2{\cal P}^2]^{1/2}}
 \delta(t-\xi_\pm) \frac{q{\cal K}_a}{q\cal P}\bigg[ 
 \nonumber \\
& \qquad
 g_{\mu\nu}^{\mathrm T}\;
 \Big(
 \phi_a^{(0)}(t,\zeta)
 \pm
 \frac{\kappa {\cal P}^2}{[(q{\cal P})^2-q^2{\cal P}^2]^{1/2}}
 \phi_a^{(1)}(t,\zeta)
+
 \frac{\kappa^2\,[{\cal P}^2]^2}{(q{\cal P})^2-q^2{\cal P}^2}
 \phi_a^{(2)}(t,\zeta)
 \Big)
 \nonumber \\
& \qquad
 +\frac{q^2\,{\cal P}_\mu^{\mathrm T}{\cal P}_\nu^{\mathrm T}}
 {(q{\cal P})^2-q^2{\cal P}^2}\,
 \Big(
 \phi_a^{(0)}(t,\zeta) \pm \frac{3 \kappa{\cal P}^2}
 {[(q{\cal P})^2-q^2{\cal P}^2]^{1/2}} \phi_a^{(1)}(t,\zeta)
 +
\frac{3\kappa^2\,[{\cal P}^2]^2}{(q{\cal P})^2-q^2{\cal P}^2}
\phi_a^{(2)}(t,\zeta) \Big)\bigg].
\label{InfCA}
\end{align}
Comparing Eqs.~(\ref{Ts_fim}) and (\ref{InfCA}), besides the
additional $\zeta-$integration, the structure of that expression
is completely analogous to the corresponding one (\ref{Ts_fim}) of
the forward case (remind $\kappa = 1/2$ and ${\cal P} \Rightarrow
2 P$),
\begin{eqnarray}
 \frac{1}{\pi}{\rm Im}\,^0 T^\twz_{{\rm nf}\,\{\mu\nu\}}\kln{q}
 &=&
  2 \int d\zeta\,\frac{q{\cal K}_a}{q\cal P}
  \bigg[
 - g_{\mu\nu}^{\mathrm T}\, \mathcal{W}_1 (x, \eta;\zeta)
 + \frac{{\cal P}_\mu^{\mathrm T} {\cal P}_\nu^{\mathrm T}}{{\cal P}^2}
 \,\mathcal{W}_2(x, \eta;\zeta)\bigg]\,.
 \label{Ts_end}
\end{eqnarray}
 Therefore, the corresponding distribution amplitudes generalize
 to the non-forward case as follows:
\begin{eqnarray}
 2x\,\mathcal{W}_{a\, \rm L}(x,\eta;\zeta)
 &=& - \frac{ x^2 {\cal P}^2}{Q^2}\,x\frac{\pd}{\pd x}
 \bigg(\frac{x}{\xi_\pm}
 \frac{\phi^{(2)}_a(\xi_\pm,\zeta)}{\sqrt{1+x^2 {\cal P}^2/Q^2}}\bigg),
 \label{WL}\\
 \frac{2\,q{\cal P}}{{\cal P}^2}\, \mathcal{W}_{a\,2}(x,\eta;\zeta)
 &=& x^2
 \frac{\pd^2}{\pd x^2}
 \bigg(\frac{x^2}{\xi_\pm^2}
 \frac{\phi^{(2)}_a(\xi_\pm,\zeta)}{\sqrt{1+x^2 {\cal P}^2/Q^2}}\bigg),
\label{W2}\\
 2x\,\mathcal{W}_{a\,1}(x,\eta;\zeta)
 &=&
 (1+x^2 {\cal P}^2/Q^2)
 \frac{2q{\cal P}}{{\cal P}^2}\,\mathcal{W}_{a\,2}(x,\eta;\zeta)
 - 2x\,\mathcal{W}_{a\,\rm L}(x,\eta;\zeta)\,,
 \label{W1}
\end{eqnarray}
having exactly the same structure as Eqs.~(\ref{for1}) --
(\ref{for3}). In fact, the last relation generalizes the (power
corrected) Callan-Gross relation to the non-forward case being
also there of the same shape as in the forward case. This
surprising result, of course, supports our conjecture that it is
an outcome of the structure of the Compton operator (in that
approximation). In principle, also here we could repeat the
remarks at the end of the last Section concerning the derivation
of that result.

Let us take the trace of expressions (\ref{InfCA}) and
(\ref{Ts_end}) and compare both with the result (\ref{FTraceX}) we
find
\begin{align}
 2 \mathcal{V}_{a\,0}(x,\eta; \zeta)
 =
 -\,3\,\mathcal{W}_{a\,1}(x,\eta; \zeta)
 +
(1 +x^2 \mathcal{P}^2/Q^2)\frac{(q{\cal P})}{x\,{\cal P}^2 }
 \,\mathcal{W}_{a\,2}(x,\eta; \zeta)
 = \mathcal{W}_{a\,L}(x,\eta; \zeta) -\, 2\,\mathcal{W}_{a\,1}(x,\eta; \zeta) \,,
\end{align}
which relates $\mathcal{V}_{a\,0}$ to $\mathcal{W}_{a\,i}$.

Now, we are in a position to discuss the remaining parts of the
symmetric Compton amplitude which in the forward case vanish, i.e.
the contributions of the last two lines of Eq.~(\ref{Ts_nonfim}).
Making use of Eqs.~(\ref{+}) and the conventions leading to
(\ref{InfCA}) for the structure functions (\ref{F3}) -- (\ref{F5})
we obtain:
\begin{eqnarray}
\label{F3x}
F^a_3(t,\zeta)
   &\equiv&
  -\int_0^1\frac{d\tau}{\tau^2}
  \bigg[F_1^a\Big(\frac{t}{\tau},\zeta\Big)
  + \frac{(q{\cal P})^2}{(q{\cal P})^2-q^2{\cal P}^2}\,
    F_2^a\Big(\frac{t}{\tau},\zeta\Big)\bigg]
  \nonumber \\
&&  + \frac{2(q{\cal P})}{[(q{\cal P})^2-q^2{\cal P}^2]^{1/2}}
  \int_0^1\frac{d\tau}{\tau^2}
  \bigg(
  \pm\,\phi_a^{(0)}\Big(\frac{t}{\tau},\zeta\Big)
  +\frac{\kappa \,{\cal P}^2}{[(q{\cal P})^2-q^2{\cal P}^2]^{1/2}} \,
  \phi_a^{(1)}\Big(\frac{t}{\tau},\zeta\Big)
  \bigg)
  \nonumber \\
&&  +\frac{\kappa \,{\cal P}^2}{[(q{\cal P})^2-q^2{\cal P}^2]^{1/2}}
  \bigg(
  \mp\,t\,\phi_a^{(0)}(t,\zeta)
  +  \frac{q{\cal P}}{[(q{\cal P})^2-q^2{\cal P}^2]^{1/2}} \,
  \phi_a^{(1)}(t,\zeta)
  \bigg),
\\
\label{F4x}
F^a_4(t,\zeta)
    &\equiv&
  3 F^a_3(t,\zeta)
  - 2 \frac{q^2{\cal P}^2}{[(q{\cal P})^2-q^2{\cal P}^2]}\,
    F^a_5(t,\zeta)
  \nonumber \\
&& - 2  \frac{\kappa \,{\cal P}^2}{[(q{\cal P})^2-q^2{\cal P}^2]^{1/2}}
  \bigg(
  \mp\,t\,\phi_a^{(0)}(t,\zeta)
  +  \frac{q{\cal P}}{[(q{\cal P})^2-q^2{\cal P}^2]^{1/2}} \,
   \phi_a^{(1)}(t,\zeta)
  \bigg) ,
\\
F^a_5(t,\zeta)
  &\equiv&
  \int_0^1\frac{d\tau}{\tau^2}\,
  F_2^a\Big(\frac{t}{\tau},\zeta\Big)\,.
  \label{F5x}
\end{eqnarray}
Since the expressions in the round brackets already have been
considered in case of the trace part we hold all the ends in our
hands which are necessary to present the wanted part
${\rm Im}\,^1 T_{\{\mu\nu\}}$ of the
non-forward Compton amplitude. After performing the
$t-$integration we obtain
\begin{align}
{\rm Im}\,^1 T^\twz_{\{\mu\nu\}\,\pm}&\kln{q}
 =\,
 \frac{ \pi}{2\kappa^2} \int d\zeta\,
 \times
 \nonumber\\
& \qquad\Bigg\{
 \Big( \frac{q{\cal K}_a}{q{\cal P}}
     - \frac{{\cal P}{\cal K}_a}{{\cal P}^2} \Big)
 g_{\mu\nu}^{\mathrm T}
  \bigg[
  -\int_0^1\frac{d\tau}{\tau^2}
  \bigg(\frac{x}{[1+x^2{\cal P}^2/Q^2]^{1/2}}\,
  F_1^a\Big(\frac{\xi_\pm}{\tau},\zeta\Big)
  + \frac{x}{[1+x^2{\cal P}^2/Q^2]^{3/2}}\,
    F_2^a\Big(\frac{\xi_\pm}{\tau},\zeta\Big)\bigg)
  \nonumber \\
&  \qquad\qquad\quad
+  \int_0^1\frac{d\tau}{\tau^2}
  \frac{2x}{[1+x^2{\cal P}^2/Q^2]}
  \bigg(
  \pm\,\phi_a^{(0)}\Big(\frac{\xi_\pm}{\tau},\zeta\Big)
  +\frac{x\,{\cal P}^2/Q^2}{2[1+x^2{\cal P}^2/Q^2]^{1/2}} \,
  \phi_a^{(1)}\Big(\frac{\xi_\pm}{\tau},\zeta\Big)
  \bigg)
  \nonumber \\
&  \qquad\qquad\quad
+\frac{x^2 \,{\cal P}^2/Q^2}{2[1+x^2{\cal P}^2/Q^2]}
  \bigg(
  \mp\,\xi_\pm\,\phi_a^{(0)}(\xi_\pm,\zeta)
  +  \frac{1}{[1+x^2{\cal P}^2/Q^2]^{1/2}} \,
  \phi_a^{(1)}(\xi_\pm,\zeta)
  \bigg)\bigg]
  \nonumber \\
& \quad\quad
- \Big( \frac{q{\cal K}_a}{q{\cal P}}
     - \frac{{\cal P}{\cal K}_a}{{\cal P}^2} \Big)
  \frac{{\cal P}_\mu^{\mathrm T}{\cal P}_\nu^{\mathrm T}}{{\cal P}^2}
  \frac{x^2 {\cal P}^2/Q^2}{1+x^2{\cal P}^2/Q^2}
  \bigg[
  \frac{3x\,F^a_3(\xi_\pm,\zeta)}{[1+x^2{\cal P}^2/Q^2]^{1/2}}
  + \frac{2x^3{\cal P}^2/Q^2}{[1+x^2{\cal P}^2/Q^2]^{3/2}}\,
  \int_0^1\frac{d\tau}{\tau^2}\,
  F_2^a\Big(\frac{\xi_\pm}{\tau},\zeta\Big)
  \nonumber \\
& \qquad\qquad\quad
  - \frac{x^2 \,{\cal P}^2/Q^2}{[1+x^2{\cal P}^2/Q^2]}
  \bigg(
  \mp\,\xi_\pm\,\phi_a^{(0)}(\xi_\pm,\zeta)
  +  \frac{1}{[1+x^2{\cal P}^2/Q^2]^{1/2}} \,
   \phi_a^{(1)}(\xi_\pm,\zeta)
 \bigg)
 \bigg]
 \nonumber \\
& \quad\quad
 - \Big(\frac{ {\cal K}_{a\mu}^{\mathrm T} {\cal P}_\nu^{\mathrm T}
 + {\cal P}_\mu^{\mathrm T} {\cal K}_{a\nu}^{\mathrm T}}{{\cal P}^2}
 - 2 \frac{q{\cal K}_a}{q{\cal P}}
 \frac{{\cal P}_\mu^{\mathrm T} {\cal P}_\nu^{\mathrm T}}{{\cal P}^2} \Big)
 \frac{x^3{\cal P}^2/Q^2}{[1+x^2{\cal P}^2/Q^2]^{3/2}}
 \int_0^1\frac{d\tau}{\tau^2}\,
  F_2^a\Big(\frac{\xi_\pm}{\tau},\zeta\Big)
\Bigg\}\,.
\end{align}
Here, for simplicity of notation, once more the dependent function
$F_3$ appears. However, the expression $x\,F_3/\sqrt{1+x^2{\cal
P}^2/Q^2}$ (fifth line) exactly coincides with the expression in
angular brackets which multiplies $g_{\mu\nu}^{\mathrm T}$ (lines
two to four). After that observation and, according to the
consideration of the expression for
 ${\rm Im}\,^0 T^\twz_{\{\mu\nu\}\,\pm}\kln{q}$,
  relating $F_1$ and $F_2$ to $\mathcal{W}_{a\,1}(\zeta)$ and
 $\mathcal{W}_{a\,2}(\zeta)$,  also using
 $(1+x^2{\cal P}^2/Q^2)/(x^2 /Q^2)\equiv (\mathcal{P}^\mathrm{T})^2$,
 we end up with the following expression:
\begin{align}
 {\rm Im}\,^1 T^\twz_{\{\mu\nu\}\,\pm}\kln{q}
 &=\,
 2\pi \int d\zeta
 %
 \Bigg\{
 \Big( \frac{q{\cal K}_a}{q{\cal P}}
     - \frac{{\cal P}{\cal K}_a}{{\cal P}^2} \Big)
 \bigg[
 g_{\mu\nu}^{\mathrm T}
 - 3
  \frac{{\cal P}_\mu^{\mathrm T}{\cal P}_\nu^{\mathrm T}}{({\cal P}^{\mathrm T})^2}
  \bigg] \times
  \nonumber\\
&  \qquad\qquad\quad\qquad\qquad
  \bigg[
  \int_0^1\frac{d\tau}{\tau^2}
  \bigg(
   \mathcal{W}_{a\,1}\Big(\frac{\xi_\pm}{\tau},x;\zeta\Big)
  + \frac{Q^2}{x^2{\cal P}^2}\,
   \mathcal{W}_{a\,2}\Big(\frac{\xi_\pm}{\tau},x;\zeta\Big)
   \bigg)
  \nonumber \\
&  \qquad\qquad\quad\qquad\qquad
 + \int_0^1\frac{d\tau}{\tau^2}
  \bigg(
  2 \mathcal{V}_{a\,0}\Big(\frac{\xi_\pm}{\tau},x;\zeta\Big)
 +\frac{x\, {\cal P}^2}{Q^2}\,
  \mathcal{V}_{a\,1}\Big(\frac{\xi_\pm}{\tau},x;\zeta\Big)
  \bigg)
+ \frac{x\, {\cal P}^2}{2\,Q^2}\,
  \mathcal{V}_{a\,1}(\xi_\pm,x;\zeta)
  \bigg]
  \nonumber \\
& \quad\qquad\qquad
 + 2\Big( \frac{q{\cal K}_a}{q{\cal P}}
     - \frac{{\cal P}{\cal K}_a}{{\cal P}^2} \Big)
  \frac{{\cal P}_\mu^{\mathrm T}{\cal P}_\nu^{\mathrm T}}{({\cal P}^{\mathrm T})^2}
  \bigg[
   \int_0^1\frac{d\tau}{\tau^2}\,
   \mathcal{W}_{a\,2}\Big(\frac{\xi_\pm}{\tau},x;\zeta\Big)
  + \frac{x\, {\cal P}^2}{2\,Q^2}\, \mathcal{V}_{a\,1}(\xi_\pm,x;\zeta)
 \bigg]
 \nonumber \\
& \quad\qquad\qquad
 + \Big(\frac{ {\cal K}_{a\mu}^{\mathrm T} {\cal P}_\nu^{\mathrm T}
 + {\cal P}_\mu^{\mathrm T} {\cal K}_{a\nu}^{\mathrm T}}{{\cal P}^2}
 - 2\, \frac{q{\cal K}_a}{q{\cal P}}
 \frac{{\cal P}_\mu^{\mathrm T} {\cal P}_\nu^{\mathrm T}}{{\cal P}^2} \Big)
 \int_0^1\frac{d\tau}{\tau^2}\,
  \mathcal{W}_{a\,2}\Big(\frac{\xi_\pm}{\tau},x;\zeta\Big)
\Bigg\}\,,
\label{TSext}
\end{align}
where $\mathcal{W}_{a\,1}$ and $\mathcal{W}_{a\,2}$ are given by
(\ref{W1}) and (\ref{W2}), and $\mathcal{V}_{a\,0}$ and $\mathcal{V}_{a\,1}$
are given by (\ref{H0}) and (\ref{H1}), respectively.

Taking the trace of this expression the angular bracket in the
first line vanishes and the pre-factors of the $\tau-$integrals in
the last two lines coincide up to sign so that only the term
\begin{align}
2\pi \Big( \frac{q{\cal K}_a}{q{\cal P}}
     - \frac{{\cal P}{\cal K}_a}{{\cal P}^2} \Big)\,
\frac{x {\cal P}^2}{Q^2}\, \mathcal{V}_{a\,1}(\xi_\pm,x;\zeta)
\end{align}
survives as it should be, cf.~Eq.~(\ref{FTraceX}). However, as for
the case of $\mathcal{V}_{a\,0}(\xi_\pm,x;\zeta)$, we are able to
relate $\mathcal{V}_{a\,1}(\xi_\pm,x;\zeta)$ to the structure
functions $\mathcal{W}_{a\,i}(\xi_\pm,x;\zeta)$. First, we observe
that
\begin{eqnarray}
 x\frac{\pd}{\pd x}\phi^{(2)}_a(\xi_\pm,\zeta)
 =
 \mp \frac{\xi_\pm \, \phi^{(1)}_a(\xi_\pm,\zeta)}{[1+x^2{\cal P}^2/Q^2]^{1/2}}
\,.
\end{eqnarray}
Then, according to the definition (\ref{H1}), making repeated use of
Eqs.~(\ref{rel1}) -- (\ref{rel4}), with the help of (\ref{WL}) we get
\begin{eqnarray}
\hspace{-1cm}
 \mathcal{V}_{a\,1}(\xi_\pm,x;\zeta)
 &=&
 x\frac{\pd}{\pd x}\frac{x \phi^{(1)}_a(\xi_\pm,\zeta)}{\sqrt{1+x^2{\cal P}^2/Q^2}}
 =
 \mp x\frac{\pd}{\pd x} \left[
 \frac{x}{\xi_\pm}\, x\frac{\pd}{\pd x} \phi^{(2)}_a(\xi_\pm,\zeta)
 \right]
 \nonumber\\
 \hspace{-1cm}
 &=&
 \mp x\frac{\pd}{\pd x} \left[
 \sqrt{1+x^2{\cal P}^2/Q^2}\,x \frac{\pd}{\pd x}
 +
 \frac{x\,\xi_\pm\,{\cal P}^2/Q^2}{2\sqrt{1+x^2{\cal P}^2/Q^2}}
 \right]
 \frac{x\, \phi^{(2)}_a(\xi_\pm,\zeta)}{\xi_\pm\,\sqrt{1+x^2{\cal P}^2/Q^2}}
 \nonumber\\
 \hspace{-1cm}
 &=&
 \pm \,x\frac{\pd}{\pd x} \frac{2\,Q^2}{\mathcal{P}^2}\left[
 \frac{\sqrt{1+x^2{\cal P}^2/Q^2}}{x}\,W_{a\,L}(\xi_\pm,x;\zeta)
- \frac{x\,\xi_\pm\,{\cal P}^2/Q^2}{2\sqrt{1+x^2{\cal P}^2/Q^2}}
\int_x \frac{\d y}{y^2}\,\mathcal{W}_{a\,L}(\xi_\pm(y),y;\zeta)
 \right]\!,
\end{eqnarray}
so that, finally using Eq.~(\ref{rel1}) again, we find
 \begin{align}
 \frac{x}{2} \frac{\mathcal{P}^2}{Q^2} \mathcal{V}_{a\,1}(\xi_\pm,x;\zeta)
 =
 [1+ \hbox{\large$\frac{1}{2}$}x \xi_\pm {\cal P}^2/Q^2]\,\bigg[&\,
 x\frac{\pd}{\pd x}\, \mathcal{W}_{a\,L}(\xi_\pm,x;\zeta)
 -
 \frac{1-\hbox{\large$\frac{1}{2}$}x\xi_\pm\,{\cal P}^2/Q^2}{[1+x^2{\cal P}^2/Q^2]}
 \mathcal{W}_{a\,L}(\xi_\pm,x;\zeta)
 \nonumber\\
 &
 -
 \frac{x^3 \mathcal{P}^2/Q^2}{[1+x^2{\cal P}^2/Q^2]^2}
 \int_x \frac{\d y}{y^2}\,\mathcal{W}_{a\,L}(\xi_\pm(y),y;\zeta)
 \bigg].
 \end{align}

Also here, like in the case of the antisymmetric part of the
Compton amplitude, we find that those contributions to the
non-forward case, whose kinematic factors will vanish in the
forward limit, are completely determined by the non-forward
generalizations of the forward distribution amplitudes! In
addition, the absorptive part of the symmetric off-forward Compton
amplitude is determined by the two structure functions
$\mathcal{W}_{a\,1}(\xi_\pm,x;\zeta)$ and
$\mathcal{W}_{a\,L}(\xi_\pm,x;\zeta)$ only!

This finishes our consideration of the symmetric part of the Compton
amplitude.

\newpage

%% file: M_Conclusions3.tex
\section{Conclusions}
\renewcommand{\theequation}{\thesection.\arabic{equation}}

\setcounter{equation}{0}


The complete twist-2 part of the Compton amplitude has been
studied including all target mass corrections. Starting from the
well-known expression of the Compton operator in coordinate space
we performed its relatively complicated Fourier transform in a
manner which allows to reveal its intrinsic structure. Whereas in
a previous work \cite{BM01} the Fourier transform is taken for the
matrix elements directly we transformed the operator expression
first and formed the matrix elements afterwards. This procedure
allows besides a clear separation of theoretically different
contributions the consideration of different kinematical
decompositions of the matrix element and, in principle, the
consideration of other external states. As a result we derived a
closed expressions for the Compton amplitude in terms of iterated
generalized parton distribution amplitudes, Eqs.~(\ref{Tas500})
and (\ref{Ts_nonf}) (For the symmetric part the polynomial
contributions which do not influence the absorptive part are
suppressed). This appears to be much better than the expressions
(\ref{OAS2}) and (\ref{OS2}) being completely integrated out.

Specializing to the forward case the generalized Wandzura-Wilczek
relation of Bl\"umlein-Tkabladze \cite{BT99} and the extended
Callan-Gross relation of Georgi-Politzer \cite{GP76} have been
explicitly recovered; the latter is in accordance with a remark by
\cite{BM01}. In addition, we were able to show that both relations
for the imaginary part can be generalized also to the non-forward
case when introducing suitable new parameters $t=z_+ +\eta z_-,\,
\zeta = z_-/t $. In terms of these variables, with $t$ being
replaced by generalized Nachtmann variables $\xi_\pm$, the
non-forward relations between the generalized structure functions
are of the same shape as in the forward case but appear as
superposition of amplitudes for different values of $\zeta$,
cf.~Eqs.~(\ref{Tas7}) and (\ref{Ts_end}). In addition, we were
able to show that those contributions which will vanish in the
forward case are completely determined by the generalizations of
the non-vanishing ones to the non-forward case,
cf.~Eqs.~(\ref{g02}) and (\ref{TSext}), together with the
non-forward expressions for the trace part, Eq.~(\ref{FTraceX}).
Thereby, the relevant structure functions are uniquely expressed
by appropriately defined generalized $n-$th order distribution
amplitudes of $\Phi^{(n)}_a(\xi_\pm,x;\zeta)$, cf.~Eqs.~(\ref{G1})
-- (\ref{G0}) and Eqs.~(\ref{WL}) -- (\ref{W1}), but, even more
interesting, only three independent structure functions, $g_1,
\mathcal{W}_1$ and $\mathcal{W}_L$, determine the absorptive part
of the complete amplitude at twist-2. Let us also emphasize that
analogous structures will appear also for more general matrix
elements, especially those with an additional meson in the final
state \cite{BEGR02}. Namely, the conclusion depends mainly on the
possibility of introducing, among others, a variable $t$ measuring
the collinear momentum. These surprising results are a strong hint
that they follow from a hidden structure of the Compton operator
itself and not from taking its matrix elements.

The Fourier transform of the twist-2 Compton operator has been
performed mainly using its local decomposition and re-summing the
result to its non-local expression. In principle, it is the same
procedure as has been developed in Ref.~\cite{BM01}. In order to
control this procedure we performed the Fourier transform of the
trace part of Compton amplitude directly and we found coinciding
results. Also the remaining parts of the amplitude could be
transformed in this way but, surely, it will more tedious than in
the considered example. Whether Fourier transforming the local or
the non-local operators -- or its matrix elements -- it turned out
necessarily that the vector $ P_1 z_1+P_2 z_2 $ for any $0 \leq
z_i \leq 1$ must be time-like. This is an important support
restriction for the generalized distribution amplitudes -- which
in the literature, usually, is not mentioned. We have the strong
impression that this condition is not only a mathematical one but
could be of physical relevance. This rests on the observation that
this condition is important for getting the correct sign for the
imaginary part as well as that the generalized Nachtmann variable
also makes sense for non-asymptotic momenta as it should be!

One drawback of our calculation concerns the current conservation.
As known for the twist-2 part of the non-forward Compton amplitude
\cite{BR01}, current conservation is valid only in a very
restricted sense or, hardly speaking, is even not fulfilled. This
is well-known \cite{dWW} and, therefore, also here it has not been
expected. For the forward case current conservation can be read
off directly from the resulting kinematical decomposition. Let us
also remark, that a generalization of the Wandzura-Wilczek
relation for the complete amplitude, as it is the case for the
leading terms \cite{BR01}, we did not obtain (of course, the
consideration of the complete amplitude remained an open problem).
Surely, such a relation exists and may be obtained using
dispersion relations, but this study is outside the present work.

Finally, let us remark that the target mass corrections resulting
from the twist-2 part of the Compton amplitude alone are hardly of
phenomenological relevance. They are to be completed by additional
competing target mass corrections from the higher twist operators.
However, the methods being advocated here, may be applied also to
these objects.

\vspace*{.3cm}

\noindent {\bf \large Acknowledgement}

\noindent The authors gratefully acknowledge many useful
discussions with J. Bl\"umlein; especially we very much benefitted
from him about the mass-corrected WW- and CG-relations in
Refs.~\cite{BT99,GP76}. J.Eilers gratefully acknowledges the
Graduate College "Quantum field theory" at Center for Theoretical
Studies of Leipzig University for financial support. D. Robaschik
thanks for kind hospitality during various stays at Institute of
Theoretical Physics and financial support by Graduate College.

%% file: M_Append3.tex
\begin{appendix}

\section{Fourier transform of propagator and Compton operator}
\renewcommand{\theequation}{\thesection.\arabic{equation}}
\setcounter{equation}{0}

\subsection{Fourier transform of the propagator function; helpful relations}

First, let us give a derivation of the Fourier transformation of the propagator function. We start with the well known result (cf., e.g.,~\cite{BP77}, Eq.~(9.716))
\begin{eqnarray}
\int \d^4 x\, \e^{iqx}\, \frac{1}{(x^2 -\im\epsilon)^\lambda}
=
-\,\im \pi^2\,\frac{\Gamma(2-\lambda)}{\Gamma(\lambda)}
\frac{4^{2-\lambda}}{(q^2 + \im\epsilon)^{2-\lambda}}\,.
\end{eqnarray}
Applying $(-\im\, u \pd_q)^n$ on both sides, one obtains
\begin{eqnarray}
\int \d^4 x\, \e^{iqx}\, \frac{(ux)^n}{(x^2 -\im\epsilon)^\lambda}
&=&
-\,\im^{n+1} \pi^2\,\frac{\Gamma(n+1)}{\Gamma(\lambda)}
\frac{4^{2-\lambda+n/2}}{(q^2 + \im\epsilon)^{n+2-\lambda}}
\sum_{s=0}^{\left[\frac{n}{2}\right]}
\frac{\Gamma(n+2-\lambda-s)}{\Gamma(s+1)\Gamma(n-2s+1)}
\left(-\frac{q^2}{4}\right)^s (u^2)^s (uq)^{n-2s}
\nonumber\\
&=&
-\,\im^{n+1} \pi^2\,\frac{\Gamma(n+1)\Gamma(2-\lambda)}{\Gamma(\lambda)}
\frac{4^{2-\lambda+n/2}}{(q^2 + \im\epsilon)^{n+2-\lambda}}
\left(\hbox{\large$\frac{1}{2}$}\sqrt{u^2\,q^2}\right)^n
C_n^{2-\lambda}\left(\frac{uq}{\sqrt{u^2\,q^2}}\right)
\nonumber\\
&=&
-\,\im^{n+1} \pi^2\,\frac{\Gamma(n+1)}{\Gamma(\lambda)}
2^{2-\lambda}\h_n^{2-\lambda}(u,q)
\,.
\end{eqnarray}

Obviously, taking $\lambda =2$ one gets Eq.~(\ref{f1}); then,
applying $-\im\,\pd_q^\alpha$ one gets Eq.~(\ref{f2}).
Thereby we have used the first of the following collection of various
relations which are obeyed by the generalized functions $\h_n^\nu(u,q)$:
\begin{eqnarray}
 \partial_\alpha^u \h_n^\nu
&=&
 q_\alpha \h_{n-1}^{\nu+1} - u_\alpha \h_{n-2}^{\nu+1}\,,
\label{h1}\\
 \partial_\alpha^q \h_n^\nu
&=&
 - q_\alpha \h_{n}^{\nu+1} + u_\alpha \h_{n-1}^{\nu+1}\,,
\label{h2}\\
 \kln{uq} \, \h_n^\nu
&=&
 \kln{n+1} \, \h_{n+1}^{\nu-1} + u^2 \, \h_{n-1}^\nu\,,
\label{h5}\\
 2\kln{n+\nu} \, \h_n^{\nu}
&=&
 q^2\,\h_{n}^{\nu+1} - u^2 \, \h_{n-2}^{\nu+1}\,,
\label{h6}
\\
\Box_u \h_n^{\nu} &=& 2(\nu -1)\h_{n}^{\nu+1}\\
\left( q \U^{n+1}\right) \, \h^1_n &=&
\frac{q^2}{2} \, \h^1_{n+1}\,,
\label{u4} \\
 {2\kln{n+\nu}}\,\U_\beta^{n+\nu} \, \h_n^\nu
&=&
 q^2 u_\beta \, \h^{\nu+1}_n - u^2 q_\beta \, \h^{\nu+1}_{n-1}\,,
\label{u6}\\
 \partial_\beta^q \, q^2 \, \h^\nu_n
&=&
2\kln{n+\nu-1}
 \left(\U_\beta^{n+\nu-1} \, \h^\nu_{n-1} - q_\beta \, \h^\nu_n\right)
 \qquad \quad \text{for} \qquad  n \geq 1\,.
 \label{u7}
\end{eqnarray}

In addition, the following integral representations of some
fractions will be used below:
\begin{alignat}{2}
\frac{1}{n+1} &= \int_0^1 \d\tau \; \tau^n\,,&&
\label{tau1}\\
\frac{1}{\kln{n+1}^2} &= \int_0^1 \d\tau_1 \int_0^1 \d\tau_2 \;
 \kln{\tau_1 \tau_2}^n =
 - \int_0^1 \d\tau \; \ln\tau \;\, \tau^n
\quad \,,&&
\label{tau2}\\
\frac{1}{n\kln{n+1}} &= \int_0^1 \frac{\d\tau_1}{\tau_1} \int_0^1
\d\tau_2 \; \kln{\tau_1 \tau_2}^n = \int_0^1 \frac{\d\tau}{\tau}
\; \kln{ 1-\tau  } \;\, \tau^{n}& \qquad
\text{for}& \quad
n\geq 1\,,
\label{tau3}\\
\frac{1}{n\kln{n+1}^2} &= \int_0^1 \frac{\d\tau_1}{\tau_1}\int_0^1
\d\tau_2  \int_0^1 \d\tau_3 \; \kln{\tau_1 \tau_2 \tau_3}^n =
\int_0^1 \frac{\d\tau}{\tau} \; \kln{ 1-\tau + \tau\ln\tau } \;\,
\tau^{n}& \qquad \text{for}& \quad n\geq 1\,.
\label{tau4}
\end{alignat}

\subsection{Fourier transformation of the Compton operator}
Let us first perform the
required Fourier transformation of the unsymmetrized Compton
operator. Thereby, in order to make the similarity between the
vector and axial vector case obvious, only the unsymmetrized
operators have been written, but the final (anti)symmetrized
expressions can easily be obtained due to the $\kappa-$dependence.
We also omit the subscript $(5)$ due to $\gamma_5$.

Proceeding in the same manner as for the trace part we obtain in
the first step:
\begin{eqnarray}
\hat T^{\twz}_{\alpha\beta}(q)&:=&
 \int \frac{\d^4 \! x}{2\pi^2} \; \e^{\im qx} \frac{x_\alpha}{\kln{x^2
 - \im\epsilon}^2} \; O_\beta^\twz\kln{\kappa x, - \kappa x}
 = \sum_{n=0}^\infty \frac{\kln{\im\kappa}^n}{n!} \int \frac{\d^4 \! x}{2\pi^2} \;
 \e^{\im qx} \frac{x_\alpha}{\kln{x^2 - \im\epsilon}^2} \,
 O_{\beta n}^\twz\kln{x}
\nonumber\\
&=&\sum_{n=0}^\infty \frac{\kln{\im\kappa}^n}{\kln{n+1}!}
 \int \frac{\d^4 \! x}{2\pi^2} \; \e^{\im qx} \frac{x_\alpha}{\kln{x^2 - \im\epsilon}^2} \;
 \partial_\beta^x \; H_{n+1}\kln{x^2,\Box_x}
 \int \D^4 u\; O_\rho\!\kln{u} \; x^\rho \kln{ux}^n
\nonumber\\
 &=& \sum_{n=0}^\infty \frac{\kln{\im\kappa}^n}{\kln{n+1}!}
 \int \D^4 u \; O_\rho\!\kln{u} \; \partial^\rho_u \;
 H_{n+1}\kln{u^2,\Box_u} \; u_\beta \int \frac{\d^4 \! x}{2\pi^2}
\; \e^{\im qx} \frac{x_\alpha \; \kln{ux}^n }{\kln{x^2 -
\im\epsilon}^2}\,,
\nonumber
\end{eqnarray}
{then, the result (\ref{f2}) of the Fourier transformation
is taken and in the next line the relations
(\ref{u3}), (\ref{h4}) and (\ref{h6}) are used, thereby writing
the contribution for $n=0$ separately,}
\begin{eqnarray}
 &=& \frac{1}{2} \sum_{n=0}^\infty \frac{\kln{-\kappa}^n}{n+1}
 \int \D^4 u \; O_\rho\!\kln{u} \; \partial^\rho_u \;
 H_{n+1}\kln{u^2,\Box_u} \; u_\beta \kln{q_\alpha \mathbf{h}_n^1 -
 u_\alpha \mathbf{h}_{n-1}^1 }
\nonumber\\
& =& \frac{1}{2} \sum_{n=0}^\infty \frac{\kln{-\kappa}^n}{n+1}
 \int \D^4 u \; O_\rho\!\kln{u} \; \partial^\rho_u \KLn{
\U^{n+1}_\beta \kln{q_\alpha \, \h^1_n - \U^n_\alpha \, \h^1_{n-1}
} }
\nonumber\\
& =& \frac{1}{4} \int \D^4 u \; O_\rho\!\kln{u}\;
 \partial^\rho_u \, \partial_\alpha^q  \,
 \kln{2 \, u_\beta \, \ln\kln{q^2/q^2_0}
 - \sum_{n=1}^\infty \frac{\kln{-\kappa}^n}{n\kln{n+1}} \; q^2 \;
 \U^{n+1}_\beta \h^1_n}\,,
\nonumber
\end{eqnarray}
and finally, the derivation w.r.t. $q^\alpha$ is done observing (\ref{u6}) and (\ref{h2}),
\begin{eqnarray}
\hat T^{\twz}_{\alpha\beta}(q) &=& g_{\beta\rho} \;
\frac{q_\alpha}{q^2} \; \int \D^4 u \; O^\rho\!\kln{u}
  + \frac{1}{8} \sum_{n=1}^\infty \frac{\kln{-\kappa}^n}{n\kln{n+1}^2}
  \int \D^4 u \; O_\rho\!\kln{u} \;
\partial^\rho_u \Big\{q^2 \, q_\alpha u_\beta \kln{q^2 \, \mathbf{h}_n^3
- 4 \, \mathbf{h}_n^2 }
\nonumber \\
 \label{OUNS}
 && \qquad 
 + u^2 q^2 \, u_\alpha q_\beta \, \mathbf{h}_{n-2}^3
 + u^2 q^2 g_{\alpha\beta}\,
 \mathbf{h}_{n-1}^2 -u^2 \, q_\alpha q_\beta \kln{q^2
 \mathbf{h}_{n-1}^3 - 2 \, \mathbf{h}_{n-1}^2 } - \kln{q^2}^2
 u_\alpha u_\beta \, \mathbf{h}_{n-1}^3 \Big\}.
\end{eqnarray}

Now, for the axial vector, let us take the antisymmetric part of
the expression (\ref{OUNS}) also observing the
correct symmetry w.r.t. $\kappa$. Obviously, only the first two
terms in the bracket of Eq.~(\ref{OUNS}) contribute:
\begin{eqnarray}
 \hat T^{\twz}_{[\alpha\beta]}(q) &=&
 \int \frac{\d^4 \! x}{2\pi^2} \; \e^{\im qx} \frac{1
 }{\kln{x^2
 - \im\epsilon}^2} \; x_{[\alpha}
 \left(O_{\beta]}^{5\,\twz} \kln{\kappa x, - \kappa x}
 +
 O_{\beta]}^{5\,\twz} \kln{- \kappa x, \kappa x}\right)
\nonumber\\
 &=&
\frac{q^2}{4} \sum_{n=0}^\infty \frac{\kappa^n
\kln{1+(-1)^n}}{\kln{n+1}^2} \int \D^4 u \; O_{5\,\rho}\!\kln{u}
\;
\partial^\rho_u \; \KLn{ q_{\kleo{\alpha}}u_{\okle{\beta}}  \; \h^2_n }
\nonumber\\
&=&  {q^2} \int_0^1 \d\tau_1 \int _0^1 \d\tau_2 \int \D^4 u
\;\left( O_{5\rho} \kln{{u}} + O_{5\rho} \kln{{-u}}\right) \,
\partial^\rho_u \kln{ \frac{q_{\kleo{\alpha}}u_{\okle{\beta}}}{
[(q + \kappa\tau_1\tau_2u)^2 + i\epsilon]^2 }},
    \nonumber
\end{eqnarray}
where in the second line we used relation (\ref{h6}), and in the
last line we used relations (\ref{tau2}) and (\ref{h7}); in
addition, we have taken the symmetry w.r.t. $u$ instead of
$\kappa$. This proves our equation (\ref{OAS1}).

Now, we consider the symmetric part of the expression
(\ref{OUNS}) thereby multiplying already with
${S_{\mu\nu|}}^{\alpha\beta}$. This leads to the following result
\begin{eqnarray}
&&
  {S_{\mu\nu|}}^{\alpha\beta} \int \frac{\d^4 \! x}{2\pi^2} \;
  \e^{\im qx} \frac{1}{\kln{x^2 - \im\epsilon}^2} \,\, x_{\alpha}
O_{\beta}^\twz\kln{\kappa x, -\kappa x}
\nonumber\\
&&
 =
 \left[ 2\,g_{\rho\klso{\mu}} q_{\okls{\nu}} - g_{\mu\nu}\,q_\rho \right]
 \,\frac{1}{q^2} \,\int \D^4 u \; O^\rho\kln{u}
  -  \frac{q^2}{4}\sum_{n=1}^\infty \frac{\kln{-\kappa}^n}{n\kln{n+1}^2}
  \int \D^4 u \; O^\rho\kln{u}
\nonumber\\
&&
 \quad \times \; \partial_\rho^u \; \kls{ \kln{ g_{\mu\nu}
 - \frac{q_\mu q_\nu}{q^2} } \Big( 2 \, u^2 \, \h^2_{n-1}
+  \left[(uq)^2 - u^2 q^2 \right] \h^3_{n-1} \Big)  + q^2 \kln{
u_\mu - \frac{q_\mu \kln{uq}}{q^2} } \kln{ u_\nu - \frac{q_\nu
\kln{uq}}{q^2} } \h^3_{n-1} }
\nonumber\\
&&
 = \KLn{ 2 \, g_{\rho\klso{\mu}} q_{\okls{\nu}} - g_{\mu\nu} \, q_\rho } \,
 \int \D^4 u \; O^\rho\kln{u}
 + \, 2 \, \int_0^1 \frac{\d\tau}{\tau} \; \kln{1-\tau
 + \tau \ln\tau} \int \frac{\D^4 u}{\kappa^4\tau^4} \; O^\rho
\kln{\frac{u}{\kappa\tau}}
\nonumber\\
&&
 \quad \times \; \partial_\rho^u \, \frac{2}{[(q + u)^2 + i\epsilon]^3} \;
 \kls{ \kln{ q^2 \, g_{\mu\nu} - q_\mu q_\nu }  \,
 \KLn{ \left[(u q)^2- u^2q^2\right]  + \hbox{\large$\frac{1}{2}$} u^2 (q + u)^2  }
 + \,  \kln{ u_\mu {q^2}- {q_\mu \kln{uq}} }
       \kln{ u_\nu {q^2}- {q_\nu \kln{uq}} }}\,.
        \nonumber
\end{eqnarray}
For the first equality 
relations (\ref{h5}) and (\ref{h6}) are used, and for the second
equality, after shifting the summation according to $n \rightarrow n+1$,
the relations (\ref{tau4}) and (\ref{h7}) have been used.
This proves our equation (\ref{OS1}).

\section{Fourier transform of the trace part: Nonlocal }
\renewcommand{\theequation}{\thesection.\arabic{equation}}
\setcounter{equation}{0}

The aim of this Appendix is to show, at least for the simple case
of the trace of the Compton amplitude with mass corrections, that
the Fourier transform, as has been performed in the Section III.B
by using the local expansion, i.e., matrix elements of local
operators, really coincides with the Fourier transform performed
by using the nonlocal expression, as has been given in
Ref.~\cite{GLR01}, thereby avoiding the explicit use of infinite
sums whose convergence has not been proved.

The trace of the Compton amplitude is given by the following
expression:
\begin{eqnarray}
\langle P_2,S_2 |\hat T^{\twz}_{\rm trace}(q)|P_1,S_1 \rangle
 &=&
 -\,2\im \int \frac{\d^4 \! x}{2\pi^2} \; \frac{\e^{\im qx}}{\kln{x^2 - \im\epsilon}^2} \;
  \,\langle P_2,S_2 |\left(O^\twz\kln{\kappa x, - \kappa x} -
 O^\twz\kln{-\kappa x,  \kappa x}\right)|P_1,S_1 \rangle
 \nonumber\\
&=& -\,2\int D {\mathbb Z} \Phi_a({\mathbb Z}, \mu^2) K^a_\nu({\mathbb
P},{\mathbb S}) \int \frac{\d^4 \! x}{2\pi^2} \; \frac{\e^{\im
qx}}{\kln{x^2 - \im\epsilon}^2} \;
 \Big[x^{\nu}(2 + \Pi \partial_{\Pi})-
 \hbox{$\frac{1}{2}$} \im \Pi^\nu x^2\Big]
(3 + \Pi \partial_{\Pi})
\nonumber\\
& & \sqrt{\pi} \Big(\sqrt{(x\Pi)^2 - \Pi^2 x^2}\Big)^{-3/2}
J_{3/2}\Big(\hbox{$\frac{1}{2}$} \sqrt{(x\Pi)^2-\Pi^2 x^2}\Big)
e^{\im x\Pi/2 },
\end{eqnarray}
where in the first line, Eq.~(\ref{T_as0}) of Section IV, the
scalar twist-2 operator $ O^{\twz} \equiv x^{\alpha} O^{\twz}_\alpha $
has been introduced  and, in the second line, for the
matrix element $\langle P_2,S_2 |\left(O^\twz\kln{\kappa x, -
\kappa x} - O^\twz\kln{-\kappa x,  \kappa x}\right)|P_1,S_1
\rangle$ we used the expression (2.19) of Ref.~\cite{GLR01}, cf.~also
Eq.~(\ref{NLO}) and observe $x\pd \sim t \pd_t$. In
addition, we used the abbreviations (\ref{abk1}) -- (\ref{abk3}),
e.g.,
\begin{eqnarray}
&& \Pi_\mu = \kappa {\mathbb{P_\mu Z}}, \qquad {\mathbb{PZ}} = P_1
z_1 + P_2 z_2,
\nonumber\\%
&& \Phi_a ({\mathbb Z}, \mu^2) = f^{\twz}_a({\mathbb Z}, \mu^2) -
f^{\twz}_a(-{\mathbb Z}, \mu^2), \nonumber
\end{eqnarray}
as well as replaced $x\partial_x \rightarrow \Pi \partial_{\Pi}$;
$K^a_\nu({\mathbb P},{\mathbb S})$
denote, as the in previous sections, kinematical factors
$\overline u(P_2, S_2) \gamma_\nu u(P_1,S_1)$, $\overline u(P_2,
S_2) \sigma_{\nu \mu}p^\mu u(P_1, S_1)$ or others arising in the
kinematical decomposition of the matrix element.

Already here we can see that there might occur some difficulties.
Namely, wanting to avoid exponential growing of this expression
the argument of the Bessel function $J_{3/2}(\sqrt{(x\Pi)^2- \Pi^2
x^2}/2)$ must be real. For arbitrary values of $x$ this can be
fulfilled for time-like ${\mathbb{P Z}}$ only. This requirement
could lead to a strong support restriction for the generalized
parton distribution $\Phi_a({\mathbb Z}) $. However, the Bessel
function results from an infinite summation and in application to
Compton scattering only a finite number of powers of $x^2$ might
be important, so this conclusion is possibly to restrictive.

For time-like ${\mathbb{P Z}}$ we choose, without restriction of
generality, a special Lorentz frame, ${\Pi} = (\Pi_0 \equiv \kappa
p, 0, 0, 0)$. With this assumption all calculations can be
performed straightforward.

To proceed with the Fourier transformation we use the following
representations,
\begin{eqnarray}
      \frac{i}{4\pi^2} \frac{1}{x^2 - i\epsilon} &=&
      \int \frac{d^4k}{(2\pi)^4} e^{-ikx} \frac{1}{k^2 +i\epsilon};
      \\
      \frac{1}{2\pi^2} \frac{x^{\nu}}{(x^2 - i\epsilon)^2} &=&
      \int \frac{d^4k}{(2\pi)^4} e^{-ikx} \frac{k^{\nu}}{k^2 +i\epsilon};
\end{eqnarray}
\begin{eqnarray}
\hspace{-.8cm} \sqrt{\pi}\, {\rho}^{-3/4}
J_{3/2}\Big(\hbox{$\frac{1}{2}$}\sqrt{\rho}\Big) &=& \int
d^4\ell\, e^{-\im \ell x} \left[\int \frac{d^4 x'}{(2\pi)^4}\,
e^{\im \ell x'} \frac{1}{8 \Gamma(2)}\int_{-1}^1 d\tau (1 -
\tau^2) \exp\big\{\im \tau \sqrt{\rho'}/2\big\}\right]
 \nonumber\\
&=& \frac{1}{\pi (\Pi^2)^{3/2}}\int d^4 \ell\, e^{-\im \ell x}\,
\delta(\ell_0)\,
        \Theta\left(\hbox{$\frac{1}{4}$}\Pi^2 - {\vec{\ell}}^{\,\,2}\right),
\\
{\rm with} \qquad \rho &: =& (x\Pi)^2- \Pi^2 x^2 = (\kappa p)^2\,
{\vec{x}}^{\,\,2}\geq 0,
\end{eqnarray}
where the first two are the well-known Fourier transforms of these
distributions; the third relation is obtained by using the Poisson
integral representation for the Bessel function (cf.,
Ref.~\cite{PBM}, Eq. II.7.12.7), performing the corresponding
Fourier transformation which, in the last line, is inverted. This
is obtained in two steps, namely, using
\begin{eqnarray}
\int \frac{d^4 x}{(2\pi)^4}\, e^{\im \ell x} \exp\big\{\im \tau
|\kappa p| \sqrt{{\vec{x}}^{\,\,2}}/2\big\} =
\frac{\im\delta(\ell_0)}{|\vec{\ell}\,|(\pi \kappa p)^2} \left\{
\frac{1}{(\tau + 2 |\vec{\ell}\,|/|\kappa p|+ \im \epsilon)^2}
 - \frac{1}{(\tau - 2 |\vec{\ell}\,|/|\kappa p|+ \im \epsilon)^2}
\right\},
\end{eqnarray}
and
\begin{eqnarray}
\int_{-1}^1 d\tau \,(1 - \tau^2) \left\{\frac{1}{(\tau
+y+\im\epsilon)^2}-\frac{1}{(\tau-y+\im\epsilon)^2}\right\} =
-\,4\pi \im\, y\, \Theta(1-y)\Theta(1+y) = -\,4\pi \im\, y\,
\Theta(1-y^2).
\end{eqnarray}

Putting together these formulae we obtain for the Compton
amplitude
\begin{eqnarray}
\langle P_2,S_2 |\hat T^{\twz}_{\rm trace}(q)|P_1,S_1
\rangle &=& -\frac{2}{\pi} \int D {\mathbb Z}\, \Phi_a({\mathbb
Z})\, K^a_\nu({\mathbb P},{\mathbb S}) \int d^4x
\,\frac{1}{(2\pi)^4}\int d^4k \left[k^{\nu}(2 + \Pi
\partial_{\Pi})-\Pi^\nu\right] (3 + \Pi \partial_{\Pi})
\nonumber\\
& & \qquad \frac{1}{k^2 + i\epsilon}\int d^4\ell\,
\frac{\delta(\ell_0)}{(\Pi^2)^{3/2}}\,
\Theta\left(\hbox{$\frac{1}{4}$}\Pi^2
-{\vec{\ell}}^{\,\,2}\right)
\exp\left[ix(q-k-\ell+\Pi/2)\right] \\
&=& -\frac{2}{\pi} \int D {\mathbb Z}\, \Phi_a({\mathbb Z})\,
K^a_\nu({\mathbb P},{\mathbb S}) (2 + \Pi\partial_{\Pi}) \Big\{ (3
+ \Pi \partial_{\Pi}) I^{\nu}_1(q,\Pi) - \Pi^\nu I_2(q,\Pi)
\Big\},
\end{eqnarray}
where the following integrals have been introduced:
\begin{eqnarray}
I^{\nu}_1(q,{\Pi}) &=& \int d^4k \int d^4\ell \,\frac{k^{\nu}}{k^2
+i\epsilon}\, \delta(\ell_0)\, \delta^4\left(q-k-\ell+ \Pi/2
\right) \,\Theta\left(\hbox{$\frac{1}{4}$}\Pi^2
-{\vec{\ell}}^{\,\,2} \right) \frac{1}{(\Pi^2)^{3/2}}
\nonumber\\
&=& \int d^3 \ell \frac{(q-\ell+\Pi/2 )^{\nu}} {(q_0
+\sqrt{\Pi^2}/2)^2 -(\vec{q} -{\vec{\ell}}\;)^{2} + i\epsilon}
\,\Theta\left(\hbox{$\frac{1}{4}$}\Pi^2
-{\vec{\ell}}^{\,\,2}\right) \frac{1}{(\Pi^2)^{3/2}}
\nonumber \\
&=& (q^{\nu}+\Pi^{\nu} /2) I_2 +
(q^{\nu}-\frac{q\Pi}{\Pi^2}\,\Pi^\nu)I_3 ,
\\
I_2(q,\Pi) &=& \int d^4k \int d^4\ell\, \frac{1}{k^2 +i\epsilon}\,
\delta(\ell_0)\, \delta^4\left( q-k-\ell+ \Pi /2 \right)\,
\Theta\left( \hbox{$\frac{1}{4}$}\Pi^2 -{\vec{\ell}}^{\,\,2}
\right) \frac{1}{(\Pi^2)^{3/2}}
\nonumber\\
&=& \int d^3 \ell\, \frac{1}{(q_0 + \sqrt{\Pi^2}/2)^2 - (\vec{q} -
\vec{\ell}\;)^2 +i\epsilon }\, \Theta
\left(\hbox{$\frac{1}{4}$}\Pi^2 -{\vec{\ell}}^{\,\,2}\right)
\frac{1}{(\Pi^2)^{3/2}},
\\
I_3 (q,{\Pi}) &=& \frac{-1}{\sqrt{\Pi^2} \left[(q\Pi )^2 -
q^2\,\Pi^2\right]} \int d^3 \vec{\ell} \frac{\vec{\ell}\,
\vec{q}}{(q_0 + \sqrt{\Pi^2}/2)^2 - (\vec{q} -\vec{\ell})^2 +
i\epsilon} \,\Theta \left(\hbox{$\frac{1}{4}$}\Pi^2
-{\vec{\ell}}^{\,\,2}\right). \nonumber
\end{eqnarray}
Concerning the introduction of $I_3$ we observe that only the
projection of $\vec\ell$ onto the direction of $\vec q$, whose
covariant notation reads $q^{\nu} -\Pi^\nu (q\Pi)/\Pi^2$,
contributes to the corresponding integral.
The integral $ I_2$ takes the value
\begin{eqnarray}
I_2(q,\Pi) &=& -\frac{\pi}{\Pi^2} \left\{ 1 - \frac{q\Pi +\Pi^2/2
}{\Pi^2 }\,\ln \frac{(q +\Pi )^2 }{q^2}
+\frac{1}{2} \left(\frac{\sqrt \Delta}{\Pi^2} +\frac{(q +\Pi ) \Pi
}{\Pi^2 }\,\frac{(q \Pi) }{\sqrt \Delta}\right) \ln \frac{q^2 +
q\Pi  + \sqrt\Delta }{q^2 + q\Pi  - \sqrt\Delta} \right\},
\nonumber
\end{eqnarray}
where $\Delta := (q\Pi)^2 - q^2 \Pi^2 = \Pi^2 {\vec q}^{\,\,2}\geq
0$. Now, we observe the following relations:
\begin{eqnarray}
(2 + \Pi \partial_{\Pi})\Big[(3 + \Pi
\partial_{\Pi})(q^{\nu}+\Pi^{\nu} /2) -  \Pi^\nu \Big] I_2(q,\Pi)
&=& (q^{\nu}+\Pi^{\nu} /2)\, (2 + \Pi \partial_{\Pi})(3 + \Pi
\partial_{\Pi})I_2(q,\Pi)
\nonumber\\
(2 + \Pi \partial_{\Pi})(3 + \Pi \partial_{\Pi})I_2(q,\Pi) &=&
\frac{\pi}{(q+\Pi )^2},
\end{eqnarray}
which leads to a comparatively simple expression for the Compton
amplitude in terms of $ I_3$:
\begin{eqnarray}
-\frac{2}{\pi} \int D {\mathbb Z} \Phi_a({\mathbb Z})
K^a_\nu({\mathbb P},{\mathbb S}) \left\{\left(q^{\nu} + \Pi
^{\nu}/2\right)
 \frac{\pi}{(q+\Pi )^2}
+\Big(q^{\nu}-\frac{(q\Pi)}{\Pi^2}\Pi^\nu \Big) (2 + \Pi
\partial_{\Pi})(3 + \Pi \partial_{\Pi})\, I_3(q,\Pi) \right\}
\label{TraceZ}
\end{eqnarray}
For the integral $I_3$  and its derivatives one gets
\begin{eqnarray}
\hspace{-.5cm}
 I_3 (q,\Pi)
 &=&
 \frac{\pi}{2}
 \left\{
 \left(\frac{2}{\Pi^2} +\frac{q^2 + q \Pi}{2\Delta}\right)
 -\frac{2 q\Pi+\Pi^2}{(\Pi^2)^2}\ln \frac{(q+\Pi )^2 }{q^2 }
 \right.
\nonumber\\
& & \hspace{-.5cm} \quad\left. +\frac{1}{\sqrt
\Delta}\left(\frac{1}{(\Pi^2 )^2}\left[\Delta + \left(q\Pi +
 \Pi^2 /2\right)^2\right] -\frac{(q^2 + q\Pi )^2}{4\Delta}
 \right)
\ln \frac{q^2 +q\Pi  - \sqrt\Delta}{q^2 +q\Pi  + \sqrt\Delta}
\right\}
\nonumber\\
\hspace{-.5cm} (2 + \Pi \partial_{\Pi})(3 + \Pi
\partial_{\Pi})I_3(q,\Pi) &=& \frac{\pi\,\Pi^2}{2\Delta} \left\{
\frac{q^2 + q\Pi}{(q+\Pi )^2} -\frac{q^2}{2\sqrt{\Delta}}
\ln\frac{q^2 + q\Pi  + \sqrt\Delta}{q^2 + q\Pi  - \sqrt\Delta}
\right\}.
\end{eqnarray}
After insertion of this expression into (\ref{TraceZ}) one finally
obtains
\begin{align}
&\langle P_2,S_2 |\hat T^{\twz}_{\rm
trace}(q)|P_1,S_1 \rangle =
 -\,2 \int D {\mathbb Z} \Phi_a({\mathbb Z}) K^a_\mu({\mathbb P},{\mathbb S})
\times
\nonumber\\
&   \qquad \qquad \qquad \frac{q^2}{2\,\Delta}\left\{ \Pi ^{\mu}
\left[\frac{(q\Pi)}{2\,\Delta^{1/2}} \ln\frac{q^2 + q\Pi +
\Delta^{1/2}}{q^2 + q\Pi - \Delta^{1/2}} -
\frac{q\Pi+\Pi^2}{(q+\Pi)^2}\right]
- q^{\mu}\left[ \frac{\Pi^2 }{2\,\Delta^{1/2}} \ln\frac{q^2 + q\Pi  +
\Delta^{1/2}}{q^2 + q\Pi  - \Delta^{1/2}} -\frac{q\Pi }{q^2} +
\frac{q\Pi+\Pi^2}{(q+\Pi )^2}\right] \right\}
\nonumber\\
\intertext{}
 & \qquad \qquad
 = -\,q^2 \int D {\mathbb Z} \Phi_a({\mathbb Z}) K^a_\mu({\mathbb P},{\mathbb S})
\frac{\Pi^\mu q^\nu - \Pi^\nu q^\mu}{\Delta}\left\{ \frac{q_\nu +
\Pi_\nu}{(q + \Pi)^2 + i\epsilon} - \frac{q_\nu}{q^2 + i\epsilon}
+ \frac{1}{2\,\Delta^{1/2}} \ln\frac{q^2 + q\Pi +
\Delta^{1/2}}{q^2 + q\Pi - \Delta^{1/2}} \right\}. \label{TRnl}
\end{align}
Reminding relation (\ref{gda0}) for the connection between $u$ and
$\mathbb{PZ}$, one observes that the result (\ref{TRnl}) of
Fourier transforming the nonlocal expression really coincides with
the corresponding result, Eqs.~(\ref{FTrace1}) and (\ref{I_00}) of
Section III, after having replaced $u \rightarrow \kappa u$.
Obviously, the latter method, which is obtained by summing up the
Fourier transforms of the local parts looks much simpler than the
procedure presented in this Appendix.

\section{Computation of antisymmetric part of the Compton amplitude}
 \renewcommand{\theequation}{\thesection.\arabic{equation}}
\setcounter{equation}{0}

In this Appendix we sketch the explicit computation of the antisymmetric
part of the Compton amplitude (\ref{Tas1}):
\begin{eqnarray}
T^\twz_{[\mu\nu]}\kln{q} &=&
- \int_0^1 d\tau \ln\tau
\int D {\mathbb Z} \; %
\Phi_{5a} ({\mathbb Z}, \mu^2)
\epsilon_{\mu\nu}^{\phantom{\mu\nu}\alpha\beta}\, q_{\alpha}\, q^2 %
{\cal K}^a_\rho({\mathbb P},{\mathbb S})\;\partial_{\Pi}^\rho %
\frac{\Pi_\beta}{\big[\left( q + \tau \Pi \right)^2  + \im \epsilon\big]^2} %
\end{eqnarray}
where, in accordance with (\ref{tau2}), we re-formulated the double
integral in Eq.~(\ref{Tas1}) according to
\begin{eqnarray}
\label{double}
 \int_0^1 d\tau_1 \int_0^1 d\tau_2 \,F(\tau_1\tau_2)
 =
 - \int_0^1 d\tau \ln\tau \;F(\tau)\, .
\end{eqnarray}

Again, as in the case of the trace part, in order not to be confronted with the product of two $\delta$--functions, we like to lower the power of the denominator
in the expression (\ref{Tas6}). Furthermore, in order to facilitate the computation we rewrite the differentiation w.r.t. $\Pi^\rho = \kappa{\mathbb{PZ}}^\rho$ as differentiation w.r.t. $\tau$ and $\Pi^2$,
\begin{eqnarray}
    \partial_{\Pi}^\rho
    &=& q^\rho \frac{\partial}{\partial(q\Pi)}
    + 2 \Pi^\rho \frac{\partial}{\partial\Pi^2}
    =
    \frac{q^\rho}{(q\Pi)} \tau \frac{\partial}{\partial\tau}
    - 2\left(\frac{q^\rho}{(q\Pi)} - \frac{\Pi^\rho}{\Pi^2} \right)
    \Pi^2 \frac{\partial}{\partial\Pi^2}\,.
    \label{diffPi}
\end{eqnarray}
Performing the partial $\tau$--integration and, afterwards,
going back to the double integral by using (\ref{double}), we get
\begin{eqnarray}
    T^\twz_{[\mu\nu]}\kln{q}
    &=&
    - \int_0^1 d\tau \ln\tau\, \int D {\mathbb Z} \; %
    \Phi_{5a} ({\mathbb Z},\mu^2)\,
    \epsilon_{\mu\nu}^{\phantom{\mu\nu}\alpha\beta}\,
    \nonumber\\
    && \qquad \times
    \left\{ q_{\alpha}\,{\cal K}^a_\beta({\mathbb P},{\mathbb S})\;
    +
    q_{\alpha}\,\Pi_\beta\, {\cal K}^a_\rho({\mathbb P},{\mathbb S})\;
    \left[
    \frac{q^\rho}{(q\Pi)} \tau \frac{\partial}{\partial\tau}
    - 2\left( q^\rho \frac{\Pi^2}{(q\Pi)}- \Pi^\rho \right)
    \frac{\partial}{\partial\Pi^2}\,
    \right]\right\}
    \frac{q^2}{\big[\left( q + \tau \Pi \right)^2  + \im \epsilon\big]^2}
    \nonumber\\
    &=&\;
    \int D {\mathbb Z} \; %
    \Phi_{5a} ({\mathbb Z},\mu^2) \, %
    \epsilon_{\mu\nu}^{\phantom{\mu\nu}\alpha\beta}\,
    q_{\alpha}\, {\cal K}^a_\beta\;%
    \int_0^1 d\tau_1 \int_0^1 d\tau_2 %
    \frac{q^2}{\big[\left( q + \tau_1\tau_2 \Pi \right)^2  + \im \epsilon\big]^2}
    \nonumber \\
    &\;\;+&
    \int D {\mathbb Z} \; %
    \Phi_{5a} ({\mathbb Z},\mu^2)\,
    \epsilon_{\mu\nu}^{\phantom{\mu\nu}\alpha\beta}\,
    q_{\alpha}\,\Pi_\beta\, \frac{(q{\cal K}^a)}{(q\Pi)}
    \int_0^1 d\tau
    \frac{q^2}{\big[\left( q + \tau \Pi \right)^2  + \im \epsilon\big]^2}
    \nonumber\\
    &\;\;-&
    \int D {\mathbb Z} \; %
    \Phi_{5a} ({\mathbb Z},\mu^2)\,
    \epsilon_{\mu\nu}^{\phantom{\mu\nu}\alpha\beta}\,
    q_{\alpha}\,\Pi_\beta\, \frac{(q{\cal K}^a)}{(q\Pi)}
    \int_0^1 d\tau_1 \int_0^1 d\tau_2 %
    \frac{q^2}{\big[\left( q + \tau_1\tau_2 \Pi \right)^2  + \im \epsilon\big]^2}
    \nonumber \\
    &\;\;-&\!
    2\! \int\! D {\mathbb Z} \; %
    \Phi_{5a} ({\mathbb Z},\mu^2)\,
    \epsilon_{\mu\nu}^{\phantom{\mu\nu}\alpha\beta}\,
    q_{\alpha}\,\Pi_\beta
    \Big( (q{\cal K}^a) \frac{\Pi^2}{(q\Pi)}- (\Pi{\cal K}^a) \Big)
    \frac{\partial}{\partial\Pi^2}
    \int_0^1\!\! d\tau_1\! \int_0^1 \!\! d\tau_2 \,%
    \frac{q^2}{\big[\left( q + \tau_1\tau_2 \Pi \right)^2
    + \im \epsilon\big]^2}\,,
\end{eqnarray}
Now using the integral (\ref{I(0,2)}),
\begin{eqnarray}
    \int_0^1 d\tau \frac{q^2}{\big[R(\tau)  + \im \epsilon\big]^2}
    = \frac{- q^2}{2[(q\Pi)^2 - q^2\Pi^2]}
    \left\{
    \frac{q\Pi + \Pi^2}{R(1)+\im \epsilon} - \frac{q\Pi}{R(0)+\im \epsilon}
    + \Pi^2 \int_0^1 d\tau    \frac{1}{ R(\tau)  + \im \epsilon}
    \right\}\,,
    \nonumber
\end{eqnarray}
as well as
\begin{eqnarray}
    \int_0^1 d\tau_1 \int_0^1 d\tau_2
    \frac{q^2}{\big[ R(\tau_1\tau_2)  + \im \epsilon\big]^2}
    =
    \frac{ - q^2}{2[(q\Pi)^2 - q^2\Pi^2]}
    \int_0^1 \frac{d\tau_1}{\tau_1}
    \left\{
    \frac{q\Pi+\tau_1\Pi^2}
    {R(\tau_1)+\im \epsilon}-\frac{q\Pi}{R(0)+\im \epsilon}
    +  \int_0^1 d\tau_2
    \frac{\tau_1 \Pi^2}{ R(\tau_1\tau_2)  + \im \epsilon }
    \right\}\,,
    \nonumber 
\end{eqnarray}
together with the differentiation w.r.t. $\Pi^2$ of the latter one, we obtain
\begin{eqnarray}
  T^\twz_{[\mu\nu]}\kln{q}
    &=&
    \int D {\mathbb Z} \; %
    \Phi_a ({\mathbb Z},\mu^2)\,
    \epsilon_{\mu\nu}^{\phantom{\mu\nu}\alpha\beta}\,
    \frac{ - q^2}{2[(q\Pi)^2 - q^2\Pi^2]}
    \times
    \nonumber\\
    &&  \left\{
    q_{\alpha}\,{\cal K}^a_\beta\;
    \int_0^1 \frac{d\tau_1}{\tau_1}
    \left(
    \frac{q\Pi + \tau_1\Pi^2}{R(\tau_1)+\im \epsilon} - \frac{q\Pi}{R(0)+\im \epsilon}
    +  \int_0^1 d\tau_2
    \frac{\tau_1 \Pi^2}{ R(\tau_1\tau_2)  + \im \epsilon }
    \right)
    \right.
    \nonumber\\
    && +
    q_{\alpha}\,\Pi_\beta\, \frac{(q{\cal K}^a)}{(q\Pi)}
    \left[\left(\frac{q\Pi + \Pi^2}{R(1)+\im \epsilon} - \frac{q\Pi}{R(0)+\im \epsilon}
    + \Pi^2 \int_0^1 d\tau_1
    \frac{1}{R(\tau_1)  + \im \epsilon}\right)\right.
    \nonumber\\
    && \hspace{2cm}
    - \int_0^1 \frac{d\tau_1}{\tau_1}
    \left.\left(
    \frac{q\Pi + \tau_1\Pi^2}{R(\tau_1)+\im \epsilon} - \frac{q\Pi}{R(0)+\im \epsilon}
    +  \int_0^1 d\tau_2
    \frac{\tau_1 \Pi^2}{R(\tau_1\tau_2)  + \im \epsilon}
    \right) \right]
    \nonumber\\
    && -2 q_{\alpha}\,\Pi_\beta\,
    \left( (q{\cal K}^a) \frac{\Pi^2}{(q\Pi)}- (\Pi{\cal K}^a) \right)
    \times
    \nonumber\\
    && \qquad \left[
    \frac{q^2}{[(q\Pi)^2 - q^2\Pi^2]}
    \int_0^1 \frac{d\tau_1}{\tau_1}
    \left(
    \frac{q\Pi + \tau_1\Pi^2}{R(\tau_1)+\im \epsilon} - \frac{q\Pi}{R(0)+\im \epsilon}
    + \int_0^1 d\tau_2
    \frac{\tau_1 \Pi^2 }{R(\tau_1\tau_2)  + \im \epsilon}
    \right)\right.
    \nonumber\\
    && \qquad +  \int_0^1 d\tau_1
    \left(
    \frac{1}{R(\tau_1)+\im \epsilon}
    + \int_0^1 d\tau_2
    \frac{1}{R(\tau_1\tau_2)  + \im \epsilon}
    \right)
    \nonumber\\
    && \left. \left.
    \qquad -  \int_0^1 d\tau_1
    \left(
    \frac{\tau_1 q\Pi + \tau_1^2\Pi^2}{\big[R(\tau_1)+\im
    \epsilon\big]^2}
    +  \int_0^1 d\tau_2
    \frac{\tau_1^2\tau_2^2 \Pi^2}{\big[ R(\tau_1\tau_2) + \im \epsilon\big]^2}
    \right)\right]
    \right\}.
    \label{TAS2}
\end{eqnarray}
Again, using the integral (\ref{I(1,2)}) together with
\begin{eqnarray}
 \int_0^{\tau_1} d\tau \frac{\tau^2\Pi^2}{[R(\tau)+ \im\epsilon]^2} &=&
 \frac{-1}{2[(q\Pi)^2 - q^2\Pi^2]}
 \left(
 \frac{q^2\, q\Pi + \left(2 (q\Pi)^2 - q^2 \Pi^2\right)\tau_1}{R(\tau_1) + \im \epsilon}
 - \frac{q^2\,q\Pi}{R(0) + \im \epsilon}
 + q^2\,\Pi^2\int_0^{\tau_1} d\tau \frac{1}{R(\tau)+ \im\epsilon}
 \right)\,,
 \nonumber
\end{eqnarray}
for the last line in Eq.~(\ref{TAS2}) we obtain
\begin{eqnarray}
    && \frac{1}{2}\left(\frac{1}{R(1) + \im \epsilon} -
    \int_0^1 d\tau_1 \frac{1}{R(\tau_1)+ \im\epsilon}
    \right)
    \nonumber\\&&
    + \frac{1}{2}\int_0^1 \frac{d\tau_1}{\tau_1}
    \frac{1}{[(q\Pi)^2 - q^2\Pi^2]}\left(
    \frac{q^2\,q\Pi + (q\Pi)^2 \tau_1
    +[(q\Pi)^2 - q^2 \Pi^2]\tau_1}{R(\tau_1) + \im \epsilon}
    - \frac{q^2\,q\Pi}{R(0) + \im \epsilon}
    + q^2\,\Pi^2\int_0^{1} d\tau_2 \frac{\tau_1}{R(\tau_1\tau_2)+ \im\epsilon}
    \right).\nonumber
\end{eqnarray}

At this stage we observe that three different kinds of expressions
occur, namely, those without any $\tau$-integration, with a single
and with a double integral, $\int_0^1 d\tau_1 {R^{-1}(\tau_1)}$
and $\int_0^1d\tau_1 d\tau_2 {R^{-1}(\tau_1\tau_2)}$.
Re-scaling the variable $\Pi$ by $\tau_1\Pi$ and $\tau_1\tau_2\Pi$,
respectively, we are led to the generalized distribution amplitudes,
Eq.~(\ref{GDA}), up to order two.

Putting together the various contributions for the three different
kinematic structures we finally get
\begin{eqnarray}
   T^\twz_{[\mu\nu]}\kln{q}
    &=&
    \frac{1}{2}\int D {\mathbb Z} \,
    \epsilon_{\mu\nu}^{\phantom{\mu\nu}\alpha\beta}\,
    \frac{ - q^2}{[(q\Pi)^2 - q^2\Pi^2]}
    \times
    \nonumber\\
    && \quad \Bigg\{
    q_{\alpha}\,{\cal K}^a_\beta
    \left[\Phi_a^{(1)}({\mathbb Z})
    \left(\frac{q\Pi + \Pi^2}{R(1)+\im \epsilon}
    - \frac{q\Pi}{R(0)+\im \epsilon}\right)
    + \Phi_a^{(2)}({\mathbb Z})\frac{\Pi^2}{R(1)+\im \epsilon}
    \right]
    \nonumber\\
    && \quad
    + q_{\alpha}\,\Pi_\beta\, \frac{(q{\cal K}^a)}{(q\Pi)}
    \left[
    \left(
    \Phi_a^{(0)}({\mathbb Z})-\Phi_a^{(1)}({\mathbb Z})
    \right)
    \left(
    \frac{q\Pi}{R(1)+\im \epsilon} - \frac{q\Pi}{R(0)+\im \epsilon}
    \right)\right.
    \nonumber\\
    && \quad \qquad\qquad\qquad
    -
    \frac{3q^2\Pi^2}{(q\Pi)^2 - q^2\Pi^2}\Phi_a^{(1)}({\mathbb Z})
    \left(
    \frac{q\Pi}{R(1)+\im \epsilon} - \frac{q\Pi}{R(0)+\im \epsilon}
    \right)
    \nonumber\\
    && \quad \qquad\qquad\qquad
    -
    \left.
    \frac{3(q\Pi)^2}{(q\Pi)^2 - q^2\Pi^2}
    \left(
    \Phi_a^{(1)}({\mathbb Z}) + \Phi_a^{(2)}({\mathbb Z})
    \right)
    \frac{\Pi^2}{R(1)+\im \epsilon}
    \right]
    \nonumber\\
    && \quad
    + q_{\alpha}\,\Pi_\beta\,\frac{(\Pi{\cal K}^a)}{\Pi^2}
    \left\{
    \frac{3q^2\Pi^2}{(q\Pi)^2 - q^2\Pi^2}
    \left[
    \Phi_a^{(1)}({\mathbb Z})
    \left(\frac{q\Pi + \Pi^2}{R(1)+\im \epsilon}
    - \frac{q\Pi}{R(0)+\im \epsilon}\right)
    + \Phi_a^{(2)}({\mathbb Z})\frac{\Pi^2}{R(1)+\im
    \epsilon}
    \right]\right.
    \nonumber\\
    && \quad\qquad\qquad\qquad
    +
    \left.\left(
    \Phi_a^{(0)}({\mathbb Z})
    +3 \Phi_a^{(1)}({\mathbb Z})
    +2 \Phi_a^{(2)}({\mathbb Z})
    \right)
    \frac{\Pi^2}{R(1)+\im \epsilon}
    \right\}\Bigg\}\,.
    \label{Tas3}
\end{eqnarray}
This result of the Fourier transform, after some reorganization,
already has been given by Eqs.~(\ref{Tas500}) -- (\ref{Tas50}), where
we already observed that because of the symmetry of the distribution
amplitudes the terms containing $1/R(0)$ vanish.

\section{Computation of the symmetric part of the Compton amplitude}
\renewcommand{\theequation}{\thesection.\arabic{equation}}
\setcounter{equation}{0}

In this Appendix we sketch the explicit computation of the symmetric part of the Compton amplitude (\ref{Ts1}),
\begin{eqnarray}
 T^\twz_{\{\mu\nu\}}\kln{q}
 &=&
 2\int_{0}^1 d\tau \int_0^1 d\sigma \int_0^1 d\rho
 \int D {\mathbb Z} \; \Phi_a({\mathbb Z}, \mu^2) \;
 {\cal K}^a_\rho({\mathbb P},{\mathbb S})\; \left[q^2\right]^3
 \partial_{\widetilde\Pi}^\rho
 \left\{
  \frac{2 A_{\mu\nu}^{\mathrm T}(q,\widetilde\Pi)}{\big[
  \tilde R(\tau) + \im \epsilon \big]^3}
  + \frac{B_{\mu\nu}^{\mathrm T}(q,\widetilde\Pi)}{\big[
  \tilde R(\tau) + \im \epsilon \big]^2}
\right\} , 
\nonumber
\end{eqnarray}
with the abbreviations (\ref{abkS1}) and (\ref{abkS2}). The
integrations over $\tau,\,\sigma$ and $\rho$ will be performed
partially such that finally we get multiple integrals over
$1/[R(\tau) + \im\epsilon]$ alone.

First, let us perform the $\tau-$integration, which gets simplified if the
integration is taken for the range $ -1 \leq \tau \leq 1 $ (thereby the $R(0)-$terms are cancelled):
\begin{align}
T^\twz_{\{\mu\nu\}}\kln{q}
 =&
 \int_0^1 d\sigma \int_0^1 d\rho
 \int D {\mathbb Z} \; \Phi_a({\mathbb Z}, \mu^2) \;
 {\cal K}^a_\rho({\mathbb P},{\mathbb S})\; \left[q^2\right]^3
 \partial_{\widetilde\Pi}^\rho \times
 \nonumber\\
 &
 \left\{
 \frac{A_{\mu\nu}^{\mathrm T}(q,\Pi)}{[(q\Pi)^2 - q^2 \Pi^2]}
 \left[- \left(\frac{q\tilde\Pi + \tilde\Pi^2}{[\tilde R(1) + \im \epsilon]^2}
 + \frac{- q\tilde \Pi + \tilde\Pi^2}{[\tilde R(-1) + \im \epsilon]^2}\right)
 \right.\right.
 \nonumber\\
 &
 \left.\qquad\qquad\qquad\qquad
 + \frac{3}{2}\frac{\Pi^2}{[(q\Pi)^2 - q^2 \Pi^2]}
 \left(\frac{q\tilde\Pi + \tilde\Pi^2}{\tilde R(1) + \im \epsilon}
 +\frac{- q\tilde\Pi + \tilde\Pi^2}{\tilde R(-1) + \im \epsilon}
 + \tilde\Pi^2 \int_{-1}^1\frac{\d\tau}{\tilde R(\tau) + \im \epsilon} \right)
 \right]
 \nonumber\\
 &
 ~~- \frac{B_{\mu\nu}^{\mathrm T}(q,\Pi)}{[(q\Pi)^2 - q^2 \Pi^2]}
 \left.\left(\frac{q\tilde\Pi + \tilde\Pi^2}{\tilde R(1) + \im \epsilon}
 + \frac{- q\tilde\Pi + \tilde\Pi^2}{\tilde R(-1) + \im \epsilon}
 + \Pi^2 \int_{-1}^1\frac{\d\tau}{\tilde R(\tau) + \im \epsilon}
 \right)
 \right\}\,.
 \nonumber
\end{align}

Now, observing
\begin{eqnarray}
\int_0^1 \d\sigma \int_0^1 \d\rho f(\sigma\rho) =
- \int_0^1 \d\tau' \ln\tau' f(\tau')
\label{logint}
\end{eqnarray}
with $\tilde\Pi \equiv \tau'\Pi$ as well as
\begin{eqnarray}
\frac{1}{2}\tau'\frac{\pd}{\pd\tau'}\frac{1}{R(\pm\tau')+ \im\epsilon} =
- \frac{\tilde\Pi^2 \pm q\tilde\Pi}{[R(\pm\tau')+ \im\epsilon]^2}
\qquad {\rm and} \qquad
\int_{0}^1 \d\tau' \ln \tau'\,
\frac{\partial}{\partial\tau'}\,\frac{1}{R(\tau') + \im \epsilon}
=
-\int_{0}^1 \frac{\d\tau'}{\tau'}\frac{1}{R(\tau') + \im \epsilon}\,,
\end{eqnarray}
we can reduce the remaining second power (modulo a surface term
being compensated by the replacement $\tilde\Pi \rightarrow -
\tilde\Pi$ below). Furthermore, taking into account the explicit
form (\ref{abkS1}) of $A_{\{\mu\nu\}}^{\mathrm T}$ and
$B_{\{\mu\nu\}}^{\mathrm T}$ we get
\begin{eqnarray}
T^\twz_{\{\mu\nu\}}\kln{q}
 &=&
 -\frac{q^2}{2}
 \int D {\mathbb Z} \; \Phi_a({\mathbb Z}, \mu^2) \,
\Bigg\{
 \int_{0}^1 \d\tau' \ln \tau'\,
 {\cal K}^a_\rho\, \partial_{\tilde \Pi}^\rho\,\times
 \nonumber\\
& &\bigg[
\bigg(g_{\mu\nu}^{\mathrm T}
+\frac{3}{2}
 \frac{\tilde\Pi_{\mu}^{\mathrm T}\tilde\Pi_{\nu}^{\mathrm T}q^2}
      {(q\tilde\Pi)^2 - q^2 \tilde\Pi^2}
\bigg)
 \frac{\tilde\Pi^2}{(q\tilde\Pi)^2 - q^2 \tilde\Pi^2}
 \bigg(
 \frac{q\tilde\Pi + \tilde\Pi^2}{\tilde R(\tau') + \im \epsilon}
 + \tilde\Pi^2 \int_{0}^1\frac{\d\tau}{\tilde R(\tau\tau') + \im \epsilon}
 \bigg)
 + (\tilde\Pi \rightarrow - \tilde\Pi)
\bigg]
  \nonumber\\
& & -
\int_{0}^1 \d\tau' \,
 {\cal K}^a_\rho\, \partial_{\tilde \Pi}^\rho\,
\bigg[
 \bigg(g_{\mu\nu}^{\mathrm T}+
 \frac{\tilde\Pi_{\mu}^{\mathrm T}\tilde\Pi_{\nu}^{\mathrm T}q^2}
      {(q\tilde\Pi)^2 - q^2 \tilde\Pi^2}
 \bigg)
 \frac{1}{\tilde R(\tau') + \im \epsilon}
 + (\tilde\Pi \rightarrow - \tilde\Pi)
\bigg]
\Bigg\}\,,
 \nonumber
 \end{eqnarray}
and, again, using the expression (\ref{diffPi}) for the derivative w.r.t.
$\Pi$ one obtains
\begin{eqnarray}
T^\twz_{\{\mu\nu\}}\kln{q}
 &=&
 \frac{q^2}{2}
 \int D {\mathbb Z} \, \Phi_a({\mathbb Z}, \mu^2) \,
 \int_{0}^1 \frac{\d\tau'}{\tau'}
\Bigg\{
\bigg[
\Big({\cal K}_{a\mu}^{\mathrm T} \Pi_\nu^{\mathrm T}
 + \Pi_\mu^{\mathrm T} {\cal K}_{a\nu}^{\mathrm T}
 - 2 \frac{q{\cal K}_a}{q\Pi} \Pi_\mu^{\mathrm T} \Pi_\nu^{\mathrm T} \Big)
 \frac{q^2}{(q\Pi)^2 - q^2\Pi^2}
 \nonumber\\
& &\qquad
 +
 \bigg(
 \frac{q{\cal K}_a}{q\Pi}\tau'\frac{\pd}{\pd\tau'}
 -
 2\Big( \frac{q{\cal K}_a}{q\Pi} - \frac{\Pi{\cal K}_a}{\Pi^2} \Big)
 \Pi^2 \frac{\pd}{\pd\Pi^2}
 \bigg)
 \bigg( g_{\mu\nu}^{\mathrm T}
 +
 \frac{\Pi_{\mu}^{\mathrm T}\Pi_{\nu}^{\mathrm T}q^2}{(q\Pi)^2 - q^2\Pi^2}
 \bigg)
 \bigg]
 \frac{1}{R(\tau') + \im \epsilon}
  \nonumber\\
& &
-
\ln \tau'
 \bigg[
 3 \Big({\cal K}_{a\mu}^{\mathrm T} \Pi_\nu^{\mathrm T}
 + \Pi_\mu^{\mathrm T} {\cal K}_{a\nu}^{\mathrm T}
 - 2 \frac{q{\cal K}_a}{q\Pi} \Pi_\mu^{\mathrm T} \Pi_\nu^{\mathrm T} \Big)
 \frac{q^2}{(q\Pi)^2 - q^2\Pi^2}
 \nonumber\\
& &\qquad
 +
 \bigg(
 \frac{q{\cal K}_a}{q\Pi}\tau'\frac{\pd}{\pd\tau'}
 -
 2\Big( \frac{q{\cal K}_a}{q\Pi} - \frac{\Pi{\cal K}_a}{\Pi^2} \Big)
 \Pi^2 \frac{\pd}{\pd\Pi^2}
 \bigg)
 \bigg( g_{\mu\nu}^{\mathrm T}
 +
 3 \frac{\Pi_{\mu}^{\mathrm T}\Pi_{\nu}^{\mathrm T}q^2}{(q\Pi)^2 - q^2\Pi^2}
 \bigg)
 \bigg]\times
 \nonumber\\
& &\qquad\qquad
 \frac{\Pi^2}{(q\Pi)^2 - q^2\Pi^2}
  \left(
 \frac{\tau' q\Pi + \tau'^2\Pi^2}{R(\tau') + \im \epsilon}
 + \tau'^2\Pi^2 \int_{0}^1\frac{\d\tau}{ R(\tau\tau') + \im \epsilon}
 \right)
 + (\Pi \rightarrow - \Pi)
\Bigg\}\,. \nonumber
\end{eqnarray}

Now, the homogeneous derivatives w.r.t.~$\tau'$ and $\Pi^2$ have
to be performed which, despite being straightforward, are quite
tedious. Thereby, the following relations are useful to observe
(eventually modulo surface terms which cancel each other finally):
\begin{align}
\Pi^2\frac{\pd}{\pd\Pi^2} \frac{1}{R(\tau) + \im \epsilon}
 &=
 -\frac{\tau^2 \Pi^2}{[R(\tau) + \im \epsilon]^2}
 =
 \frac{1}{2} \tau \frac{\pd}{\pd\tau}\frac{1}{R(\tau) + \im \epsilon}
 +
 \frac{1}{2} \frac{\tau q\Pi}{[R(\tau) + \im \epsilon]^2}\,,
 \nonumber\\
\int_{0}^1 \d\tau \frac{\tau \Pi^2}{[ R(\tau ) + \im \epsilon]^2}
 &=
 \frac{\Pi^2}{2[(q\Pi)^2 - q^2 \Pi^2]}
 \left(
 \frac{q^2 + q\Pi}{[R(1) + \im \epsilon]}
 +
 q\Pi \int_{0}^1 \frac{\d\tau}{[ R(\tau) + \im \epsilon]}
 \right)\,,
  \nonumber\\
 \int_{0}^{\tau'} \d\tau\,
 \frac{\tau^2 [\Pi^2]^2}{[ R(\tau)+\im\epsilon]^2}
 &=
 -\frac{\Pi^2}{2[(q\Pi)^2 - q^2 \Pi^2]}
 \left(
 \frac{q\Pi(q^2+\tau'\,q\Pi)+\tau'[(q\Pi)^2-q^2\Pi^2]}{R(\tau')+\im\epsilon}
 +
 \tau'\,q^2\Pi^2 \int_{0}^1 \d\tau\frac{1}{[ R(\tau\tau') + \im \epsilon]}
 \right)\,,
 \nonumber\\
 -\int_{0}^1\d\tau {\tau}^2 \ln \tau \frac{\pd}{\pd\tau}&
  \frac{1}{ R(\tau )+\im\epsilon}
 =
 \int_{0}^1 \d\tau\,(1 + 2\ln \tau)
 \frac{\tau}{R(\tau)+\im\epsilon}\,.
 \nonumber
 \end{align}
In addition, replacing the single integrals containing the
logarithm $\ln\tau'$  by double integrals according to relation
(\ref{logint}) we finally get ($\Delta \equiv (q\Pi)^2 -
q^2\Pi^2$)
\begin{align}
T^\twz_{\{\mu\nu\}}\kln{q}
 =&~~
 \frac{q^2}{2}
 \int D{\mathbb Z}\,\Phi_a({\mathbb Z}, \mu^2)\,\frac{q{\cal K}_a}{q\Pi}
 \times
 \nonumber\\
& \qquad \Bigg\{
g_{\mu\nu}^{\mathrm T}
 \bigg[\frac{1}{R(1)+\im\epsilon}
 + \frac{\Pi^2}{\Delta}
   \int_0^1\d\sigma \frac{\sigma (q\Pi+\sigma\Pi^2)}{R(\sigma)+\im\epsilon}
 + \frac{\Pi^2}{\Delta}
   \int_0^1\d\sigma \int_0^1\frac{\d\tau}{\tau}
   \frac{(\sigma\tau)^2\Pi^2}{R(\sigma\tau)+\im\epsilon}
 \bigg]
 \nonumber\\
& \qquad +
 \Pi_{\mu}^{\mathrm T}\Pi_{\nu}^{\mathrm T}\frac{q^2}{\Delta}
 \bigg[
 \frac{1}{R(1)+\im\epsilon}
 + \frac{3\,\Pi^2}{\Delta}
   \int_0^1\d\sigma \frac{\sigma (q\Pi+\sigma\Pi^2)}{R(\sigma)+\im\epsilon}
 + \frac{3\,\Pi^2}{\Delta}
   \int_0^1\d\sigma \int_0^1\frac{\d\tau}{\tau}
   \frac{(\sigma\tau)^2\Pi^2}{R(\sigma\tau)+\im\epsilon}
 \bigg]
 \Bigg\}
 \nonumber\\
& + \frac{q^2}{2}
 \int D{\mathbb Z}\,\Phi_a({\mathbb Z}, \mu^2)\,
 \bigg(\frac{q{\cal K}_a}{q\Pi}+ \frac{\Pi{\cal K}_a}{\Pi^2}\bigg)
 \frac{\Pi^2}{\Delta}\times
 \nonumber\\
& \qquad \Bigg\{
 g_{\mu\nu}^{\mathrm T}
 \bigg[\frac{q^2+q\Pi}{R(1)+\im\epsilon}
 + \int_0^1\d\rho \frac{\rho (q\Pi-\rho\Pi^2)}{R(\rho)+\im\epsilon}
 - \frac{\Pi^2}{\Delta}
   \int_0^1\d\rho\int_0^1\d\sigma \bigg(
   \frac{(\rho\sigma)^2[4(q\Pi)^2-q^2\Pi^2]}{R(\rho\sigma)+\im\epsilon}
   \nonumber\\
&  \qquad\qquad  +
   \frac{\rho\sigma(q\Pi)[2(q\Pi)^2+q^2\Pi^2]/\Pi^2}{R(\rho\sigma)+\im\epsilon}
   \bigg)
   - \frac{\Pi^2}{\Delta}
   \int_0^1\d\rho\int_0^1\d\sigma \int_0^1\frac{\d\tau}{\tau}
   \frac{(\rho\sigma\tau)^2[4(q\Pi)^2-q^2\Pi^2]}
   {R(\rho\sigma\tau)+\im\epsilon}
 \bigg]
 \nonumber\\
& \qquad +
 \Pi_{\mu}^{\mathrm T}\Pi_{\nu}^{\mathrm T} \frac{q^2}{\Delta}
  \bigg[\frac{q^2+q\Pi}{R(1)+\im\epsilon}
 - \int_0^1\d\rho \frac{2q^2-\rho q\Pi +3\rho^2\Pi^2}{R(\rho)+\im\epsilon}
 - \frac{3\Pi^2}{\Delta}
   \int_0^1\d\rho\int_0^1\d\sigma \bigg(
   \frac{(\rho\sigma)^2[4(q\Pi)^2+q^2\Pi^2]}{R(\rho\sigma)+\im\epsilon}
   \nonumber\\
&  \qquad\qquad +
   \frac{\rho\sigma(q\Pi)[2(q\Pi)^2+3q^2\Pi^2]/\Pi^2}{R(\rho\sigma)+\im\epsilon}
   \bigg)
   - \frac{3\Pi^2}{\Delta}
   \int_0^1\d\rho\int_0^1\d\sigma \int_0^1\frac{\d\tau}{\tau}
   \frac{(\rho\sigma\tau)^2[4(q\Pi)^2+q^2\Pi^2]}
   {R(\rho\sigma\tau)+\im\epsilon}
 \bigg] \Bigg\}
 \nonumber\\
& + \frac{q^2}{2}
 \int D{\mathbb Z}\,\Phi_a({\mathbb Z}, \mu^2)\,
 \Big({\cal K}_{a\mu}^{\mathrm T} \Pi_\nu^{\mathrm T}
 + \Pi_\mu^{\mathrm T} {\cal K}_{a\nu}^{\mathrm T}
 - 2 \frac{q{\cal K}_a}{q\Pi} \Pi_\mu^{\mathrm T} \Pi_\nu^{\mathrm T} \Big)
 \frac{q^2}{\Delta}\,\times
  \nonumber\\
&\qquad\qquad \int_0^1\d\rho \bigg[
   \frac{1}{R(\rho)+\im\epsilon}
 + \frac{3\,\Pi^2}{\Delta}
   \int_0^1\d\sigma
   \frac{\rho\sigma (q\Pi+\rho\sigma\Pi^2)}{R(\rho\sigma)+\im\epsilon}
 + \frac{3\,\Pi^2}{\Delta}
   \int_0^1\d\sigma \int_0^1\frac{\d\tau}{\tau}
   \frac{(\rho\sigma\tau)^2\Pi^2}{R(\rho\sigma\tau)+\im\epsilon}
 \bigg]\,.
 \end{align}
Rescaling the distribution amplitudes in the same manner as in the antisymmetric case, one obtains the expression (\ref{Ts_nonf}).

%

\end{appendix}